\pdfoutput=1
\documentclass{elsarticle}
\usepackage{bm, color, graphicx, fullpage}
\usepackage{natbib, nicefrac, ifpdf}
\usepackage{amsmath, amssymb, amsfonts, amsthm}
\usepackage[hidelinks]{hyperref}

\allowdisplaybreaks
\DeclareMathOperator{\sign}{sign}

\def\diff{\ensuremath{\textrm{d}}}
\def\bmx{\ensuremath{\bm{x}}}

\def\bmi{\ensuremath{\bm{i}}}
\def\bmj{\ensuremath{\bm{j}}}
\def\bmk{\ensuremath{\bm{k}}}

\def\OO{\ensuremath{\textrm{O}_2}}

\def\m{\ensuremath{\textrm{m}}}
\def\cm{\ensuremath{\textrm{cm}}}
\def\mm{\ensuremath{\textrm{mm}}}
\def\um{\ensuremath{\mu\textrm{m}}}
\def\ns{\ensuremath{\textrm{ns}}}
\def\GB{\ensuremath{\textrm{GB}}}
\def\ps{\ensuremath{\textrm{ps}}}

\def\kV{\ensuremath{\textrm{kV}}}
\def\C{\ensuremath{\textrm{C}}}

\begin{document}
\begin{abstract}
  We review a scalable two- and three-dimensional computer code for low-temperature plasma simulations in multi-material complex geometries. Our approach is based on embedded boundary (EB) finite volume discretizations of the minimal fluid-plasma model on adaptive Cartesian grids, extended to also account for charging of insulating surfaces. We discuss the spatial and temporal discretization methods, and show that the resulting overall method is second order convergent, monotone, and conservative (for smooth solutions). Weak scalability with parallel efficiencies over 70\% are demonstrated up to 8192 cores and more than one billion cells. We then demonstrate the use of adaptive mesh refinement in multiple two- and three-dimensional simulation examples at modest cores counts. The examples include two-dimensional simulations of surface streamers along insulators with surface roughness; fully three-dimensional simulations of filaments in experimentally realizable pin-plane geometries, and three-dimensional simulations of positive plasma discharges in multi-material complex geometries. The largest computational example uses up to $800$ million mesh cells with billions of unknowns on $4096$ computing cores. Our use of computer-aided design (CAD) and constructive solid geometry (CSG) combined with capabilities for parallel computing offers possibilities for performing three-dimensional transient plasma-fluid simulations, also in multi-material complex geometries at moderate pressures and comparatively large scale.
\end{abstract}

\title{An adaptive Cartesian embedded boundary approach for fluid simulations of two- and three-dimensional low temperature plasma filaments in complex geometries}
\author[RM]{Robert Marskar}  
\ead[RM]{robert.marskar@sintef.no}
\address[RM]{SINTEF Energy Research}

\maketitle

\section{Introduction}
Electrical discharges in gaseous media are of substantial significance in a wide range of applications. In the high-voltage industry, for example, electrical discharges in the form of coronas, streamers, and leaders, can be particularly detrimental and lead to permanent damage on electrical equipment. The damage is usually induced by a hot electrical arc which (i) permanently decomposes the molecular structure of the gas, or (ii) leads to irreparable thermal damage on solid surfaces. For example, arcing may lead to electrode welding, or decomposition of conductive particles on the surfaces of insulators that degrade their performance over time. In other applications, electrical discharges are used as catalysts. Streamers - which are fast transients that evolve due to self-enhanced electric fields at their tips - have found their use in sterilization of polluted gases and liquids by initiating chemical reactions or breaking up molecules \cite{1167639, 18870, 4504897, 55956, Nair2004, Grymonpre2001}. In the aerodynamics community, dielectric barrier discharges are used to control the gas flow in the boundary layer on airplane wings \cite{Boeuf2005, Moreau2007, Soloviev2009}. Excellent reviews on electrical discharges and their applications can be found in \cite{Ebert2006,Bogaerts2002} and references therein. 

Computer simulations of electrical discharges are challenging for several reasons. Firstly, electrical discharges are inherently nonlinear phenomena which evolve transiently with exponential increases in mass densities, and the discharge chemistry is often complicated, requiring inputs of rate constants that may be more or less uncertain. Secondly, electrical discharges can evolve as filamentary plasmas with dynamic ranges that span several orders of magnitude. In air at standard atmospheric conditions, for example, a streamer channel is shielded by a space charge layer that requires a numerical resolution on the micrometer scale, while the streamer head itself moves as an ionization wave which can stretch over several centimeters. Numerical solutions can usually not be obtained on coarse grids due to the strong nonlinearities of the problem, and mesh resolution requirements are often quite extreme. Integration of the equations of motion require concurrent updates of electrodynamic, fluid (or kinetic), and radiative transport equations, possibly even coupled with nonlinear boundary conditions on electrodes and insulators. For this reason, software development of low-temperature plasma codes for large scale simulations is a challenge for researchers and industry alike, and production runs of three-dimensional discharges in engineering-relevant geometries may even be a challenge for todays largest parallel computers. Few results on three-dimensional propagation of streamers have therefore been presented to date. No reported results have so far combined 3D simulations of filamentary plasmas with the geometrical complexities encountered in e.g. applied high-voltage engineering, which can feature multiple types of materials, for example electrodes and insulators, which require separate treatment with respect to boundary conditions.

In three dimensions, all results reported in the scientific literature to date rely on geometrical simplicity, such as plane-parallel electrodes \cite{Pancheshnyi2008, Luque2008, Nijdam2016a, Teunissen2017} or point-plane gaps \cite{Hallac2003, Papageorgiou2011, Plewa2018}. Some computer codes also rely on adaptive mesh refinement (AMR) for reducing the computational effort, a technique which becomes mandatory for large scale 3D simulations. Three-dimensional simulations have been performed by using a code based on Gerris in point-plane gaps \cite{Kolobov2012}. Similar simulations, although on a very small length scale, have also been performed by using COMSOL Multiphysics \cite{7295248}, and an in-house code that was recently reported \cite{Plewa2018}. Interactions between multiple streamers have also recently been investigated \cite{shi2017}. Concept simulations for streamer discharges in transformer oil have also been presented \cite{Jadidian2013, 6145703, Jadidian2014, 5252362}, but the simulations disregard the liquid-gas phase transition (which is observed in virtually all experiments), and must therefore be met with some skepticism. Teunissen and Ebert \cite{Teunissen2017} also report on three-dimensional streamer simulations through atmospheric-pressure nitrogen in plane-parallel $4\,\cm$ gaps. The same authors have also presented fully 3D simulations in $(24\,\cm)^3$ plane-parallel domains for oxygen-nitrogen mixtures at $133\,\textrm{mbar}$ \cite{Nijdam2016a}. Note that for streamer discharges there is an inherent scaling with pressure; at decreasing pressures the physical length and time scales of the discharge increases. Decreased pressure therefore acts as a magnifying glass that allow simulations of larger domains and longer time scales. In our opinion, the simulations in \cite{Nijdam2016a, Teunissen2017} ran quite fast, within 24 hours on two Xeon E5-2680v4 processors. The results clearly demonstrate the benefits of using AMR. 

In the two-dimensional domain (either Cartesian or cylindrically symmetric coordinates), streamer simulations in complex geometries have progressed further. Many of these results arise from in-house codes from the aerodynamics community \cite{Boeuf2005, Moreau2007, Likhanskii2008, Soloviev2009, Soloviev2014}, but commercial and open source codes are also gaining traction. Other researchers have shown interest in engineering and atmospheric applications. For example, by using a static mesh, \citet{Dubinova2016} has investigated streamer propagation along a cylindrical rod by using the Plasimo software \cite{VanDijk2009}, and the results qualitatively agree with experiments. Inception criteria for streamers near large ice particles have also been simulated \cite{Dubinova2015}, as well as various investigations of dielectric barrier discharges \cite{Celestin2009, Jansky2010, Pechereau2012, Pechereau2016}. Temporally adaptive methods have also been reported \cite{Duarte2012}, and show promise. 

Even though simulation capabilities of low-temperature plasmas have increased steadily over the past few decades, quantitative prediction using low-temperature plasma simulations remains challenging, particularly at the fidelity that is required for engineering applications. Some of these challenges are natural and arise due to a lack of physics in the mathematical models, uncertainty in rate constants for chemical reactions, or the inability of the code to scale to the desired length and time scales. Another uncertainty is the fact that computer codes used by various academic groups do not give the same answer to the same problem \cite{Bagheri2018}, demonstrating a broader need for code verification and validation efforts. 

In this paper, we review a parallel two- and three-dimensional computer code for simulations of low-temperature plasmas in complex geometries. Our work uses adaptively refined structured Cartesian grids, and includes support for internal boundary conditions for both insulators and electrodes. We embed our algorithms into Chombo \cite{chombo, ebchombo}, which is a high-performance computing (HPC) framework that runs on systems ranging from desktops to supercomputers. Our current solvers include a multifluid Poisson solver; convection-diffusion-reaction solvers, and diffusion-approximation models for the radiative transport equation (RTE). 

Our computational approach is based on structured adaptive mesh refinement (SAMR) with an embedded boundary (EB) formalism. Structured grids normally lead to good conditioning of the discretized equations and therefore resolve some of the disadvantages that are commonly encountered with 3D unstructured adaptive grids. These include a lack of efficient mesh adaption and convergence loss with skewed elements, which is especially problematic for adaptive 3D unstructured grids near curved boundaries. In addition, unstructured grids use indirect memory addressing for data accessing, often leading to bandwidth limitations of the code. SAMR resolve these issues even for geometrically complex cases, at the cost of a significant increase in algorithm complexity, difficulties which are primarily related to spatial discretization and time stepping near the boundaries. We discuss how we resolve some of these difficulties. 

The structure of this paper is as follows: In Sec.~\ref{sec:theory} we review the minimal plasma model and present the equations of motion and their coupling, and we outline our numerical methods in Sec.~\ref{sec:method}. Most of the methods discussed in Sec.~\ref{sec:method} are known in the literature, and the main algorithmic work involved in this paper is therefore a demonstration of their successful linkage. Code verification studies are presented in Sec.~\ref{sec:verification} and weak scalability studies are presented in Sec.~\ref{sec:weak_scalability} for up to 8192 cores. Demonstrations of our computer code are given in Sec.~\ref{sec:examples}, and some final remarks and a brief summary are given in Sec.~\ref{sec:remarks} and Sec.~\ref{sec:conclusions}. 

\section{Theory}
\label{sec:theory}
Typically, streamers develop in the non-uniform electric field near curved electrodes. Their spatial and temporal development occur due to field driven acceleration of electrons to kinetic energies above the required ionization energy, leading to an avalanche of electrons that leave behind positive ions. For the most part, ions can be modeled as fluids whereas the most accurate electron descriptions are provided by kinetic models (e.g. particle models) that sample the electron phase space distribution. Kinetic models capture many of the physical mechanisms of streamers (e.g. stochastic branching and run-away electrons), but the spatial and temporal resolutions need be extremely fine which limit their applicability to comparatively small spatial and temporal scales. On the other hand, electron fluid models evolve the average electron density (or momentum) without regard to the velocity distribution function (which is generally not Maxwellian) and do not share this practical restriction. In fluid models, the electron evolution is described by its drift and diffusion velocities, and the production rate of electron ion-pairs is described by rate coefficients rather than stochastic evaluation of ionization probabilities. In the simplest case, which is the local field approximation (LFA) where the electrons are in equilibrium with the electric field, the ionization coefficient is simply a function of the electric field alone. More sophisticated physics is possible by using the local energy approximation (LEA) where this coefficient is a function of the average electron energy rather than the field. In either case, when compared to kinetic models the available level of physics in electron fluid models is quite limited. However, electron fluid models are applicable to much larger spatial scales and are therefore useful in the range of applications where kinetic models are unrealistic to apply. In this paper we will consider the minimal fluid model where the electrons and ions are described only by their average densities, and the momentum equations are closed by imposing drift and diffusion velocities to be functions of the electric field. Certainly, improvements to this model are possible by using Euler equations for the electrons, as is done in \cite{Eichwald2006,Dujko2013,Markosyan2013}.

\subsection{Electrodynamic equations}
\label{subsec:electrodynamic}
For a weakly ionized plasma such as a streamer, the electric current is sufficiently small so that one may disregard magnetic field effects. The remaining Maxwell equation for the evolution of the electromagnetic field is Poisson's equation:
\begin{equation}
  \label{eq:poisson}
  \nabla\cdot\left(\epsilon\nabla\Phi\right) = -\frac{\rho}{\epsilon_0},
\end{equation}
where $\Phi$ is the electric potential, $\epsilon$ the relative permittivity, and $\rho$ the charge density. The other Maxwell equation for $\bm{E} = -\nabla\Phi$ is $\bm{J} + \epsilon_0\frac{\partial\bm{E}}{\partial t} = 0$, but we only use this for estimating a time step size which ensures numerical stability during our simulations. Eq.~\eqref{eq:poisson} must be supported by boundary conditions on the domain edges (faces in 3D) and internal surfaces. 

\subsection{Plasma-fluid equations}
The spatiotemporal evolution of electrons, ions, and neutrals is solved in the continuum approximation. For each species $\mu$ we have
\begin{equation}
  \label{eq:cdr}
  \frac{\partial n_\mu}{\partial t} + \nabla\cdot(\bm{v}_\mu n_\mu - D_\mu\nabla n_\mu) = S_\mu,
\end{equation}
where $n_\mu$ is the volumetric density of species $\mu$; $\bm{v}_\mu$ its drift velocity, $D_\mu$ its diffusion coefficient, and $S_\mu$ a source term (describing the interplay between attachment, impact ionization, photoionization etc.). Equation~\eqref{eq:cdr} is a convection-diffusion-reaction (CDR) equation which describes the evolution of individual species densities. 

Equation~\eqref{eq:cdr} is supplemented by boundary conditions describing the mass flux into or out of the computational domain. Our approach is to use outflow conditions on domain walls, and to expose the boundary conditions on embedded electrodes and dielectrics in an abstract plasma-kinetic framework which allows us to modify the surface kinetics of the plasma without affecting the underlying solver code.

\subsection{Radiative transfer}
\label{subsec:radiative_transfer}
Radiative transfer is handled by solving the radiative transfer equation (RTE) in the diffusion approximation. The RTE is
\begin{equation}
  \label{eq:RTE}
  \frac{\partial f_\nu\left(\bmx, \bm{\Omega}, t\right)}{\partial t} + c\bm{\Omega}\cdot\nabla f_\nu\left(\bmx, \bm{\Omega}, t\right) = -c\kappa_\nu(\bmx) f_\nu\left(\bmx, \bm{\Omega}, t\right) + \frac{1}{4\pi}\eta_\nu(\bmx),
\end{equation}
where $f_\nu$ is the photon distribution function (i.e. the number of photon with frequency $\nu$ at $(\bmx, t)$ traveling in direction $\bm{\Omega}$), $\kappa_\nu$ is the Beer's length for photons $\nu$, and $\eta_\nu$ is an isotropic source term (hence the factor of $4\pi$). Thus, $\eta_\nu$ is the total number of photons produced at $(\bm{x},t)$. 

\subsubsection{The multigroup approximation}
In the RTE, the frequency $\nu$ is a continuous variable. For most applications it becomes necessary to reduce the computational load by invoking the monochromatic multigroup approximation. That is, one assumes that $f_\nu$ consists of a number of frequency bands $\gamma$ where each frequency band is sufficiently sharp-line in order to individually invoke a monochromatic approximation. We take $f_\nu = \sum_\gamma f_\gamma\delta(\gamma)$ which essentially replaces Eq.~\eqref{eq:RTE} with a finite set of equations for each frequency band $\gamma$.

The multigroup RTE is solved in the diffusive $\textrm{SP}_1$ approximation (i.e. the \emph{Eddington} approximation) by closing the first order moment equation. That is, taking the first and second moments of Eq.~\eqref{eq:RTE} with respect to $\bm{\Omega}$ yields
\begin{subequations}
  \begin{align}
    \label{eq:Enu}
    \frac{\partial \Psi_\gamma}{\partial t} + \nabla\cdot\bm{F}_\gamma &= -c\kappa_\gamma \Psi_\gamma + \eta_\gamma, \\
    \label{eq:Fnu}
    \frac{\partial \bm{F}_\gamma}{\partial t} + \nabla\cdot\bm{\Pi}_\gamma &= -\kappa_\gamma \bm{F}_\gamma,
  \end{align}
\end{subequations}
where $\Psi_\gamma = \int_{4\pi}f_\gamma \diff\Omega$ is the radiative density, $\bm{F}_\gamma = c\int_{4\pi}f_\gamma\bm{\Omega} \diff\Omega$ is the radiative flux, and $\Pi_\gamma^{ij} = c\int_{4\pi}f_\gamma\Omega_i\Omega_j \diff\Omega$ is the Eddington tensor. This system is closed by assuming
\begin{equation}
  \label{eq:eddingtonTensor}
  \Pi_\gamma^{ij} = \frac{c}{3}\delta_{ij}\Psi_\gamma,
\end{equation}
which is equivalent to the Eddington approximation. Note that the underlying assumption here is that the medium is primarily a scattering medium so that the radiation moves in a diffusion-like manner. In such cases the time required for substantial increase in the radiative flux is much longer than the time required for traversing a mean free path. We can therefore impose $\partial_t\bm{F}_\gamma = 0$, and insertion of Eq.~\eqref{eq:eddingtonTensor} into Eq.~\eqref{eq:Fnu} yields
\begin{equation}
  \label{eq:diffFlux}
  \bm{F}_\gamma = -\frac{c}{3\kappa_\gamma}\nabla \Psi_\gamma,
\end{equation}
which is expected from a diffusive approximation. Insertion of Eq.~\eqref{eq:diffFlux} into Eq.~\eqref{eq:Enu} yields
\begin{equation}
  \label{eq:sp1}
  \frac{1}{c}\frac{\partial \Psi_\gamma}{\partial t} + \kappa_\gamma \Psi_\gamma - \nabla\cdot\left(\frac{1}{3\kappa_\gamma}\nabla \Psi_\gamma\right)  = \frac{\eta_\gamma}{c}. 
\end{equation}
This equation must be supplemented by appropriate boundary conditions for radiative transport. In the Eddington approximation, the appropriate outflow boundary is of the Robin type \cite{Larsen2002}:
\begin{equation}
  \label{eq:robin}
  \partial_n\Psi_\gamma + \frac{3\kappa_\gamma}{2}\frac{1 + 3r_2}{1 - 2r_1}\Psi_\gamma = \frac{g_\gamma}{\kappa_\gamma},
\end{equation}
where $r_1$, and $r_2$ are reflection coefficients and $g_\gamma$ is a surface source. In this paper, we do not consider photon reflection from boundaries nor injection of photons into the domain and the outflow boundary condition on photons on a boundary with outward (i.e. out of the gas-phase) normal vector $\bm{n}$ simplifies to
\begin{equation}
  \frac{\partial \Psi_\gamma}{\partial n} + \frac{3\kappa_\gamma}{2}\Psi_\gamma = 0,
\end{equation}
where $\partial_n \Psi_\gamma = \bm{n}\cdot\nabla \Psi_\gamma$. Thus, for free outflow of photons the boundary flux is
\begin{equation}
  \label{eq:photon_outflux}
  \bm{F}_\gamma\cdot\bm{n} = \frac{c}{2} \Psi_\gamma. 
\end{equation}

Note that the diffusion assumption does not hold in a general manner. However, a successful strategy has been to combine solutions of \eqref{eq:sp1} so that it represents the photon distribution of the exact integral solution \cite{1982TepVT..20..423Z}. We remark that diffusive approaches disregard even qualitative features (such as shadows) of the full RTE solutions. This is true even for improved diffusion models, such as the $\textrm{SP}_3$ approximation \cite{Larsen2002, Segur2006, Bourdon2007}. While we currently adopt the diffusion approximation, we can certainly foresee a necessity to account for higher-order moments of the RTE, or the use of Monte Carlo methods, sometime in the future. In fact, this will be necessary for studies of of porous media, such as packed-bed DBD reactors. Currently, direct solutions of the RTE equation \cite{Capeillere2008} in three dimensions appears infeasible, even on todays largest supercomputers. 

\subsection{Surface charge conservation}
Our final equation of motion is local conservation of charge on \emph{dielectric} surfaces. We do not solve an equivalent problem on electrodes because we assume that complete neutralization occurs on both anodes and cathodes, and that the voltage on the live and grounded electrodes are controlled by an external circuit. Charge conservation on dielectric surfaces is given by
\begin{equation}
  \label{eq:surfaceCons} 
  \frac{\partial \sigma}{\partial t} = F_\sigma,
\end{equation}
where $F_\sigma$ is the charge flux. $F_\sigma$ is always coupled to the boundary conditions for the species densities $n_\mu$ since charge must be conserved at the surface. Thus, $F_\sigma$ is simply the sum of all species fluxes at the boundary. Currently, we do not include solid state kinetics and the surface state is therefore characterized only by the net charge density, without detailed knowledge of the distribution of the ionic states that are involved on the surface.  

\subsection{Minimal plasma model}
To summarize, our plasma model consists of the following equations: 
\begin{subequations}
  \label{eq:minimal_plasma}
  \begin{align}
    \nabla\cdot\left(\epsilon_r\nabla\Phi\right) &= -\frac{\rho}{\epsilon_0}, \\
    \frac{\partial \sigma}{\partial t} &= F_\sigma, \\
    \frac{1}{c}\frac{\partial \Psi_\gamma}{\partial t} + \kappa_\gamma \Psi_{\gamma} - \nabla\cdot\left(\frac{1}{3\kappa_\gamma}\nabla \Psi_\gamma\right)  &= \frac{\eta_\gamma}{c}, \\
    \frac{\partial n_\mu}{\partial t} + \nabla\cdot\left(\bm{v}_\mu n_\mu - D_\mu\nabla n_\mu\right) &= S_\mu,
  \end{align}
\end{subequations}
where the final two equations denote \emph{sets} of equations for $\mu$ and $\gamma$.

For streamer discharges Eq.~\eqref{eq:minimal_plasma} is nonlinearly coupled through source terms and species velocities. For the purposes of this paper, this coupling is restricted to the following: 
\begin{subequations}
  \begin{align}
    \epsilon_r &= \epsilon_r(\bm{x}), \\
    \kappa_\gamma &= \kappa_\gamma(\bm{x}), \\
    \eta_\gamma &= \eta_\gamma\left(\bm{E}, n_\mu\right), \\
    \bm{v}_\mu &= \bm{v}_\mu(\bm{E}, n_\mu), \\
    D_\mu &= D_\mu(\bm{E}), \\
    S_\mu &= S_\mu\left(\bm{E}, n_\mu, \Psi_\gamma\right),
  \end{align}
\end{subequations}
where $\epsilon_r$ may additionally be discontinuous. 

\section{Numerical methods}
\label{sec:method}
SAMR exists in two separate categories, patch-based and tree-based AMR. Patch-based AMR is the more general type and contain tree-based grids as a subset; they can use refinement factors other than 2, as well as accomodate anisotropic resolutions and non-cubic patches. In patch-based AMR the domain is subdivided into a collection of hierarchically nested overlapping patches (or boxes). Each patch is a rectangular block of cells which, in space, exists on a subdomain of the union of patches with a coarser resolution. Patch-based grids generally do not have unique parent-children relations: A fine-level patch may have multiple coarse-level parents. An obvious advantage of a patch-based approach is that entire Cartesian blocks are sent into solvers, and that the patches are not restricted to squares or cubes. A notable disadvantage is that the overlapping grids inflate memory, and that additional logic is required when updating a coarse grid level from the overlapping region of a finer level. Tree-based AMR use quadtree or octree data structures that describe a hierarchy of unique parent-children relations throughout the AMR levels: Each child has exactly one parent, whereas each parent has multiple children (4 in 2D, 8 in 3D). For CPU cache performance reasons, the leaves of an octree are often cubic patches (e.g. $4^3$ or $8^3$ boxes), but the mesh can also be refined on a cell-by-cell basis. However, the use of single cell leaves becomes prohibitive at large scale for two reasons. The first is that special case must be taken in order to avoid memory inflation due to a growing tree structure. The second is that such trees, while still being SAMR, use indirect memory referencing, thus adding latency in data accessing and processing. This typically leads to poorer CPU performance since the data defined in neighboring cells may be stored on different cache lines.

\begin{figure*}[h!t!b!]
  \centering
  \includegraphics[width=0.4\textwidth]{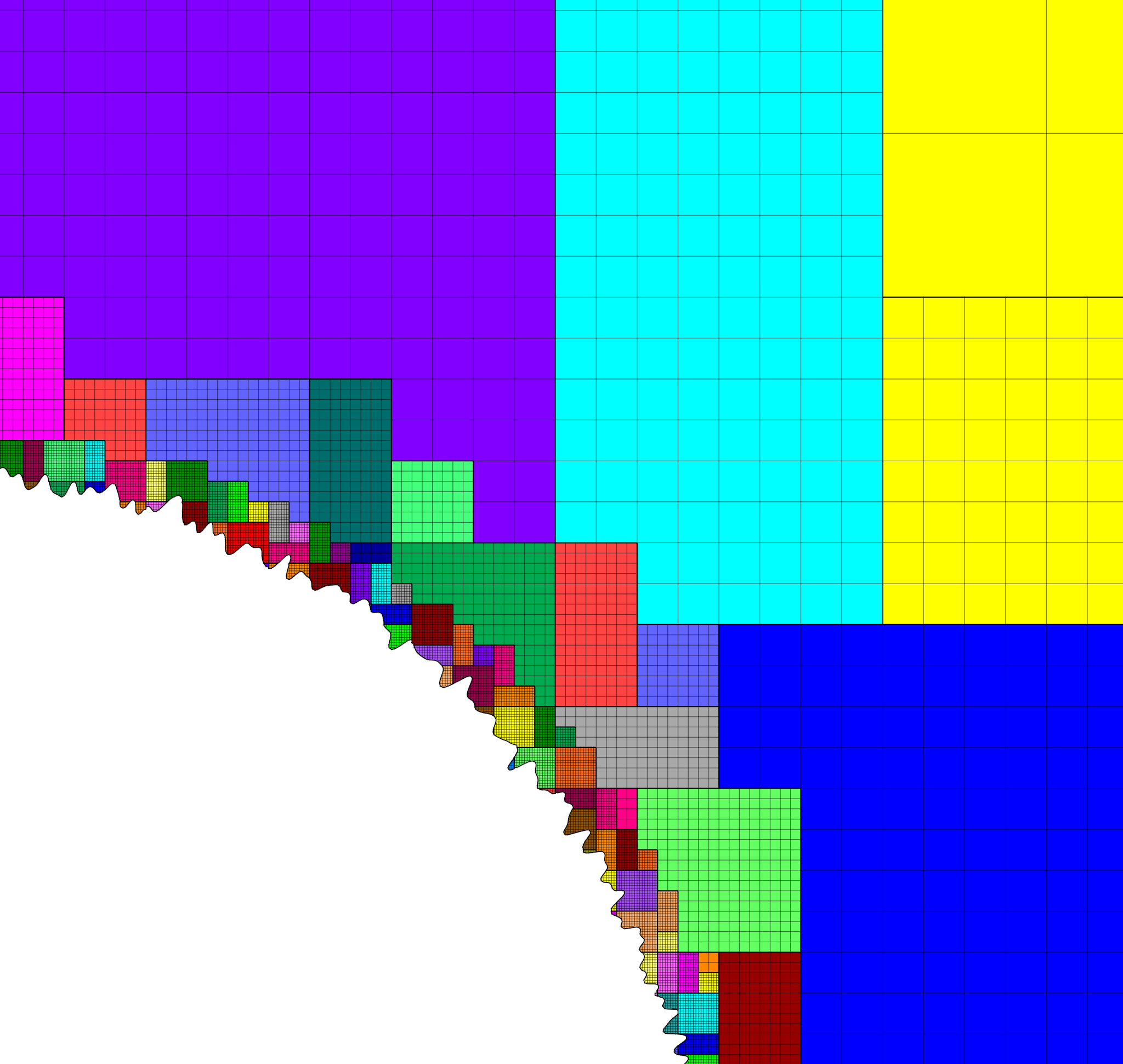}
  \hfil
  \includegraphics[width=0.4\textwidth]{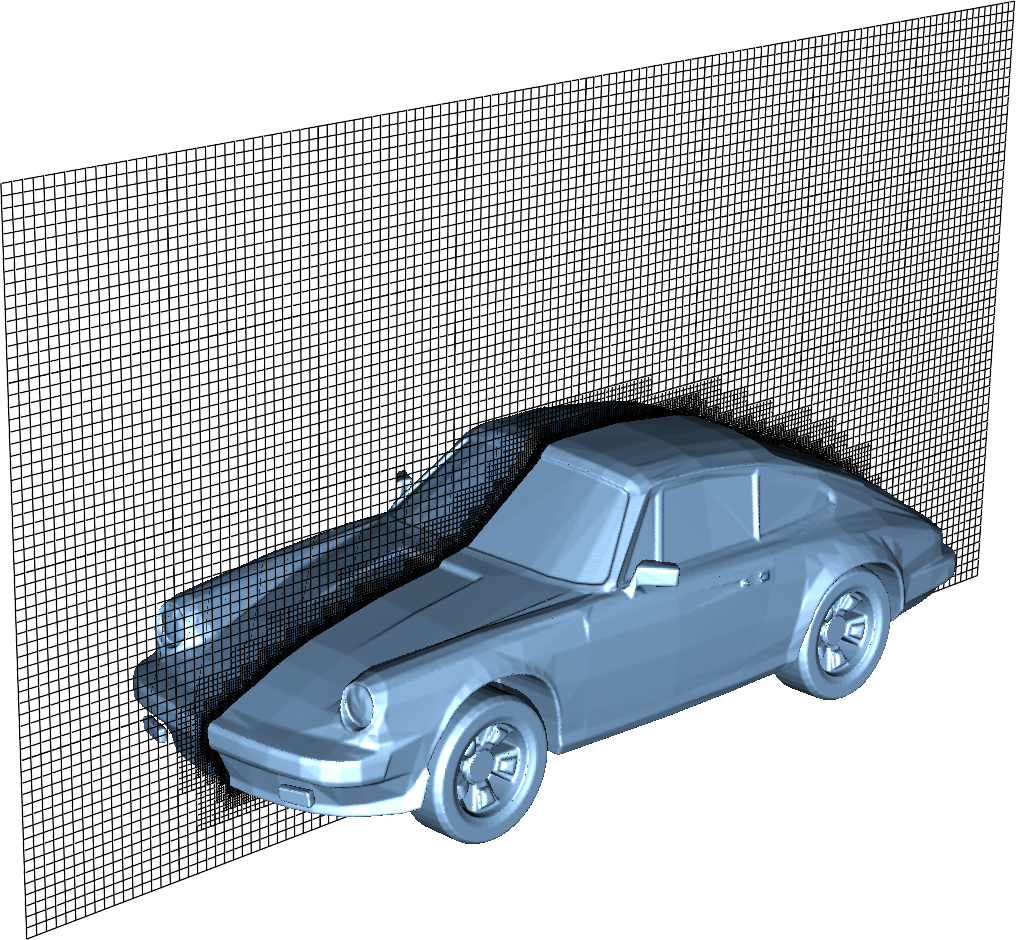}      
  \caption{Left: Patch-based refinement near a complex boundary. The colors indicate individual patches and the refinement factor between each level is 4. Right: EB data generation from a surface tessellation. The surface mesh of the car contains 5168 vertices and 10316 facets. The complete EB mesh contains three levels of refinement (factor 2 between levels) of a base mesh of $128\times64\times64$.}
  \label{fig:patch_amr}
\end{figure*}

We have implemented Eq.~\eqref{eq:minimal_plasma} into Chombo \cite{chombo,ebchombo}, which is a mixed-language C++ \& Fortran library for performing patch-based AMR that also includes support for embedded boundaries. In EB applications, the mesh is additionally described by a graph near the cut-cells that describes the connectivity and geometric moments of the cell fragments. The cells in the AMR hierarchy are related to one another by refinement and coarsening operations on that graph \cite{ebchombo}. Figure~\ref{fig:patch_amr} shows an example of patch-based refinement near a complex boundary. 

\subsection{Spatial discretization}
We now give a brief summary of the Chombo spatial discretization \cite{ebchombo}. We discretize the physical domain by a set of Cartesian cells, where some cells are cut by a level-set function $s(\bm{x})$ where $s(\bm{x}) = 0$ describes the boundary interface. The geometric information is generated by a graph on the finest level grid first, which is then coarsened throughout the hierarchy. This information is stored in a sparse data format for all levels, with a memory overhead that depends on the complexity of the geometry. Inside each cut-cell, the boundary is linearized so that it intersects through the cell as lines in 2D, and planes in 3D. Control volume faces that are cut by this surface are referred to as cut-faces, and the boundary itself is referred to as an embedded boundary (EB).

\begin{figure*}[h!t!b!]
  \centering
  \includegraphics{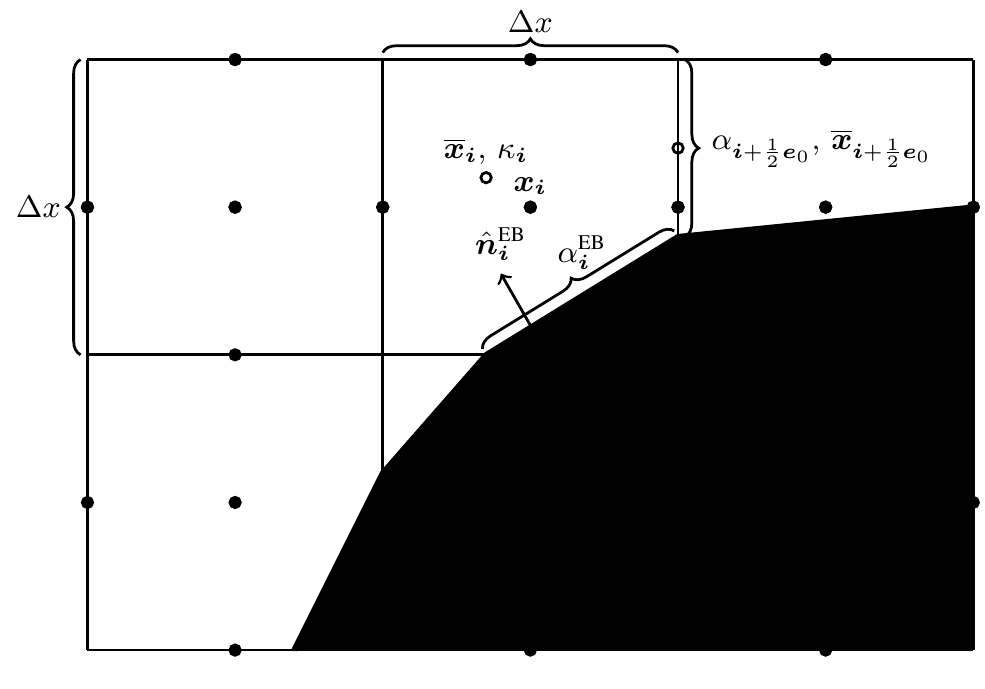}
  \caption{Cut-cell discretization.}
  \label{fig:spatial_discretization}
\end{figure*}

Figure~\ref{fig:spatial_discretization} shows the cut cells in greater detail. Here, a cell is identified by it's index $\bmi$. We will take $\bm{x}_{\bmi}$ to be the \emph{cell center}, and $\overline{\bm{x}}_{\bmi}$ to the be \emph{cell centroid}. The volume fraction is $0 \leq \kappa_{\bmi} \leq 1$: Cells with $\kappa_{\bmi} = 0$ are denoted as \emph{covered} cells and lie completely outside the computational domain; cells with volume fractions $0 < \kappa_{\bmi} < 1$ are called \emph{irregular} cells and lie partially inside the computational domain. If an irregular cell contains more than one cell fragment, it is termed a multi-valued cell. By construction, multi-valued cells occur due to graph coarsening and do not exist on the finest graph level. Cell faces are denoted by $f^\pm_d(\bmi)$ where $\pm$ indicates the high ($+$) or low ($-$) face of the cell $\bmi$ in the coordinate direction $d$. A face center for face $f^\pm_d(\bmi)$ is therefore located at $\bm{x}_{\bmi \pm \frac{1}{2}\mathbf{e}_d}$ where $\mathbf{e}_d$ is a unit vector in the $d$-direction. Face centroids are likewise denoted $\overline{\bm{x}}_{\bmi \pm \frac{1}{2}\mathbf{e}_d}$ and the corresponding area fractions are denoted by $\alpha_{\bmi\pm\frac{1}{2}\bm{e}_d}$. Finally, the EB centroid is denoted by $\overline{\bm{x}}_{\bmi}^{\textrm{EB}}$ with a corresponding area fraction $\alpha^{\textrm{EB}}_{\bmi}$ and an outward normal vector $\hat{\bm{n}}_{\bmi}^{\textrm{EB}}$. Explicitly defined expressions for these quantities can be found in \cite{ebchombo}. 

\subsection{Geometry generation}
In addition to using the native Chombo tools for geometry generation \cite{ebchombo}, we have developed tools for generating level-set functions from surface tessellations. Our surface mesh implementation uses a doubly-connected edge list (DCEL) that supports mixtures of arbitrarily-sided planar polygons, but requires that the tessellation is a two-manifold non-intersecting, watertight surface. There are large performance penalties in using surface tessellations for generating the signed distance field that Chombo requires for EB generation since we must provide the signed distance to the closest polygon, vertex, or edge. Brute force searching through all polygons is unacceptably slow, so we accellerate the polygon search by embedding the polygonal tessellation in a boundary volume hierarchy that allows orders of magnitude faster traversal through the tessellation. We use a $k$-d tree with axis-aligned boundary boxes for aggregated polygons, and the closest-feature search over the tessellation is performed by building a priority list as the tree is traversed from the root to the leaves. Figure~\ref{fig:patch_amr} shows an example of an EB mesh that was generated from polygonal data.

\subsection{Helmholtz equations}
\label{sec:helmholtz}
Next, we discuss the discretization of the Helmholtz equation
\begin{equation}
  \label{eq:helmholtz}
  \alpha a(\bm{x})\phi + \beta\nabla\cdot\left(b(\bm{x})\phi\right) = \rho.
\end{equation}
Helmholtz equation solvers are an important part of this paper. For example, the Poisson equation is represented by $\alpha = 0$, $\beta = -\epsilon_0$, $b(\bm{x}) = \epsilon_r(\bm{x})$. Furthermore, temporal discretizations of parabolic equations are also underpinned by a Helmholtz solver. For example, discretizing Eq.~\eqref{eq:sp1} by using the backward Euler method yields
\begin{equation}
  (1 + c\Delta t \kappa_\gamma)\Psi_\gamma^{n+1} - \nabla\cdot \left(\frac{c\Delta t}{3\kappa_\gamma} \nabla \Psi_\gamma ^{n+1}\right) = \Psi_\gamma^{n} - \frac{\Delta t\eta_\gamma}{c},
\end{equation}
which defines a Helmholtz problem for $\Psi_\gamma^{n+1}$. 

We follow the work in \cite{Johansen1998, McCorquodale2001, Schwartz2006, Crockett2011} and use the finite volume method for the Helmholtz equation. For ease of notation, we restrict the discussion below to the case $a=0$ which yields the Poisson equation. Extensions to the full Helmholtz problem is straightforward by adding in another diagonal term. Our implementation of the Helmholtz equation also supports multi-fluids, i.e. cases in which $b(\bm{x})$ is additionally discontinuous across a level-set surface. The multifluid problem needs additional encapsulation of a quasi-boundary condition on the interface between two materials $p$ and $p^\prime$, given by
\begin{equation}
  \label{eq:matching}
  b_p\frac{\partial \phi}{\partial n_p} +   b_{p^\prime}\frac{\partial \phi}{\partial n_{p^\prime}} = \sigma,
\end{equation}
where $\bm{n}_p$ and $\bm{n}_{p^\prime}$ are unit normals that point into each fluid, with $\bm{n}_{p^\prime} = -\bm{n}_p$, and $\sigma$ is a surface source term. Recasting Eq.~\eqref{eq:helmholtz} in integral form yields
\begin{equation}
  \label{eq:poisson_integral}
  \oint_A b(\bm{x})\nabla\phi\cdot\diff\bm{A} = \frac{1}{\beta}\int_V\rho \diff V. 
\end{equation}

\begin{figure*}[h!t!b!]
  \centering
  \includegraphics{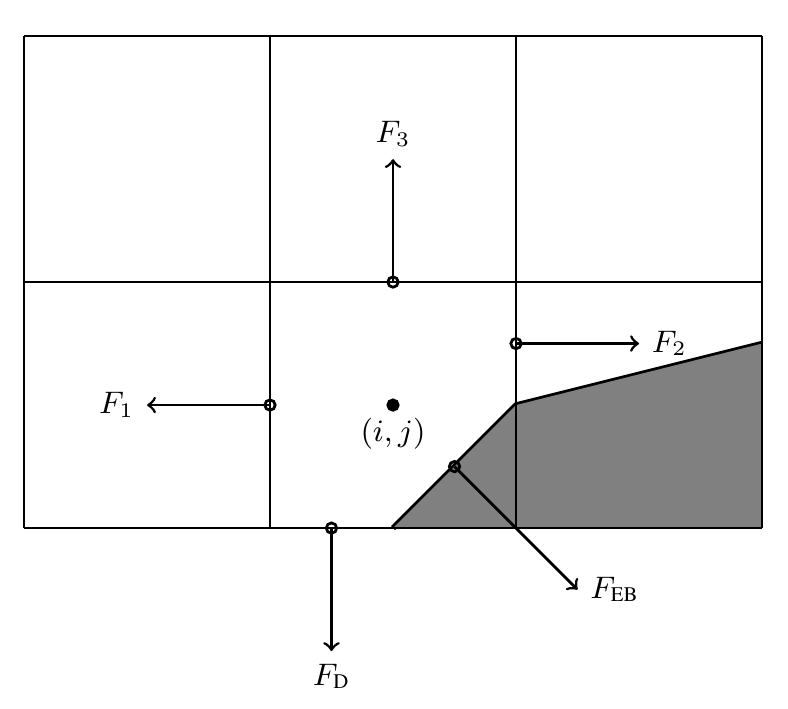}
  \caption{Cut cell at domain face.}
  \label{fig:cut_cell}
\end{figure*}

We consider the cell shown in Fig.~\ref{fig:cut_cell}. Here, the volume $V_{\bmi}$ is a cut-cell at a domain boundary. Integration of Eq.~\eqref{eq:poisson_integral} over the irregular cell yields
\begin{equation}
  \oint_A b(\bm{x})\nabla\phi\cdot\diff\bm{A} = \left(\alpha_1F_1 + \alpha_2F_2 + \alpha_3F_3 + \alpha_{\textrm{D}}F_{\textrm{D}} + \alpha_{\textrm{EB}}F_{\textrm{EB}}\right)\Delta x,
\end{equation}
where the fluxes are centroid-centered on their respective faces and $\alpha_i$ are face area fractions. The centroid fluxes are evaluated by constructing second order accurate face-centered fluxes, which are then interpolated to the respective centroids. For example, for the flux through the top face in Fig.~\ref{fig:cut_cell} we find a standard expression for second order accurate approximations of the first derivative:
\begin{equation}
  F_3 = F_{i,j+\nicefrac{1}{2}} = b_{i, j+\nicefrac{1}{2}}\frac{\phi_{i, j+1} - \phi_{i,j}}{\Delta x},
\end{equation}
For fluxes through face centroids we interpolate the face-centered fluxes. For example, the flux $F_2$ in Fig.~\ref{fig:cut_cell} is given by
\begin{equation}
  F_2 = \left[F_{i+\nicefrac{1}{2},j }(1-s) + sF_{i+\nicefrac{1}{2}, j+1}\right],
\end{equation}
where $s$ is the normalized distance from the face center to the face centroid, and $F_{i+\nicefrac{1}{2},j }$ and $F_{i+\nicefrac{1}{2}, j+1}$ are face-centered fluxes. 

Flux evaluation on coarse-fine boundaries is slightly more involved. The AMR way of handling this is to reflux the coarse side by setting the flux into the coarse cell to be the sum of fluxes from the abutting finer cells. In Chombo, this is done by precomputing a set of flux registers that hold the face centered fluxes on both sides of the coarse-fine interface (see \cite{ebchombo} for additional details). Refluxing is then a matter of subtracting the coarse flux from the divergence computation, and adding in the sum of the fine face fluxes. I.e. let $\{f_{\textrm{f}}(f_{\textrm{c}})\}$ be the set of fine faces that are obtained when coarsening of a coarse face $f_{\textrm{c}}$. In the reflux step, the divergence operator in the coarse cell is modified as
\begin{equation}
  \nabla\cdot\bm{F} \rightarrow \nabla\cdot\bm{F} + \frac{1}{\Delta x}\left(\sum_{f} F_{f} - F_c\right),
\end{equation}
where $F_{c}$ and $F_{f}$ are the coarse and fine-face fluxes, and the sum runs over all the fine faces that abut the coarse face. For further details regarding the discretization and conditioning of the Poisson equation, see \cite{Johansen1998, McCorquodale2001, Schwartz2006, Crockett2011}. 

\subsubsection{Boundary conditions}
Next, we discuss four types of boundary conditions for the Helmholtz equation: Neumann, Dirichlet, Robin, and multifluid type boundary conditions. For Neumann boundary conditions the domain and embedded boundary fluxes are specified directly. For Dirichlet boundary conditions the process is more involved. For Dirichlet conditions on domain faces we apply finite differences in order to evaluate the flux through the face. For example, for a constant Dirichlet boundary condition $\phi = \phi_0$ the face-centered flux at the bottom face in Fig.~\ref{fig:cut_cell} is, to second order,
\begin{equation}
  F_{i,j-\nicefrac{1}{2}} = -\frac{b_{i,j-\nicefrac{1}{2}}}{\Delta x}\left(3\phi_{i,j+1} -\frac{1}{3}\phi_{i,j} - \frac{8}{3}\phi_0\right)
\end{equation}
As with the flux $F_2$ on the interior face, fluxes on domain faces are also interpolated to face centroids. Thus, $F_{\textrm{D}}$ becomes
\begin{equation}
  F_{\textrm{D}} = \left[F_{i,j-\nicefrac{1}{2}}(1-t) + tF_{i-1,j-\nicefrac{1}{2}}\right],
\end{equation}
where $t$ is the distance from the face center to the face centroid.

\begin{figure*}[h!t!b!]
  \centering
  \includegraphics{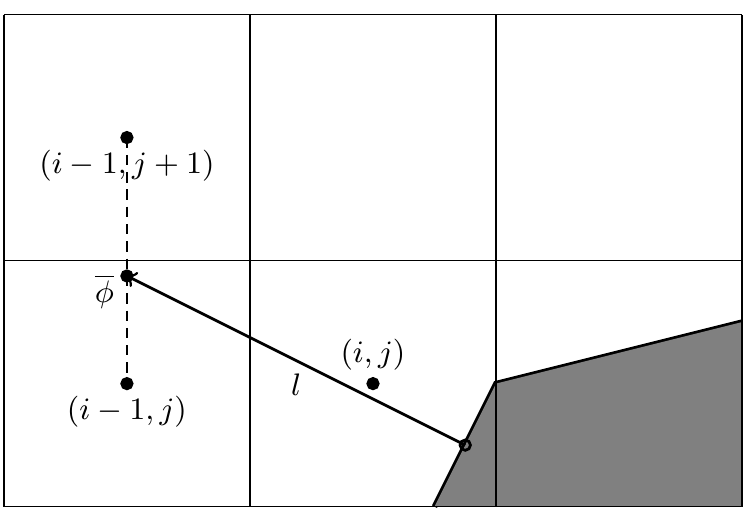}
  \caption{Ray casting at the EB for obtaining the normal gradient. }
  \label{fig:raycast}
\end{figure*}

The evaluation of Dirichlet boundary conditions on the EB is more complicated because the EB normal does not align with any of the coordinate directions. To evaluate the flux on the boundary we construct ray based or least squares based stencils for evaluating $\partial_n\phi$ (see \cite{Johansen1998} or \cite{ebchombo} for details). Regardless of which approach is used, we have
\begin{equation}
  \label{eq:bndry_stencil}
  \frac{\partial\phi}{\partial n} = w_0\phi_0 + \sum_{\bmi \in \psi}w_{\bmi}\phi_{\bmi},
\end{equation}
where $\phi_0$ is the Dirichlet value on the boundary, $w_0$ is a boundary weight and $\psi$ is a stencil that contains only interior points. The weights $w_{\bmi}$ are weights for these points. As an example, consider the flux depicted in Fig.~\ref{fig:raycast}. The first order accurate partial derivative on the boundary is given by
\begin{equation}
  \frac{\partial\phi}{\partial n} = \frac{\phi_0 - \overline{\phi}}{l},
\end{equation}
where $\overline{\phi}$ is the interpolated value at the intersection of the ray and the line that connects $\bm{x}_{i-1, j}$ and $\bm{x}_{i-1, j+1}$. Since $\overline{\phi}$ can be linearly interpolated by using these two interior points only, this is clearly in the form of Eq.~\eqref{eq:bndry_stencil}. The boundary derivative stencils are well separated from the boundary (i.e. they do not use the values of the irregular cell itself). For the Poisson equation this is a requirement in order to achieve good conditioning of the discretized system as the volume fraction approaches zero \cite{Johansen1998}. 

Higher-order approximations to the flux are built in a similar way by including more interior cells. In our experience, the best convergence results come from using second order accurate ray-based boundary stencils, which requires 3 ghost cells in the general case. If we cannot find a stencil for computing the normal derivative by ray-casting, which can occur if there aren't enough cells available, we use quadrant-based least squares for computing the normal derivative (again, see \cite{Johansen1998} or \cite{ebchombo}).

We have also implemented Robin boundary conditions of the type
\begin{equation}
  a_1\phi + a_2\frac{\partial \phi}{\partial n} = a_3,
\end{equation}
which is an appropriate type of boundary condition for the radiative transfer equation. The normal derivative is given by $\partial_n\phi = (a_3 - a_1\phi)/a_2$ so that extrapolation of $\phi$ to the boundary is sufficient for imposing the boundary flux. Our way of doing this is simply to extrapolate $\phi$ to the boundary by using either least squares or Taylor-based stencils. 

On multifluid boundaries the boundary condition is neither Dirichlet, Neumann, or Robin. Multifluid boundaries are more complex since the state at the boundary is not known, but rather depends on the solution inside both fluids. Our approach follows that of \cite{Crockett2011} where we first compute stencils for the normal derivative on each side of the boundary,
\begin{equation}
  \frac{\partial\phi}{\partial n_q} = w_0^q\phi_B + \sum_{\bmi \in \psi_q}w_{\bmi}^q\phi_{\bmi},
\end{equation}
where $q = p$ or $q=p^\prime$ and $\phi_B$ is the solution on the surface centroid, and the stencil $\psi_q$ only reaches into fluid $q$. The linear nature of this equation allows one to obtain the surface state $\phi_B$ from the matching condition Eq.~\eqref{eq:matching}, which can then be eliminated in order to evaluate $\partial\phi/\partial n_q$ on both sides of the boundary.

\subsubsection{Multigrid solver}
To solve the discretized Helmholtz equation we use the geometric multigrid (GMG) solver template that ships with Chombo \cite{ebchombo}. GMG involves smoothing of the solutions on progressively coarsened grids and is compatible with AMR. Smoothing on each level involves relaxation (e.g. Jacobi or Gauss-Seidel), which primarily reduces the magnitude of high freqency errors. Removal of low-frequency errors from the solution is much slower. Because of this, multigrid accelerates convergence by projecting the error onto a coarser grid where the error has, from the viewpoint of the grid, a shorter wavelength, making relaxation more efficient. Once a bottom grid level has been reached and an approximate bottom-level solution has been found, the error is prolongated onto a finer grid and relaxation is then re-applied. Geometric multigrid works best when the long wavelength modes of the fine grid operator are well represented as short wavelength modes on the coarse grid operator. For EB applications however, coarsening can result in the removal of finer geometric features so that the relaxation step cannot sufficiently dampen the error modes at which GMG is aimed at. Because of this, geometric multigrid for EB applications usually involve lower convergence rates between each multigrid cycle than it does for geometry-less domains and, moreover, typically involves dropping to the bottom solver sooner. Currently, we only support relaxation solvers as the bottom solver for multi-phase problems, whereas we use the built-in BiCGStab and GMRES solvers in Chombo \cite{ebchombo} for single-phase elliptic problems. In the future, we would like to use algebraic multigrid from e.g. PETSc as a bottom solver in the V-cycle in order to enhance solver efficiency for very complex geometries. In fact, this will be necessary for studies of highly porous media, such as packed bed DBD reactors \cite{VanLaer2015}. 

Our elliptic operators are embedded into the multigrid template that comes with the Chombo library, which relies on a series of C++ abstractions that implement the residual-correction form of multigrid \cite{chombo, ebchombo}. Our multifluid solver for the Poisson equation is an extension of the functionality discussed in \cite{Crockett2011} where we additionally incorporate (i) simultaneous handling of types of boundary conditions, and (ii) continuous and discontinuous permittivities. In the multigrid implementation, we use either red-black or multi-colored Gauss-Seidel relaxation. For the multi-fluid case, the boundary state is updated before each relaxation step. Our best results for the multifluid operator comes from using multi-colored (4/8 colors in 2D/3D) Gauss-Seidel relaxation. As with red-black ordering, odd colors only reach into even colors, which allows us to reduce the number of exchange operations to 2 in 2D and 4 in 3D. We remark that for patches that intersect multi-fluid boundaries, it is necessary to perform relaxation on the patch twice (once for each phase), since both patches represent individual degrees of freedom. Currently, our load-balancing algorithms do not take this into account, and so we expect that a speedup of the Poisson equation is possible by more careful estimation of computational loads. 

Finally, our convergence criteria for GMG rely on sufficient reduction of an initial residual. Let the discretized Helmholtz equation be given by $L(\phi) = \rho$ where $L(\phi)$ is the left hand side of Eq.~\eqref{eq:helmholtz} which also accounts for boundary conditions. The residual is 
\begin{equation}
  r = L\phi - \rho.
\end{equation}
We exit multigrid iteration if the residual is sufficiently small, i.e. if
\begin{equation}
  \left\Vert r\right\Vert < \lambda\left\Vert r_0\right\Vert
\end{equation}
where $r_0$ is the computed residual with $\phi = 0$ and $\lambda$ is some predefined tolerance.

\subsection{Parabolic equations}
In the presence of embedded boundaries we use implicit methods for parabolic equations (e.g. for diffusion advances). Embedded boundaries complicate the construction of explicit schemes for parabolic equations due to severe restrictions on the time step. For example, for the diffusion equation $\partial_t \phi = \nu\nabla\cdot(\nabla\phi)$, the stability constraint on the time step is
\begin{equation}
  \Delta t \leq \frac{\Delta x^2(\kappa_{\textrm{min}})^{2/\bm{D}}}{2\bm{D}\nu},
\end{equation}
where $\bm{D}$ is the spatial dimension. For cut-cells, $\kappa_{\textrm{min}}$ can come arbitrarily close to $0$, which leads to an unacceptable constraint. Although stabilization is possible by redistribution of mass in the vicinity of the embedded boundaries (see Sec.~\ref{sec:hybrid}), this is not encouraged due to the $(\Delta x)^2$ scaling. 

Our implicit method is as follows. For a parabolic equation
\begin{equation}
  \label{eq:rte_time}
  \frac{\partial \phi}{\partial t} + L\left(\phi\right) = \rho
\end{equation}
where $L$ is an elliptic operator, we use the implicit Runge-Kutta scheme by Twizell, Gumel, and Arigu (TGA) \cite{Twizell1996}. The TGA scheme discretizes Eq.~\eqref{eq:rte_time} as
\begin{equation}
  \label{eq:tga}
  \left(I - \mu_1 L\right)(I-\mu_2L)\phi^{n+1} = (I + \mu_3L)\phi^n + (I + \mu_4L)\rho^{n+1/2},
\end{equation}
whose solution requires the solution of two Helmholtz problems, one for each operator on the left hand side. The constants $\mu_i$ are
\begin{subequations}
  \begin{align}
    \mu_1 &= \frac{a - \sqrt{a^2 - 4a + 2}}{2}\Delta t, \\
    \mu_2 &= \frac{a + \sqrt{a^2 - 4a + 2}}{2}\Delta t, \\
    \mu_3 &= (1 - a)\Delta t,                           \\
    \mu_4 &= \left(\frac{1}{2} - a\right)\Delta t,      \\
    a &= 2 - \sqrt{2} - \lambda,
  \end{align}
\end{subequations}
where $\lambda$ is the machine precision. Note that the centering of the source terms in Eq.~\eqref{eq:tga} is $n+1/2$ and $n+1$, respectively. The details of the implementation of this scheme can be found in \cite{McCorquodale2001, chombo, ebchombo}. 

\subsection{Convection-Diffusion-Reaction equations}
Next, we discuss the spatial discretization of the CDR equation~\eqref{eq:cdr}. Since the discretization of the diffusion operator was given in the preceding sections, it is sufficient to discuss the discretization of the advective term $\nabla\cdot(\bm{v}\phi)$.
\subsubsection{Advective discretization}
The advective term is solved by using finite volumes, such that
\begin{equation}
  \int_V\nabla\cdot(\bm{v}\phi)\diff V = \sum_{f\in f(V)} (\bm{v}_f\cdot\bm{n}_f)\phi_f\alpha_f\Delta x^{\bm{D}-1},
\end{equation}
where $f(V)$ denotes the set of faces that enclose the control volume $V$, and $\alpha_f$ is the face aperture fraction. For our purposes, the velocity $\bm{v}$ is known so that it is sufficient to find a way to obtain the face-centered states $\phi_f$. Once these are obtained, they are interpolated to the face centroids and the discrete divergence is then computed by summing the fluxes over all cell faces, as was done in Sec.~\ref{sec:helmholtz}

Currently, we use an unsplit Godunov method with van Leer limiting in order to obtain $\phi_f$. The state at the boundary is then given by the solution to a Riemann problem with initial data given by the left and right slope-limited states; the solution to this particular problem (linear advection) is simply the upwind state. 

\subsubsection{Hybrid divergence}
\label{sec:hybrid}
A problem with cut-cell grids is that naive discretization of conservation laws $\partial_tn +\nabla\cdot (\bm{v}\phi) = 0$ yields
\begin{equation}
  \phi_{\bmi}^{k+1} = \phi_{\bmi}^{k} - \frac{\Delta t}{\kappa_{\bmi}\Delta x^{\bm{D}}}\oint(\bm{v}\phi)\cdot\diff\bm{A}_{\bmi},
\end{equation}
which gives a CFL condition on the time step
\begin{equation}
  \Delta t = \mathcal{O}\left(\frac{\Delta x}{\bm{v}_{\bmi}^{\textrm{max}}}\kappa_{\bmi}^{\frac{1}{\bm{D}}}\right).
\end{equation}
This restriction is unacceptable since $\kappa_{\bmi}$ can be arbitrarily close to zero. Various remedies to this problem exist, such as merging of small cell volumes into larger control volumes, or the use of specialized stencils. The Chombo approach \cite{ebchombo} is a simple one; we stabilize the advective discretization by computing a hybrid discretization which expands the range of influence of small control volumes to include a larger neighborhood around them. Let $D_{\bmi}^{\textrm{c}}$ be the conservative divergence of $\nabla\cdot\bm{F}$ for a cell $\bmi$, where e.g. $\bm{F} = \bm{v}\phi$. For an irregular cell $\bm{i}$ we then also compute a non-conservative divergence $D^{\textrm{nc}}_{\bmi}$ as the average of the conservative divergence in a neighborhood $\mathcal{N}_{\bmi}$ of $\bmi$:
\begin{equation}
  D_{\bmi}^{\textrm{nc}} = \frac{\sum_{\bmj\in \mathcal{N}_{\bmi}}\kappa_{\bmj}D_{\bmj}^{\textrm{c}}}{\sum_{\bmj\in \mathcal{N}_{\bmi}}\kappa_{\bmj}}.
\end{equation}
Note that there are other ways to compute $D_{\bmi}^{\textrm{nc}}$, such as monotone extrapolation also to \emph{covered faces}, in which $D_{\bmi}^{\textrm{nc}}$ is computed as if it was a regular cell . For streamer simulations, either approach is feasible. 

We then use a hybridization of these two estimates,
\begin{equation}
  D_{\bmi}^{\textrm{H}} = \kappa_{\bmi} D_{\bmi}^{\textrm{c}} + (1-\kappa_{\bmi})D_{\bmi}^{\textrm{nc}},
\end{equation}
which cancels the small denominator in $\nabla\cdot\bm{F}$ and we obtain a stable method with a standard CFL condition. However, this method fails to conserve by the mass
\begin{equation}
  \delta M_{\bmi} = \kappa_{\bmi}\left(1-\kappa_{\bmi}\right)\left(D_{\bmi}^{\textrm{nc}} - D_{\bmi}^{\textrm{c}} \right).
\end{equation}
To enforce strict conservation, the excess mass is redistributed into neighboring cells $\bmj \in \mathcal{N}_{\bmi}$
\begin{equation}
  \delta M_{\bmi} = \sum_{\bmj \in \mathcal{N}_{\bmi}}\delta M_{\bmj, \bmi},
\end{equation}
where $\delta M_{\bmi, \bmj}$ is the distributed mass from $\bmi$ to $\bmj$. For weighted redistribution we use
\begin{equation}
  \delta M_{\bmi, \bmj} = \frac{\delta M_{\bmi}\kappa_{\bmj}W_{\bmj}}{\sum_{\bmk \in \mathcal{N}_{\bmi}}\kappa_{\bmk}W_{\bmk}},
\end{equation}
where $W_{\bmi}$ are weights (for example mass). Then, for cells $\bmj \in \mathcal{N}_{\bmi}$ the final divergence is
\begin{equation}
  D_{\bmj} = D_{\bmj}^{\textrm{H}} + \delta M_{\bmi, \bmj}. 
\end{equation}
For additional information on the redistribution process, see \cite{ebchombo}. 

\subsection{Time stepping}
\subsubsection{A second order fractional step scheme with implicit diffusion}
\label{sec:rk2_tga}
Below, we provide a formulation of a second order accurate fractional step method with implicit diffusion for advancing $k\rightarrow(k+1)$ as follows: 
\begin{enumerate}
\item Advance $n^\ast = n^k + \Delta t\left[S^k - \nabla\cdot\left(\bm{v}^kn^k\right)\right]$ and $\sigma^\ast = \sigma^k + \Delta tF_\sigma^k$
\item Compute the new electric field $\bm{E}^\ast$ by solving the Poisson equation with the new space and surface charge densities $\rho^\ast$, $\sigma^\ast$.
\item Compute radiative transfer source terms $\eta^\ast = \eta\left(n^\ast, \bm{E}^\ast\right)$.
\item Obtain $\Psi_\gamma^\ast$ by solving the RTE equations
\item Compute $S^\ast = S(E^\ast,n^\ast,\Psi_\gamma^\ast, \nabla n^\ast)$ and $\bm{v}^\ast = \bm{v}\left(\bm{E}^\ast, n^\ast\right)$. Also recompute boundary fluxes on domains and internal boundaries. 
\item Advance $n^\dagger = \frac{1}{2}\left(n^k + n^\ast + \Delta t\left[S^\ast - \nabla\cdot\left(\bm{v}^\ast n^\ast\right)\right]\right)$ and $\sigma^{k+1} = \frac{1}{2}\left(\sigma^k + \sigma^\ast + \Delta tF_\sigma^\ast\right)$.
\item Obtain the new electric field $\bm{E}^\dagger$ by solving the Poisson equation with $\rho^\dagger$ and $\sigma^{k+1}$.
\item For diffusive species, obtain $n^{k+1}$ with a implicit diffusion advance (see Eq.~\eqref{eq:tga}). Otherwise, $n^{k+1} = n^\dagger$.
\item Obtain the final electric field $\bm{E}^{k+1}$ by solving the Poisson equation with $n^{k+1}$ and $\sigma^{k+1}$.
\item Compute radiative transfer source terms $\eta^\ast = \eta\left(n^{k+1}, \bm{E}^{k+1}\right)$.   
\item Obtain $\Psi_\gamma^{k+1}$ by solving the RTE equations
\end{enumerate}
In the above, steps 1 through 7 describe a consistent SSP Runge-Kutta method of order two (Heun's method). Step 8 and 9 describe an implicit diffusion advance and step 10 and 11 are the final updates for the radiative transfer equations. Note that charge injection into the domain is a part of the advective discretization, and the injected charge is therefore also redistributed in the vicinity of cells. The rationale for this design is that one avoids a possible division by $\kappa$ when normalizing the injected charge by the volume fraction of a cut cell. 

We have not yet explored the use of subcycling in time with AMR. Our experience with production runs of our solvers is that the finest AMR level typically contains a factor of ten more patches on the finest level than any other AMR level, and the benefits of subcycling in time are therefore unlikely to lead to drastic improvements in performance. Furthermore, streamer phenomena are physically unstable, and with the current operator splitting it is not clear if the coarse-grid equations can be advanced on their own with larger time steps without impacting numerical stability. In our approach, the coarse grid solutions are always synchronized with the finer levels. In other words, if a coarse grid is covered by a finer grid, the solution data on the coarse grid is the conservatively averaged version of the fine grid solution. 

Restrictions on the time step are supplied through known criteria. For advection we restrict the time step on the CFL condition on the advection part by
\begin{equation}
  \label{eq:cfl_advection}
  \Delta t \leq \frac{\Delta x}{|\bm{v}|}.
\end{equation}
In addition, we follow the physical time scale associated with the evolution of the electric field by
\begin{equation}
  \epsilon_0\frac{\partial\bm{E}}{\partial t} = \bm{J},
\end{equation}
where discretization by using the forward Euler rule yields the estimate
\begin{equation}
  \Delta t < \textrm{Min}\left(\frac{\epsilon_0 |\bm{E}|}{|\bm{J}|}\right).
\end{equation}

\section{Verification tests}
\label{sec:verification}
To verify that our code works as expected we now present standard test problems. All of our test problems use the full code framework for the verification studies. Internally, there are no differences between the numerical cases discussed here and the more complex cases discussed in Sec.~\ref{sec:examples}. 

\subsection{The square wave advective problem}
\label{sec:advection_problem}
Our first problem, which verifies that our advective schemes are conservative and monotone and shows the amount of numerical diffusion that is involved, considers the pure advection of a single scalar with initial conditions taken as a square wave
\begin{equation}
  \label{eq:square_pulse} 
  n_0(x) = \phi_0\left(\frac{x - x_0}{L}\right),
\end{equation}
where $L$ is the pulse width of the pulse
\begin{equation}
  \phi_0(\xi) = \begin{cases}
    1, & \textrm{if } |\xi| <= 1, \\
    0, & \textrm{otherwise}.
  \end{cases}
\end{equation}

Numerically, we consider a domain $(-1,-1) \times (1,1)$ with $x_0 = 0.5$ and $L = 0.2$. The velocity of the wave is taken as $v = (v_x, v_y) = (1,0)$ and we integrate for one time period, i.e. $T = L/v_x$. For this problem, where $v_x$ is constant, the problem defined by Eq.~\eqref{eq:cdr} has solution $n(x,y,t) = n_0(x-v_xt)$, representing pure translation of the initial wave.

\begin{figure}[ht]
  \centering
  \includegraphics{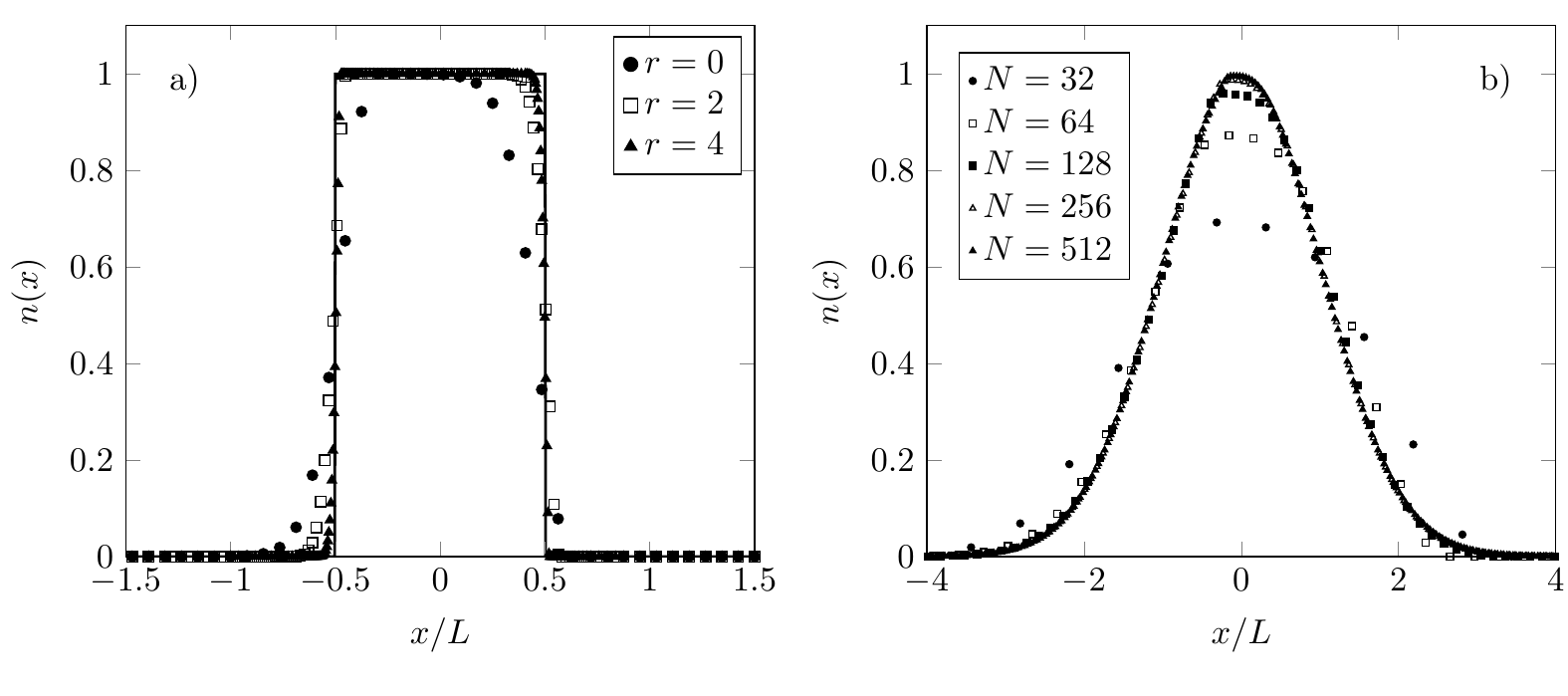}
  \caption{a) Comparison between numerically integrated solutions of the square wave problem with an analytic solution. The labels indicate the number of refinement levels used. b) Gaussian pulse advection problem. Labels represent the number of grid cells in each coordinate direction. }
  \label{fig:advection_wave_1d}
\end{figure}

Figure~\ref{fig:advection_wave_1d}a) shows the comparison between the analytic solution and the numerically computed solution for three different data sets. The data sets use a base mesh of $(128)^2$ and $0$, $2$, and $4$ levels of refinement with a factor of two between levels. In the figure, the data has been translated to the center of the analytic solution, and are plotted versus the pulse width. All three cases work as expected, and the results also show the improvement that is expected with AMR: The numerical diffusion length of a discontinuity decreases with increasing resolution. For our scheme, the diffusion length is approximately 7 cells. All three simulation cases are also conservative: Integration of the data shown in Fig.~\ref{fig:advection_wave_1d} shows that the maximum mass fluctuations are within a factor of $10^{-8}$ of the initial mass. In addition, all three cases are monotone. That is, there are no spurious oscillations or new extrema in the form of negative values or positive values exceeding the initial maxima. 

\begin{figure}[ht]
  \centering
  \includegraphics{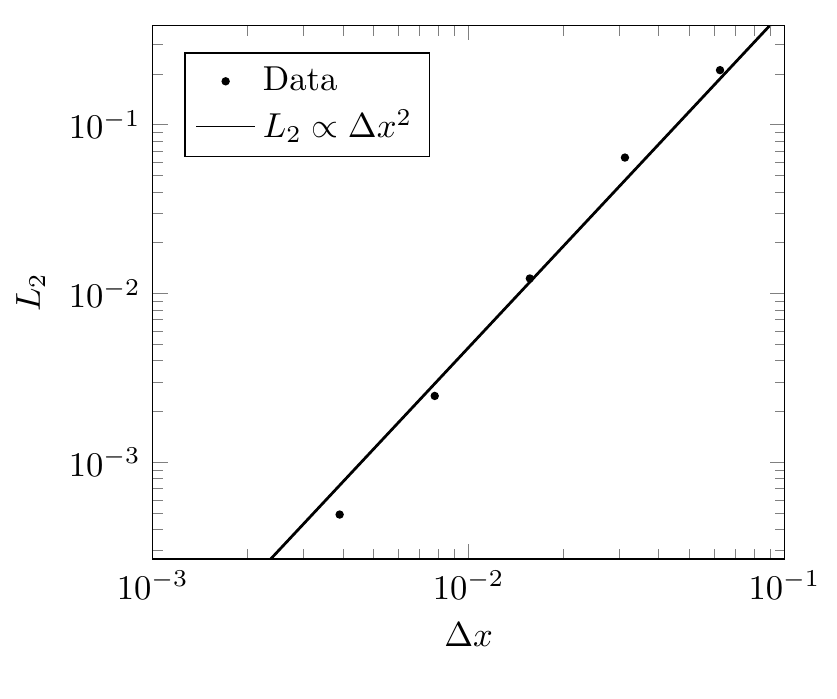}
  \caption{Convergence plot for the Gaussian wave problem.}
  \label{fig:convergence_gauss}
\end{figure}

Next, we perform a convergence test for our convection solver. We advect a Gaussian pulse
\begin{equation}
  n_1(x) = \exp\left(-\frac{x^2}{2L^2}\right)
\end{equation}
using five different resolutions. Since we want to test the convergence of our discrete scheme, we do not use AMR for these tests. The pulses have width $L=0.1$, and the cases vary in resolution from $(32)^2$ to $(512)^2$ cells. Figure~\ref{fig:advection_wave_1d}b) shows the advected pulses after one time period. We find that the numerical solutions approximate the analytic solution better when we increase the resolution. To investigate the rate of convergence of the advective solver, we compute the $L_2$ norm
\begin{equation}
  L_2(n) = \sqrt{\frac{1}{\sum_{\bmi}\kappa_{\bmi}}\sum_{\bmi}\kappa_{\bmi}\left[n_{\bmi} - n_{\textrm{e}}\left(\bm{x}_i\right)\right]^2},
\end{equation}
where $n_{\bmi}$ is the numerical solution and $n_{\textrm{e}}$ is the exact solution $n(x,t) = n_1(x-v_xT)$. Figure~\ref{fig:convergence_gauss} shows the resulting $L_2$ norm for the five cases, and we have also plotted a line $L_2\propto \Delta x^2$. The figure shows second order convergence in the $L_2$ norm. 

\subsection{The multi-fluid Poisson problem}
Our next problem considers a multi-fluid Poisson problem on a two-dimensional coaxial cable geometry. The setup has translation symmetry along the $z$-cordinate, and rotational symmetry around the $z$-axis. This problem verifies that our multi-fluid operators produce reliable solutions. We do not explicitly verify our single-phase elliptic operators, which are used for the e.g. the diffusion advancements with Euler or TGA schemes, or for the Eddington equations. In fact, we consider verification of the multi-fluid operator to be sufficient for verification purposes. The reason for this is that, internally in our code, the multi-fluid solvers are high-level C++ abstractions that couple two single-phase elliptic solvers by using the matching condition~\eqref{eq:matching}. Thus, if the multi-fluid operator is correct, we can assert with some confidence that the single-fluid operators are also correct.

\begin{figure}[h!t!b!]
  \centering
  \includegraphics{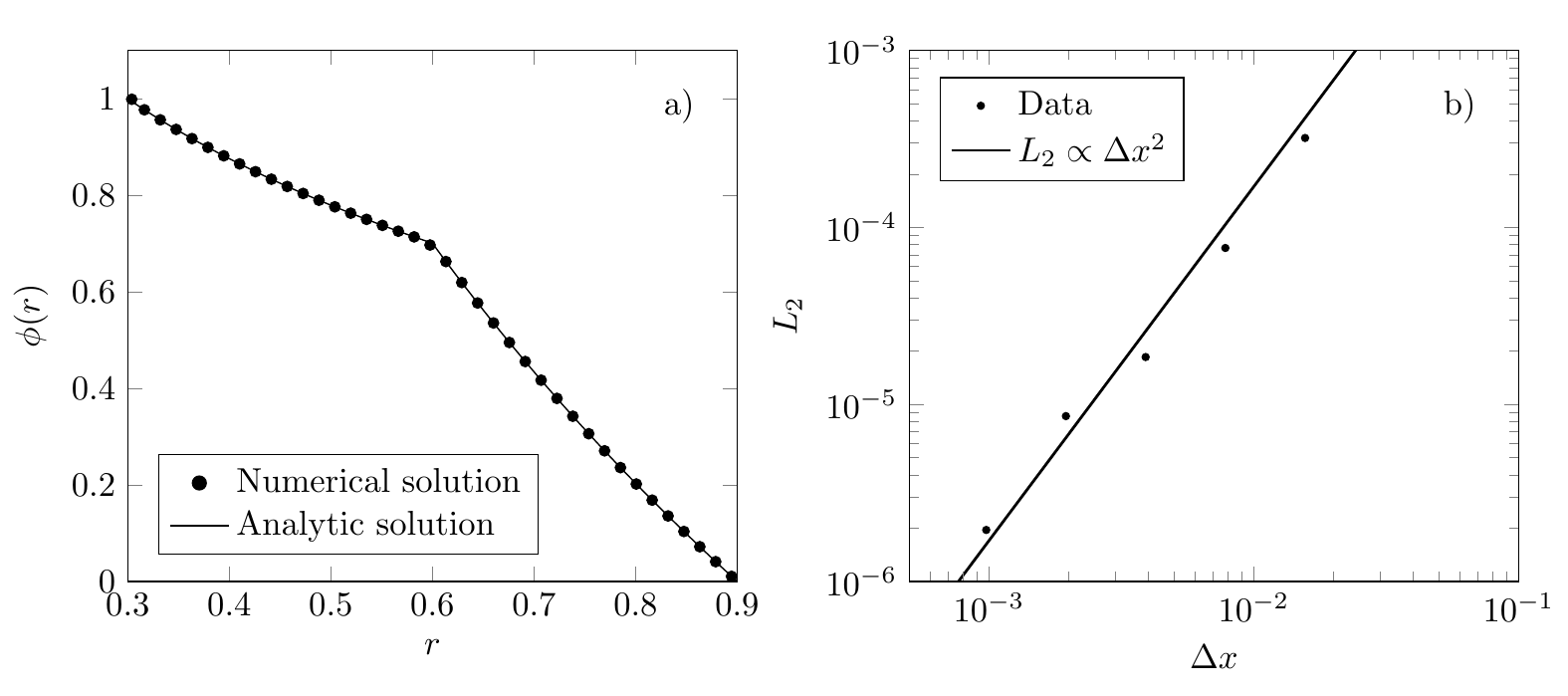}
  \caption{a) Comparison between numerically computed and analytic solutions to the multi-fluid coaxial Poisson problem. b) Convergence diagram. }
  \label{fig:poisson_coaxial}
\end{figure}

In the following we do not consider space charge, in which case the Poisson problem for the case mentioned above reduces to solving
\begin{equation}
  \label{eq:poisson_cylindrical}
  \frac{\partial}{\partial r}\left(r\epsilon\frac{\partial\phi}{\partial r}\right) = 0.
\end{equation}
We consider a geometry consisting of three concentric shells; an outer electrode, an inner electrode, and a dielectric in between the two. On the outer electrode, we take the boundary condition $\phi=0$ and on the inner electrode we take $\phi=1$. On the dielectric between the two, the boundary condition is given by Eq.~\eqref{eq:matching}. For radii, we take the outer electrode radius $R_2 = 0.9$, the inner electrode radius is $R_0 = 0.3$, and the dielectric radius is $R_1 = 0.6$. The relative permittivity of the dielectric is $\epsilon=4$ and the permittivity of the gas phase is taken to be $\epsilon_0$. We do not consider surface charge on the dielectric, in which case the exact solution to Eq.~\eqref{eq:poisson_cylindrical} is
\begin{equation}
  \label{eq:poisson_coaxial_exact}
  \phi(r) = \begin{cases}
    \displaystyle
    \phi_0 + \frac{a}{\epsilon}\ln\left(\frac{r}{R_0}\right), & R_0 \leq r \leq R_1, \\
    \displaystyle
    a\ln\left(\frac{r}{R_2}\right), & R_1 \leq r \leq R_2,
  \end{cases}
\end{equation}
where
\begin{equation}
  a = \phi_0\left[\ln\left(\frac{R_1}{R_2}\right) - \frac{1}{\epsilon}\ln\left(\frac{R_1}{R_0}\right)\right]^{-1}.
\end{equation}

We first compute the numerical solutions on a $(-1, -1)\times (1,1)$ domain using $(256)^2$ cells. Note that while the solution here is effectively one-dimensional, our numerically computed solution is 2D. This also means that the data that we extract is off-axis by half a grid cell. Figure~\ref{fig:poisson_coaxial}a) shows a comparison of Eq.~\eqref{eq:poisson_coaxial_exact} with the numerically computed solution. We find that our Poisson solver works as expected, essentially reproducing the analytic solution. 

Next, we perform a convergence study on the same problem by computing numerical solutions on mesh sizes $(128)^2$, $(256)^2$, $(512)^2$, $(1024)^2$, and $(2048)^2$. Figure~\ref{fig:poisson_coaxial}b) shows the computed norms (points) plotted together with a curve $L_2\propto \Delta x^2$. The data points show second order convergence in the $L_2$ norm. Deviations from the curve are most likely due to the data-centering that we use in our software: In our output methods, we use the cell-centered potential from the gas side in the multifluid cells, whose center may fall on either side of the embedded boundary for different resolutions, and these errors can dominate. 

\subsection{The surface charge conservation problem}
Our final test problems demonstrate charge conservation at insulating surfaces. This problem verifies that our implementation of charge transport on the gas-matter interface is physically correct. 

Our first conservation problem considers the one-dimensional case as in \ref{sec:advection_problem} but with a dielectric slab filling the half space $x > 0.0$. Extrapolated outflow conditions are applied to the advected scalar. We use a base mesh of $(128)^2$ cells and use a single level of refinement with refinement factor $2$.

\begin{figure}[h!t!b!]
  \centering
  \includegraphics{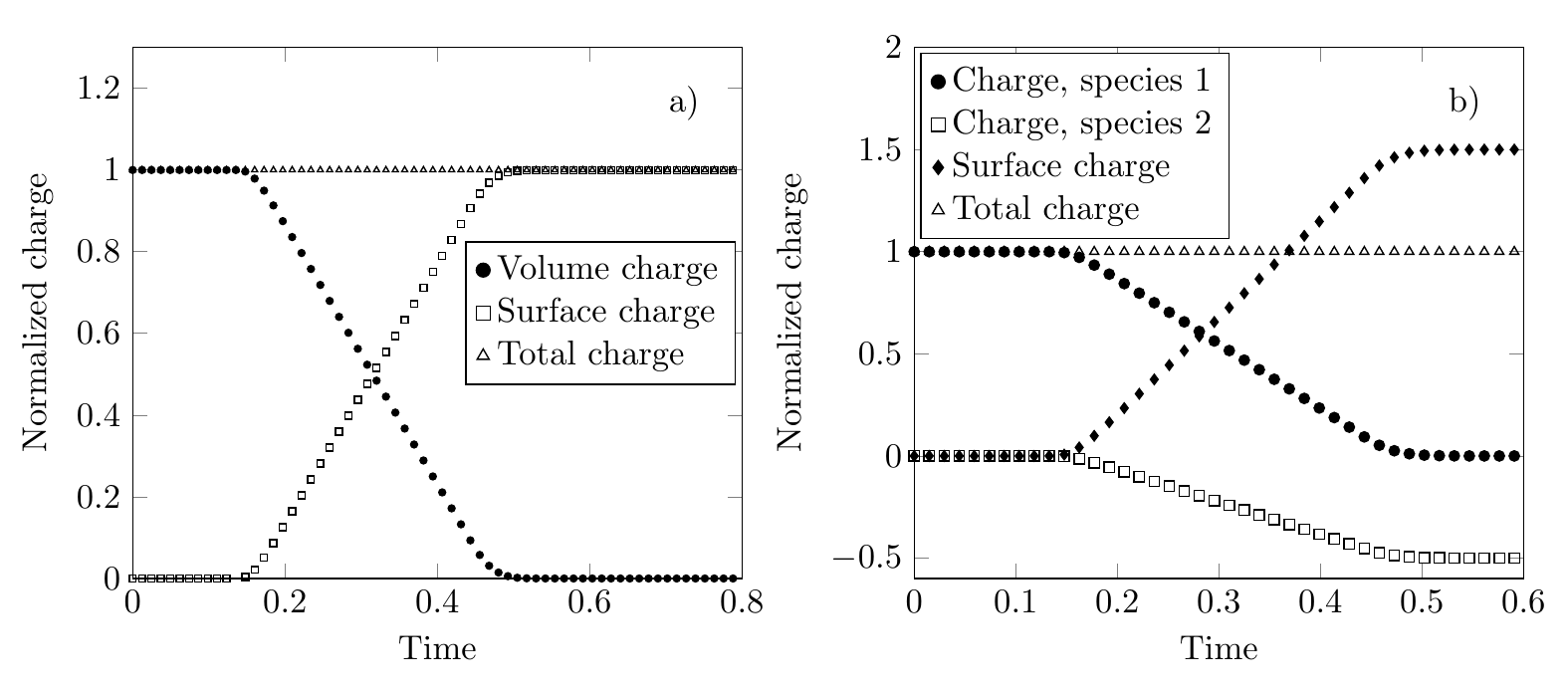}
  \caption{Charge conservation test on embedded boundaries. The figure shows the total volume charge, surface charge, and their sum. The $y$-axis has been normalized against the total initial charge.}
  \label{fig:surface_conservation}
\end{figure}

Figure~\ref{fig:surface_conservation}a) shows the evolution of the normalized volume and surface charge before and after the square wave pulse is absorbed by the insulator. The figure also shows the sum of charges. For this problem, where the pulse has a constant amplitude and velocity, one expects that the charging rate of the surface is close to linear. In the simulations, this is only achieved away from the rising and falling flanks, which is due to numerical diffusion which tends to smooth out the initial discontinuities (see Fig~\ref{fig:advection_wave_1d}). Furthermore, we find excellent charge conservation on both the surface and in the volume (to within a factor $10^{-8}$), and therefore conclude that our coupled gas-surface kinetics work as expected. 

Next, we perform a second surface conservation verification problem by coupling the boundary conditions of two advected species on the dielectric slab. We consider opposite signs of the two species' charge, i.e. the two species propagate in opposite directions. For the first species, its initial density is set to the square wave form above (recall Eq.~\eqref{eq:square_pulse}), and the other species' initial density is set to zero. On the dielectric boundary, we couple the boundary fluxes $F_1$ and $F_2$ of the two species by imposing
\begin{equation}
  F_2 = -\beta \textrm{Max}\left(0,F_1\right)
\end{equation}
where $\beta=1/2$ is the emission coefficient of the second species. In our implementation, a positive flux is directed out of the cell so that $F_2$ provides an influx only if $F_1$ provides an outflux. This problem verifies that ion bombardment coupling works as expected. In this particular problem, we expect that the dielectric boundary reflects the incoming square pulse into an outgoing square wave with half the amplitude. 

Figure~\ref{fig:surface_conservation}b) shows the integrated charges for the double advection test. The figure shows that the total charge in the domain is conserved to a very good degree (open triangles). Before the initial square wave strikes the dielectric, the total charge is contained only in the incoming species (filled circles). After the incoming square wave is absorbed by the dielectric, the ejected charge back into the gas phase is a factor of $-1/2$ times the initial charge (open squares), which is what we expect with an emission coefficient of $1/2$ and opposite signs of the species' velocities. Charge conservation then implies that the surface must have absorbed a charge of $3/2$ times the initial charge, which is what we observe (filled diamonds).

\subsection{Model verification}
We now verify that the combined algorithm is second order convergent in time by means of two-dimensional experiments. The convergence tests use Richardson extrapolation; i.e. we use a fine resolved solution as a replacement for the exact solution. We then compute the norm between the reference solution and a solution with coarser resolution. We consider a $(2\mm)^2$ domain with a blade electrode protruding $1\,\mm$ from the top domain edge. The electrode has a $100\,\um$ radius and is live with a voltage of $5\,\kV$. We impose homogeneous Neumann boundary conditions for the Poisson equation on the left and right side edges, whereas the bottom edge is grounded. The plasma kinetics that we use for this test is summarized in \ref{sec:discharge_model} and solves for three species: electrons $n_e$, positive ions $n_+$, and negative ions $n_-$, with only electrons being diffusive.

\begin{figure}[h!t!b!]
  \centering
  \includegraphics{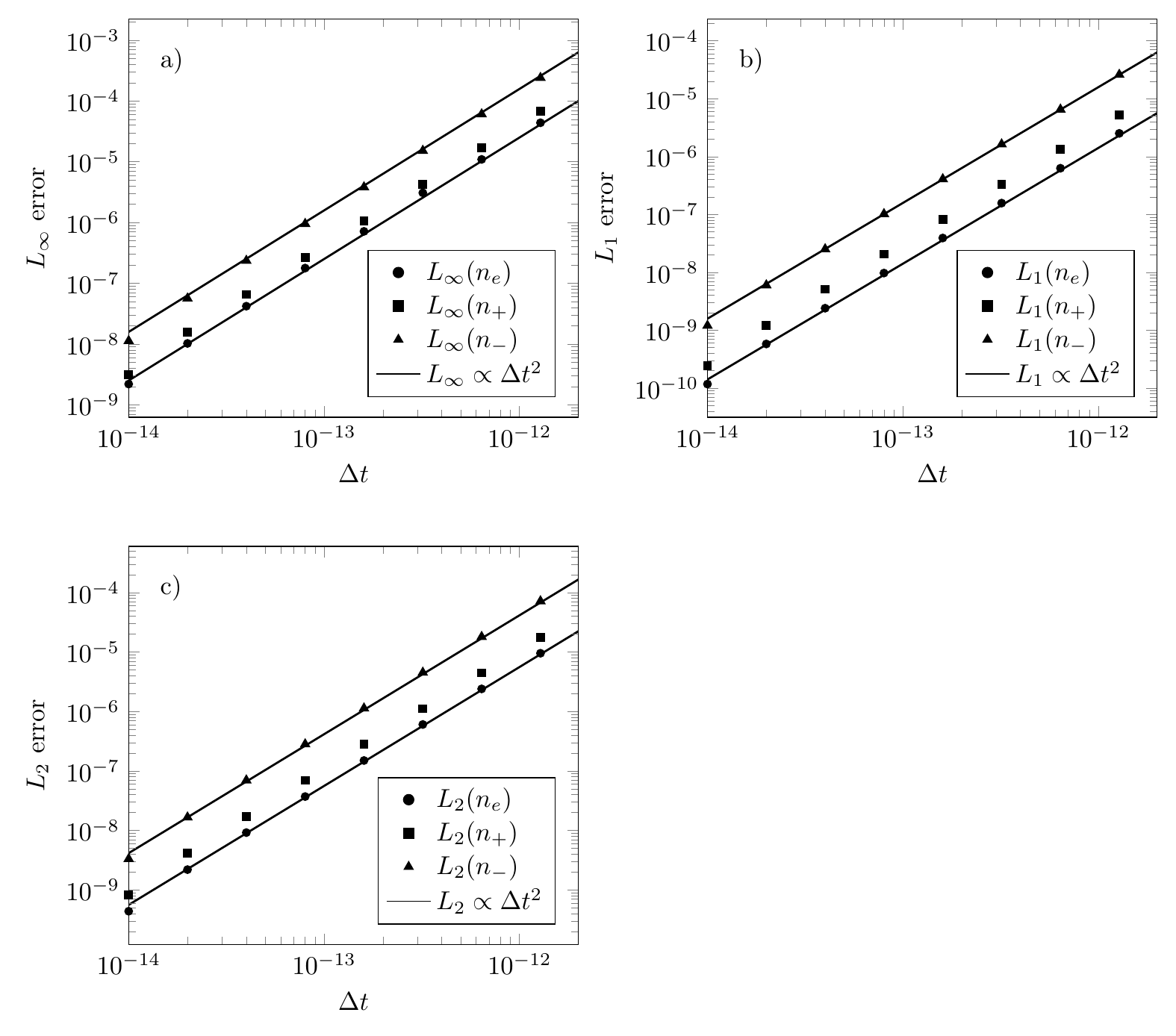}
  \caption{Temporal 2D convergence tests. The data points show the normalized error norms between the reference solution and additional solutions with coarser time steps. }
  \label{fig:convergence2d_norms}
\end{figure}

For initial conditions, we consider an electron-ion seed
\begin{align}
  n_e &= n_0\exp\left(-\frac{\left(\bm{x}-\bm{x}_0\right)^2}{R^2}\right), \\
  n_+ &= n_e, \\
  n_- &= 0,
\end{align}
where $n_0 = 10^{18}\,\m^3$, $R=100\,\um$, and $\bm{x}_0$ is the position of the electrode tip. We compute the solution error by comparison with a reference solution, which is taken to be the solution with a step size of $\Delta t_{\textrm{ref}} = 5\times10^{-15}\,\textrm{s}$ which is advanced for $1024$ time steps on a $(2048)^2$ cell grid. This is done for time steps $\Delta t_n = 2^n\Delta t_{\textrm{ref}}$, $n=1, 2, \ldots 8$ which ranges up to $256\Delta t_{\textrm{ref}}$. The relative $L_\infty$, $L_1$, and $L_2$ error norms are then computed for all three species. The results of these computations are presented in Fig.~\ref{fig:convergence2d_norms} and are plotted together with the line $L_i\propto \Delta t^2$. We find second order convergence in the $L_\infty$, $L_1$, and $L_2$ norms. 

Next, we present spatial convergence tests in 2D using the same geometry as above, and in 3D using a domain of $(1\mm)^3$ with a rod electrode protruding from the top domain edge. In 2D and 3D, the solution is advanced $16$ time steps using a step size of $\Delta t_{\textrm{ref}} = 10^{-14}\,\textrm{s}$. In 2D, the reference solution is taken to be the solution on a $(2048)^2$, and in 3D the reference solution uses a $(512)^3$ grid. The reference solution is denoted $\phi^{\Delta x}$ in both cases, and we compare this solution with coarser solutions $\phi^{2\Delta x}$ and $\phi^{4\Delta x}$. On the coarser grids, the errors are computed by comparison with a coarsening of $\phi^{\Delta x}$, i.e. the error is computed as
\begin{equation}
  E_{2\Delta x} = \phi^{2\Delta x} - A\left(\phi^{\Delta x}\right),
\end{equation}
where $A\left(\phi^{\Delta x}\right)$ indicates a coarsening of $\phi^{\Delta x}$ with a coarsening factor of 2. Likewise, $E_{4\Delta x} = \phi^{4\Delta x} - A\left(A\left(\phi^{\Delta x}\right)\right)$ indicates the error on the grid with resolution $4\Delta x$. The convergence rates are computed as
\begin{equation}
  p = \frac{\log\left(\frac{E_{4\Delta x}}{E_{2\Delta x}}\right)}{\log(2)},
\end{equation}
and results of these computations are given in Table~\ref{tab:convergence2d_norms} and Table~\ref{tab:convergence3d_norms}. We find second order convergence in all three norms in both 2D and 3D. 

\begin{table}[h!t!b!]
  \centering
  \begin{tabular}{ll|ll|l}
    Norm      & Variable & $E_{2\Delta x}$ & $E_{4\Delta x}$ & $p$ \\
    \hline
    $L_\infty$ & $n_e$ & $2.3399\times 10^{-5}$  & $1.1063\times 10^{-4}$ & $2.24$ \\
    $L_\infty$ & $n_+$ & $2.1426\times 10^{-5}$   & $1.0646\times 10^{-4}$ & $2.31$ \\
    $L_\infty$ & $n_-$ & $5.7657\times 10^{-10}$  & $2.8769\times 10^{-9}$ & $2.32$ \\
    $L_1$     & $n_e$ & $2.7712\times 10^{-7}$   & $1.3826\times 10^{-6}$ & $2.31$ \\
    $L_1$     & $n_+$ & $2.7711\times 10^{-7}$   & $1.3827\times 10^{-6}$  & $2.31$ \\
    $L_1$     & $n_-$ & $7.3172\times 10^{-12}$  & $3.6902\times 10^{-11}$ & $2.33$ \\
    $L_2$     & $n_e$ & $1.4245\times 10^{-6}$   & $7.0809\times 10^{-6}$ & $2.31$ \\
    $L_2$     & $n_+$ & $1.4240\times 10^{-6}$   & $7.0804\times 10^{-6}$ & $2.31$ \\
    $L_2$     & $n_-$ & $3.6675\times 10^{-11}$  & $1.8744\times 10^{-10}$ & $2.35$ \\ 
  \end{tabular}
  \caption{2D spatial convergence tests based on a $(2048)^2$ grid with $\Delta x = 0.976\um$. }
  \label{tab:convergence2d_norms}
\end{table}

\begin{table}[h!t!b!]
  \centering
  \begin{tabular}{ll|ll|l}
    Norm      & Variable & $E_{2\Delta x}$ & $E_{4\Delta x}$ & $p$ \\
    \hline
    $L_\infty$ & $n_e$ & $9.7570\times 10^{-4}$   & $4.7717\times 10^{-3}$ & $2.29$ \\
    $L_\infty$ & $n_+$ & $9.7156\times 10^{-4}$   & $4.7153\times 10^{-3}$ & $2.28$ \\
    $L_\infty$ & $n_-$ & $1.3212\times 10^{-8}$   & $6.7158\times 10^{-8}$ & $2.34$ \\
    $L_1$     & $n_e$ & $7.5879\times 10^{-7}$  & $3.7863\times 10^{-6}$  & $2.32$ \\
    $L_1$     & $n_+$ & $7.5695\times 10^{-7}$  & $3.7847\times 10^{-6}$  & $2.32$ \\
    $L_1$     & $n_-$ & $1.9289\times 10^{-11}$ & $9.7338\times 10^{-11}$ & $2.33$ \\
    $L_2$     & $n_e$ & $9.8833\times 10^{-6}$  & $4.8467\times 10^{-5}$  & $2.29$ \\
    $L_2$     & $n_+$ & $9.8760\times 10^{-6}$  & $4.8471\times 10^{-5}$  & $2.30$ \\
    $L_2$     & $n_-$ & $2.4219\times 10^{-10}$  & $1.2474\times 10^{-9}$ & $2.36$ \\ 
  \end{tabular}
  \caption{3D spatial convergence tests based on a $(512)^2$ grid with $\Delta x = 3.906\um$. }
  \label{tab:convergence3d_norms}
\end{table}

\section{Weak scalability}
\label{sec:weak_scalability}
\begin{figure}[h!t!]
  \centering
  \includegraphics[height=.3\textwidth]{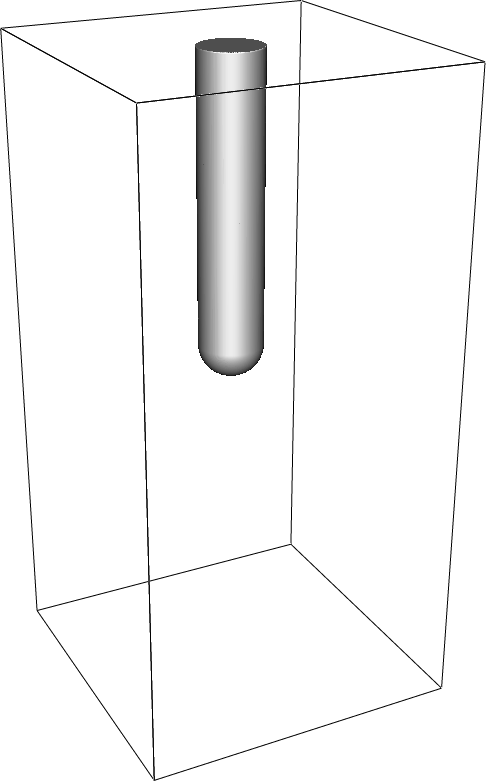}
  \hspace{1em}
  \includegraphics[height=.3\textwidth]{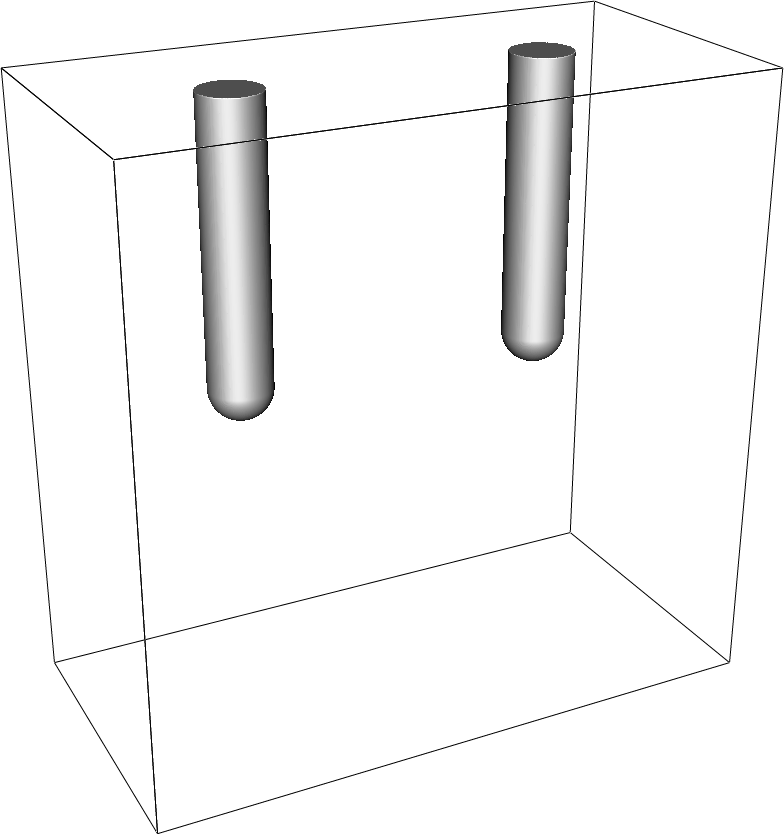}
  \hspace{1em}
  \includegraphics[height=.3\textwidth]{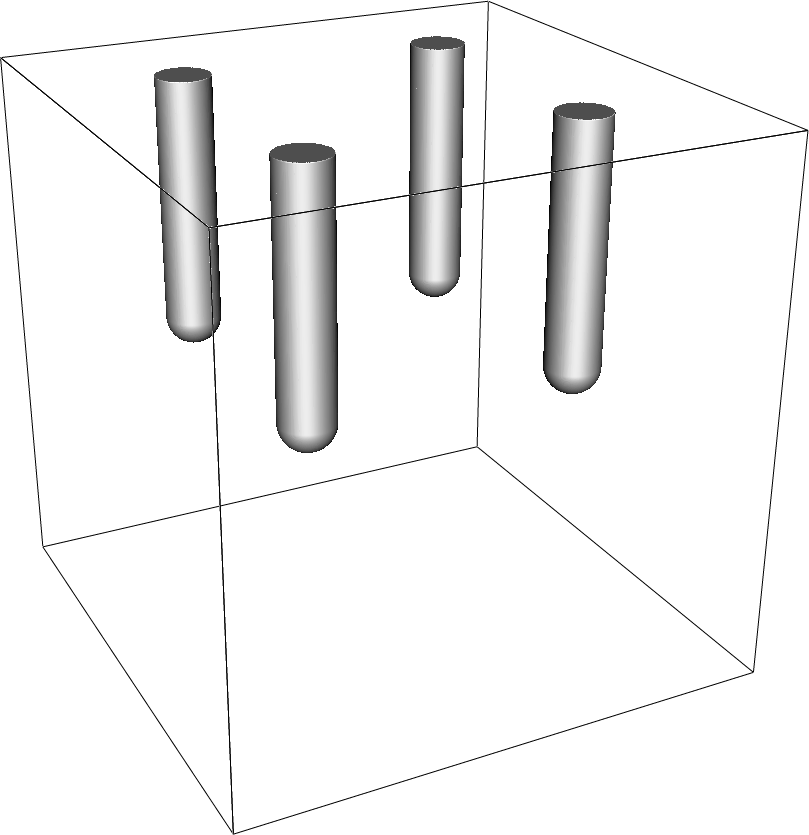}
  \caption{Geometries used for weak scalability study for concurrencies $C$. Left: $128\times128\times256$ grid. Used for $C=256, 2048$. Middle: $256\times 128\times 256$ grid. Used for $C=512, 4096$. Right: $256\times256\times256$ grid. Used for $C=1024, 8192$. All geometries are decomposed with $32^3$ patches.}
  \label{fig:replication}
\end{figure}

We now present results on weak scalability of the combined algorithm code on uniform grids. Weak scaling involves increasing the number of processor cores by the same factor as the geometry refinement. In 3D, increasing the resolution by a factor of 2 requires 8 times as many processor cores. These tests are performed by evolving the discharge model defined in \ref{sec:discharge_model} for 10 time steps and then taking the average of the execution time. We consider the three geometries in Fig.~\ref{fig:replication} with initial conditions on the charged species taken as $n_e = n_+ = 10^{10}\,\m^3$, $n_- = 0$, whereas the boundary conditions on the Poisson equation are: Live Dirichlet on the electrodes and top domain faces; homogeneous Dirichlet on the bottom domain face, and homogeneous Neumann boundary conditions elsewhere. The resolution on each geometry is then successively doubled and the core count is increased by a factor of 8. The computational domains range in size from $128\times128\times256$ to $1024\times1024\times1024$ and are decomposed with $32^3$ grid patches. We distribute the patches uniformly to all cores (i.e. the patch volumes are used as proxies for load balancing). These tests are run at core counts up to $8192$ on Fram, which is a Lenovo NeXtScale nx360M5 with dual 16-core Intel Xeon E5-2683v4 (2.1GHz) processors per node and up to 896 cores on SuperMUC Phase 2, which is a Lenovo NeXtScale nx360M5 WCT with dual 14-core Intel Xeon E5-2697v3 (2.6GHz) processors per node. On Fram, we have 4 patches per CPU core whereas on SuperMUC Phase 2 we have at maximum 3 patches per CPU core. 

\begin{figure}[h!t!b!]
  \centering
  \includegraphics{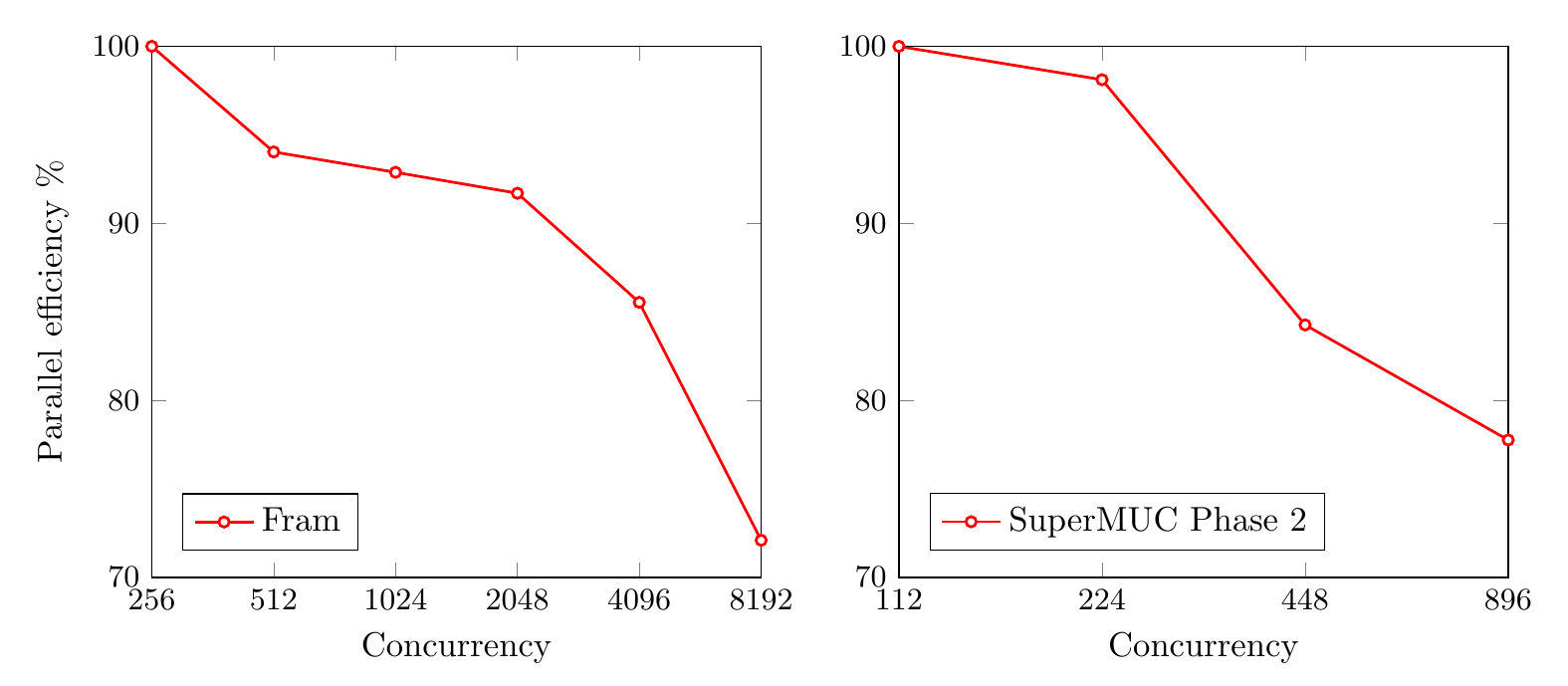}
  \caption{Weak scaling on Lenovo NeXtScale nx360M5 (Fram) and SuperMUC Phase 2. The vertical axis is the parallel efficiency and the horizontal axis is the number of processor cores. For reference, the average time per step was around 14 seconds at 256 cores on Fram and about 10 seconds at 112 cores on SuperMUC Phase 2. }
  \label{fig:weak_scalability}
\end{figure}

The results of the scalability studies on Fram and SuperMUC are shown in Fig.~\ref{fig:weak_scalability} and show reasonable scalability. On Fram, we observe a parallel efficiency above 70\% from 256 to 8192 cores and we observe about the same efficiency on SuperMUC Phase 2 from 112 cores to 896 cores. The scaling tests were not run further due to availability of each machine, but further scaling tests are likely to be performed in the future. 

\section{Numerical examples}
\label{sec:examples}
In this section we provide basic numerical examples. Note that the examples are intended as demonstrations of capabilities, and that some of these examples may be recalculated or analyzed more rigorously in the future. The examples below use refinement criteria based on resolution of space charge layers and impact ionization regions. At each regrid stage, we compute a tracer field
\begin{equation}      
  T = \frac{\left|\bm{E}\right|}{\textrm{max}\left(\left|\bm{E}\right|\right)}.
\end{equation}
We refine if
\begin{equation}
  \frac{1}{T}\left|\nabla T\right|\Delta x > \epsilon_1 \lor T > \epsilon_2.
\end{equation}
where $\epsilon_i$ are tagging thresholds. The first of these criteria is responsible for resolving gradients in the electric field, which typically arise due to space charge formation. The second test resolves the streamer head. 

Likewise, we coarsen a cell if
\begin{equation}
  \frac{1}{T}\left|\nabla T\right|\Delta x < \epsilon_3 \land T < \epsilon_4.
\end{equation}
For the simulations below, we take $\epsilon_1=\epsilon_3 = 0.1$, $\epsilon_2 = 0.8$, and $\epsilon_4 = 0.2$.

%\subsection{Remarks on grid convergence}
%We do not perform grid convergence studies for the cases below. While we consider our code to be reasonably verified following the discussion in Sec.~\ref{sec:verification}, grid convergence for practical streamer simulations is exceedingly more difficult because the computer solutions move with different velocities for different grid resolutions (see e.g. \cite{Bagheri2018}). Convergence of error norms therefore only make sense for sufficiently fine resolutions where all coarse-grained effects are gone and the solution is shown to move at the same velocity on an even finer grid. In other words, the solution must already be close to convergence. Unfortunately, for atmospheric pressure plasmas this resolution is less than $1\,\um$ (see e.g. \cite{Bagheri2018}, Fig.~7) which makes this process very difficult for practical cases, at least for large scale 3D. In fact, most streamer simulations presented in the literature to date are probably underresolved, but we remark that there are even larger variations with respect to the spatial discretization methods that are applied \cite{Bagheri2018}. 

\subsection{Creeping streamers along rough surfaces}
We first consider the propagation of positive streamers along a rough surface. The computational domain is $2\cm\times 2\cm$, with a a blade electrode protruding $1\cm$ from the top plane; the thickness and curvature of the electrode is $250\,\um$. The dielectric is represented by a planar level-set surface that sits about $1\,\mm$ away from the electrode, and whose surface has been displaced with Perlin noise \cite{Perlin2002} with an amplitude of roughly $100\,\um$. The relative permittivity of the dielectric is $4$. For initial conditions we take $n_e=n_+=10^{10}\,\m^3$. 

\begin{figure}[h!t!b!]
  \centering
  \includegraphics[width=.49\textwidth]{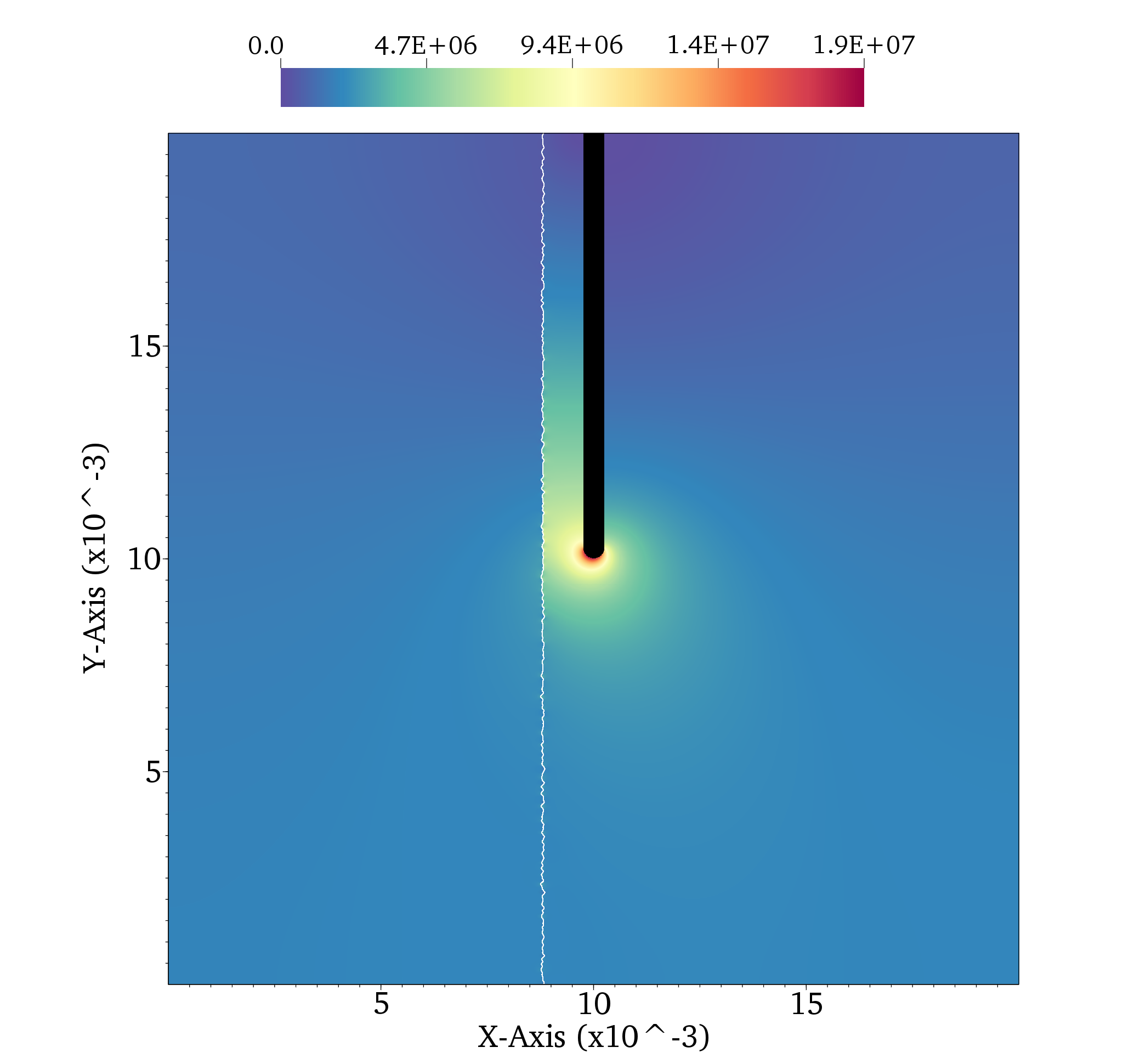}
  \caption{Initial field distribution for the DBD arrangement. The color coded data shows the electric field magnitude.}
  \label{fig:dbd_overview}
\end{figure}

In this example we use a base mesh of $128^2$ with 4 AMR levels with a refinement factor of $4$ between levels, yielding a base resolution of $312.5\,\um$ and a finest level resolution of $1.22\,\um$. Computationally, the sweet spot for this simulation is found by using a minimum box size of $16$ and a maximum box size of $32$. The integration is done with a CFL number of $0.8$. For boundary conditions on the Poisson equation, we use homogeneous Neumann conditions on the left and right side edges. We impose a voltage of $30\,\kV$ on the top domain edge and on the electrode, whereas the bottom domain edge is grounded.

\begin{figure*}[h!t!b!]
  \centering
  \includegraphics[height=.3\textheight]{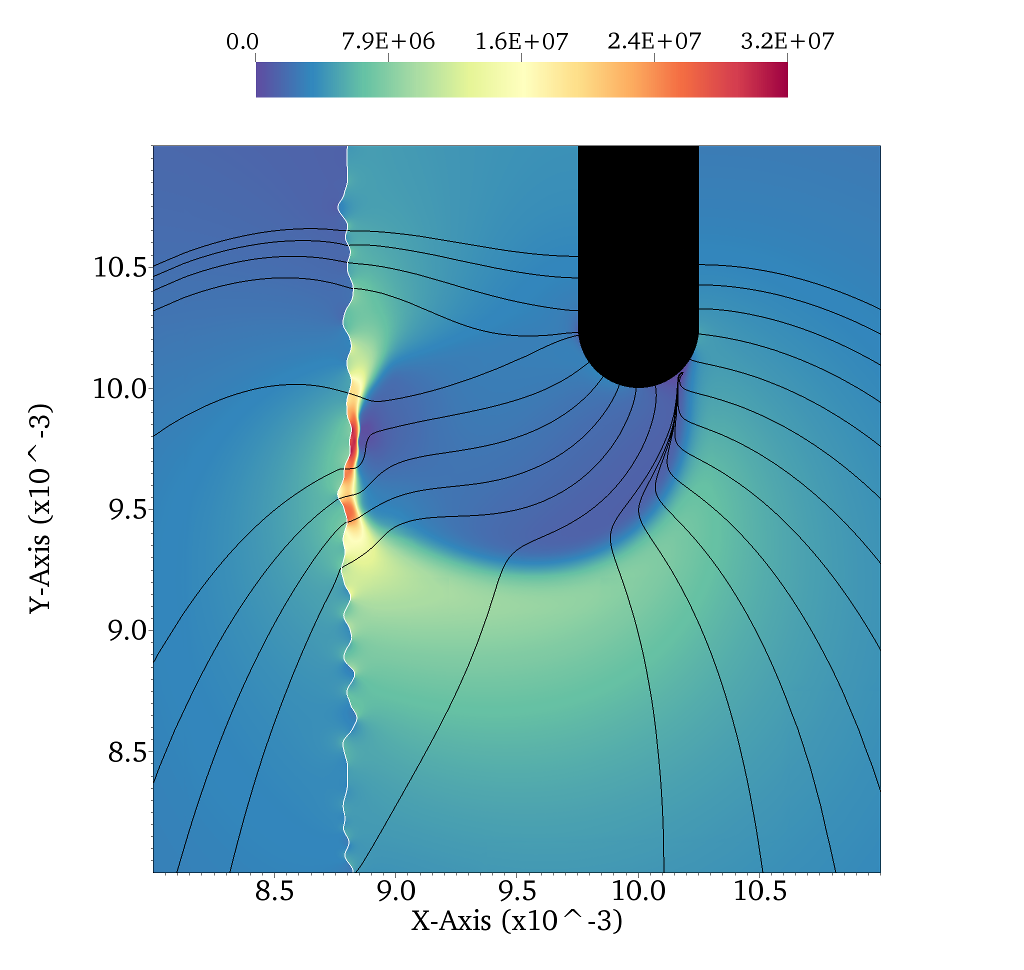}
  \includegraphics[height=.3\textheight]{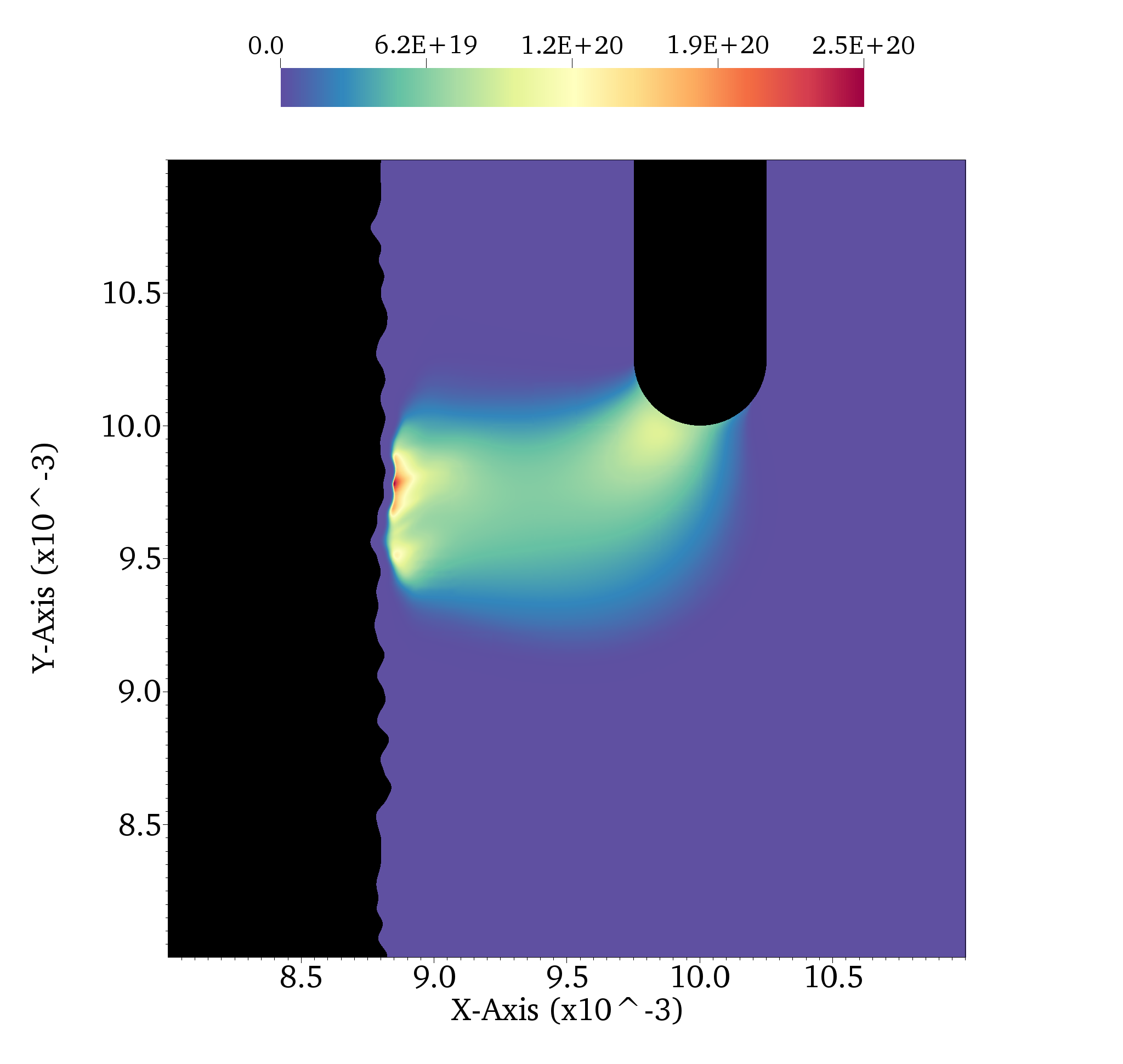} \\
  \includegraphics[height=.3\textheight]{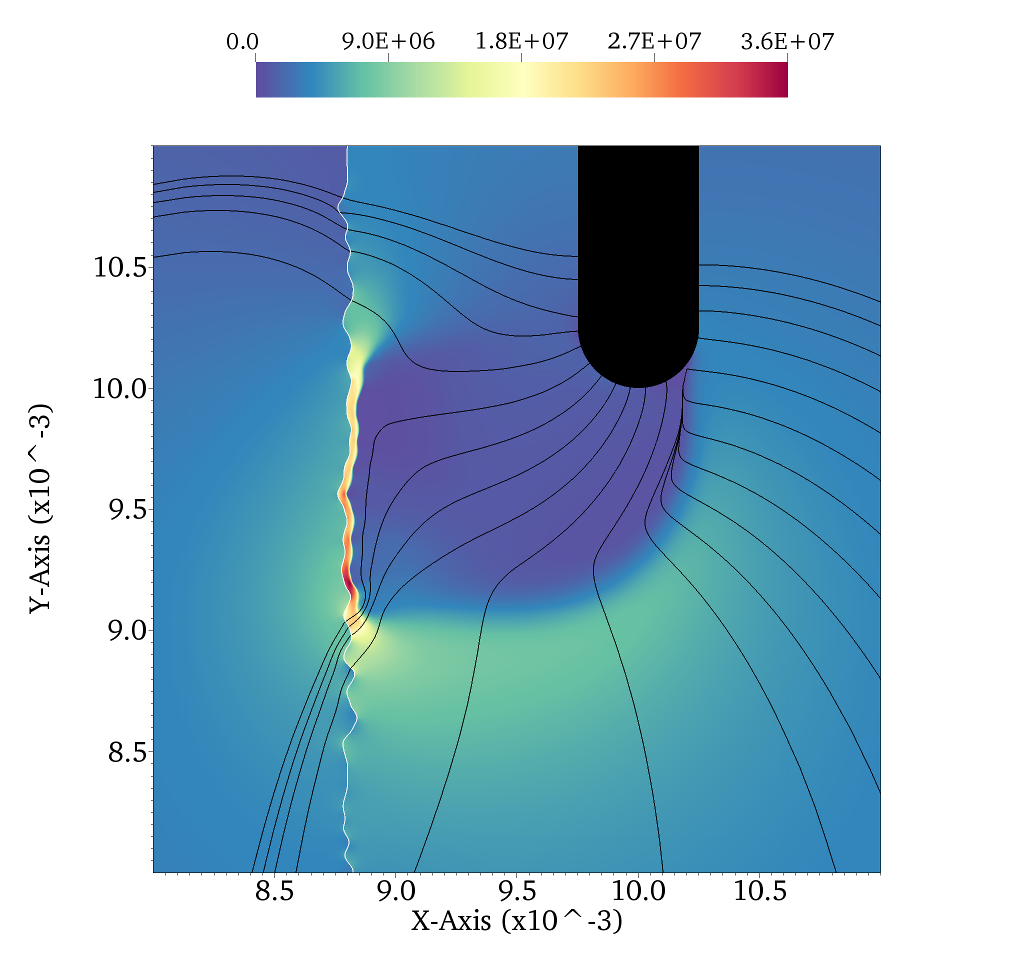}
  \includegraphics[height=.3\textheight]{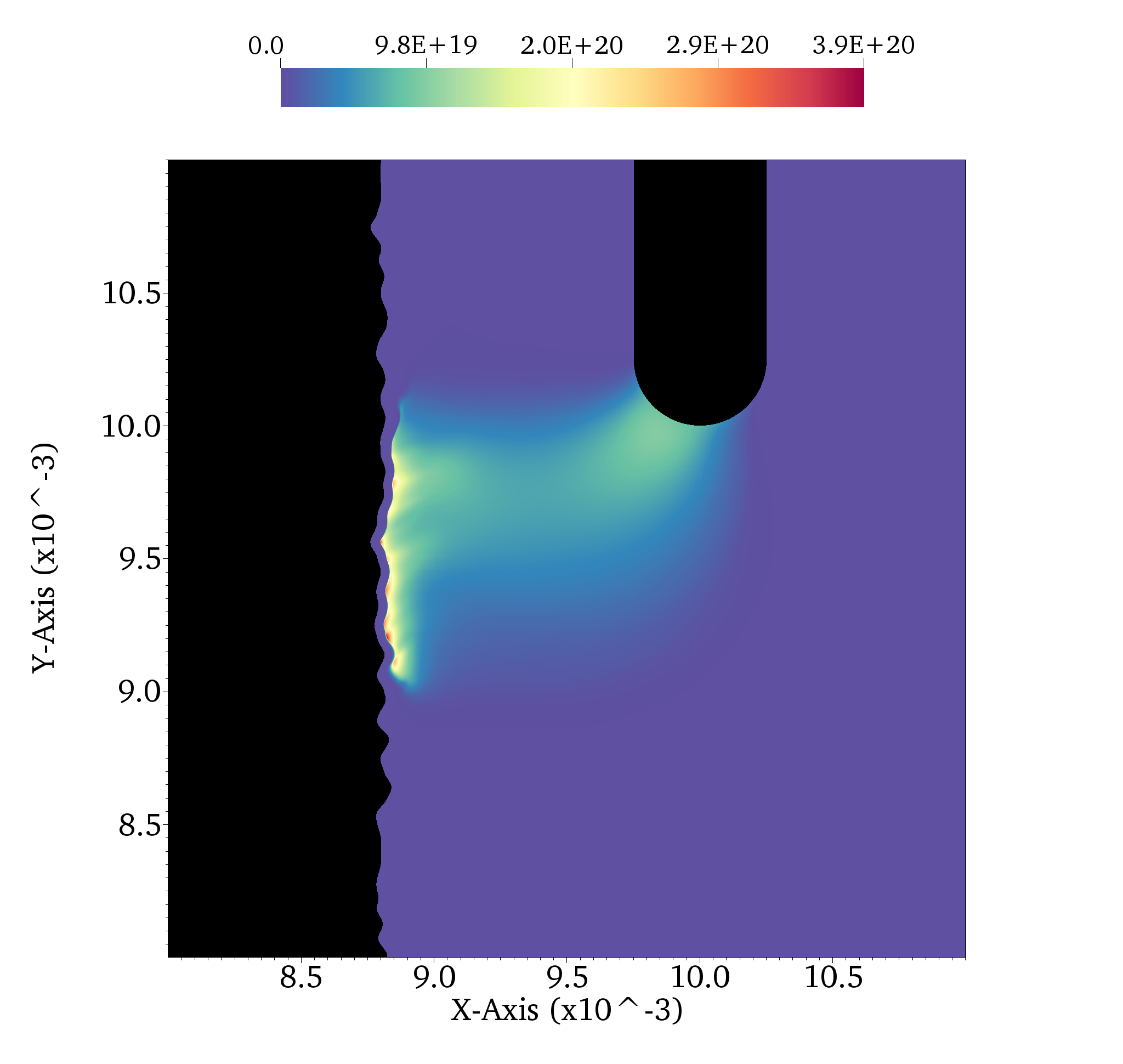} \\
  \includegraphics[height=.3\textheight]{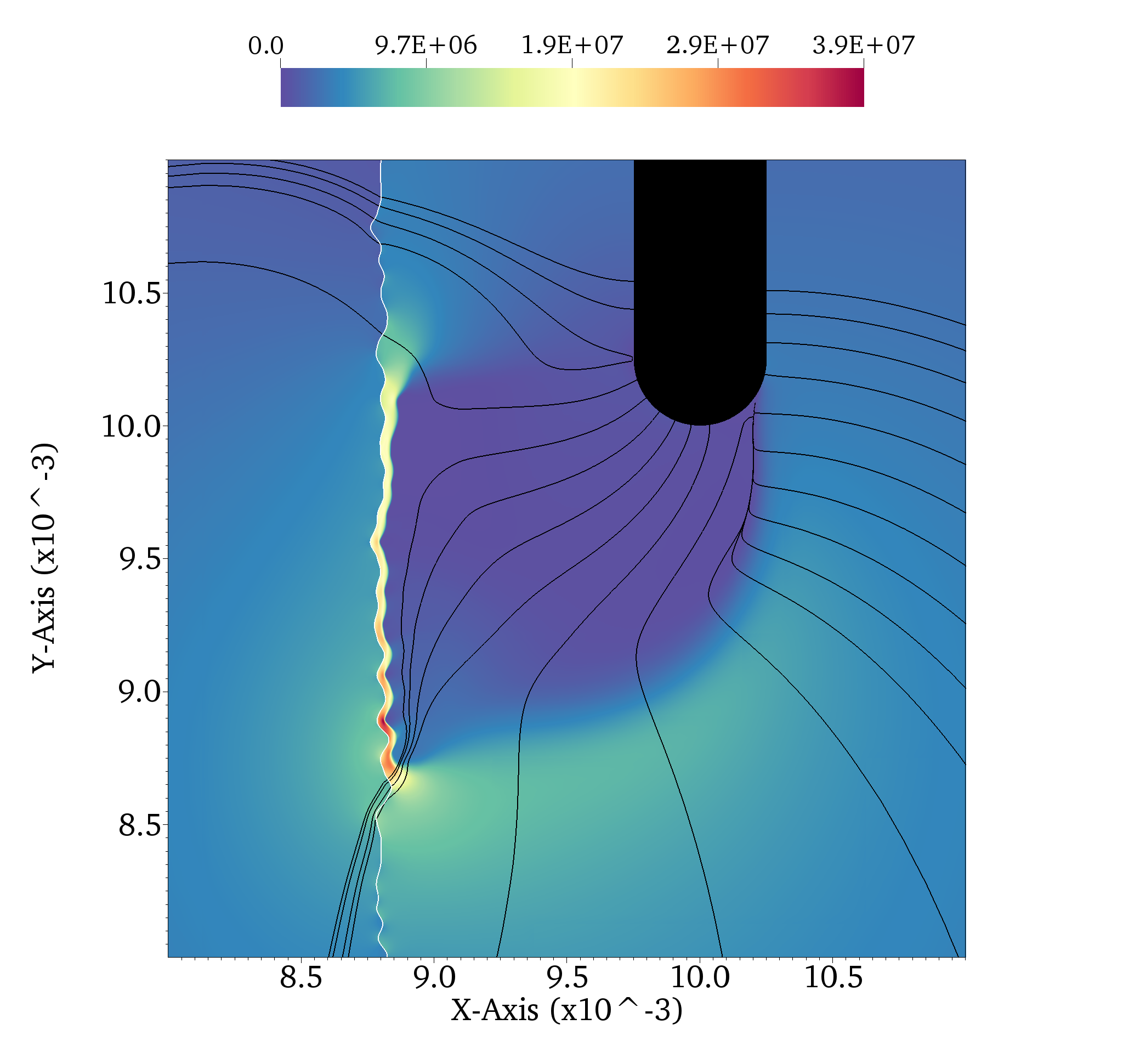}
  \includegraphics[height=.3\textheight]{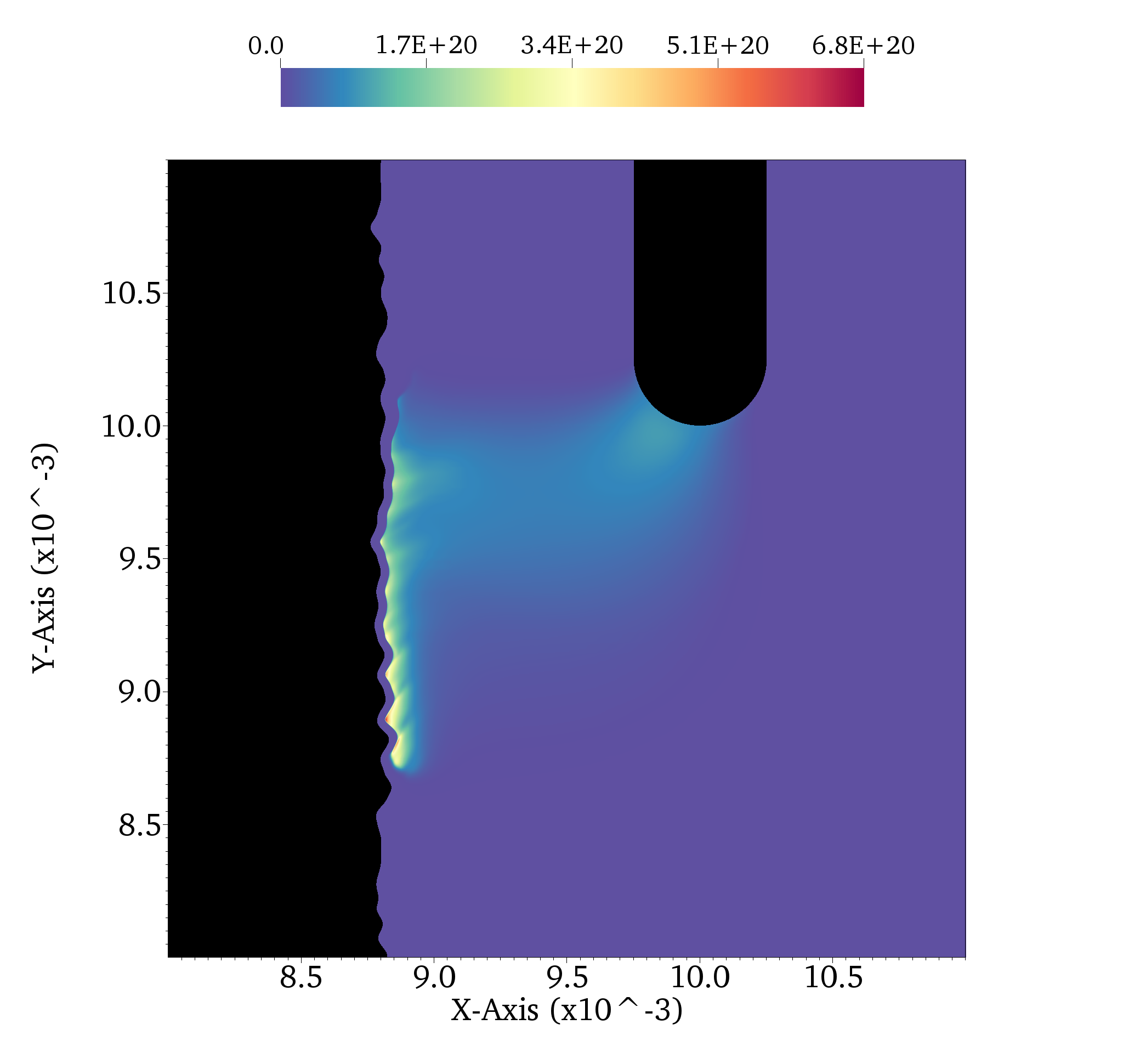}
  \caption{Snapshots of the evolution after inception. The figure shows the field magnitude (left column) and plasma density (right column). Top row: Snapshot at $1\,\ns$. Middle row: Snapshot at $1.25\,\ns$. Bottom row: Snapshot at $1.5\ns$. }
  \label{fig:dbd_evolution}
\end{figure*}

Figure~\ref{fig:dbd_overview} shows the initial field for this configuration. Arrangements such as these (often called ''triple points'') are particularly problematic in high-voltage engineering where streamer flashover is unwanted. The degraded dielectric performance occurs because the field lines that cross into the dielectric experience enhanced voltage drops in the gas-phase due polarization of the dielectric. In turn, this can lead to rapid streamer inception in the dielectric-gas gap.

\begin{figure*}[h!t!]
  \centering
  \includegraphics[width=.45\textwidth]{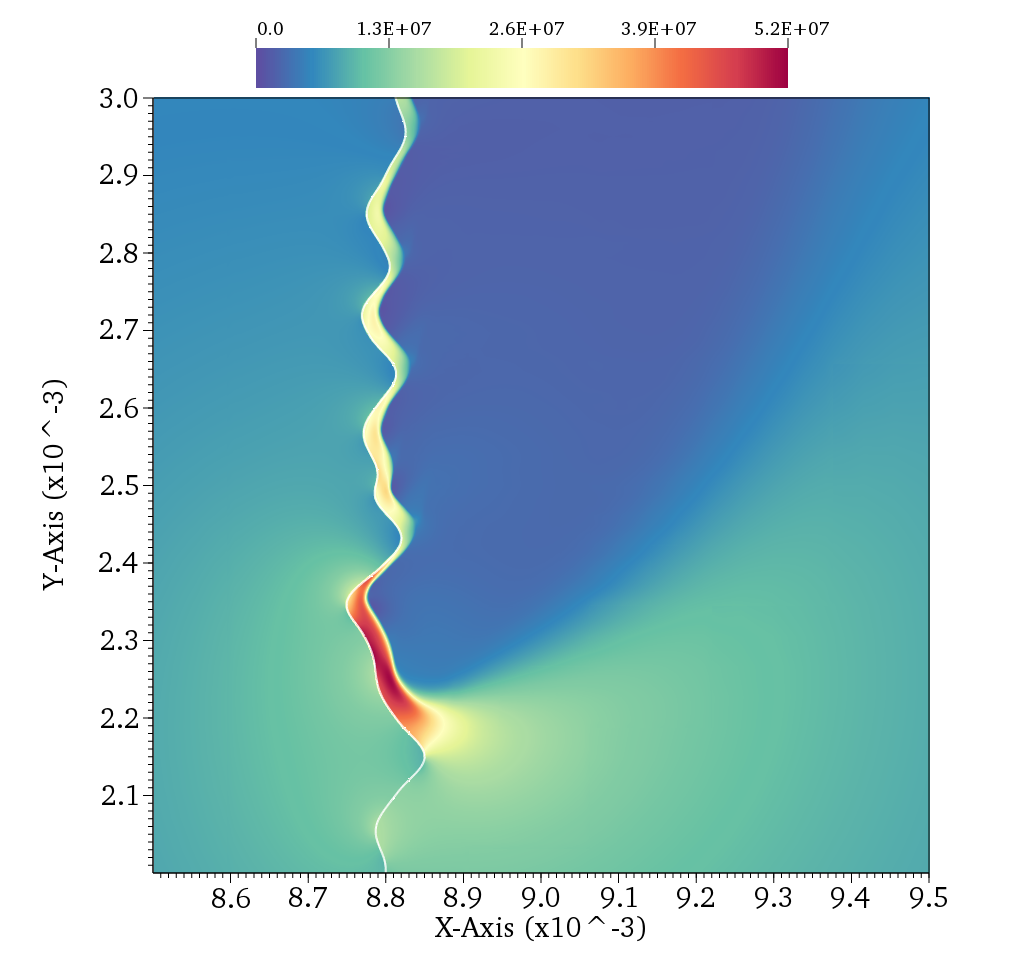}
  \includegraphics[width=.45\textwidth]{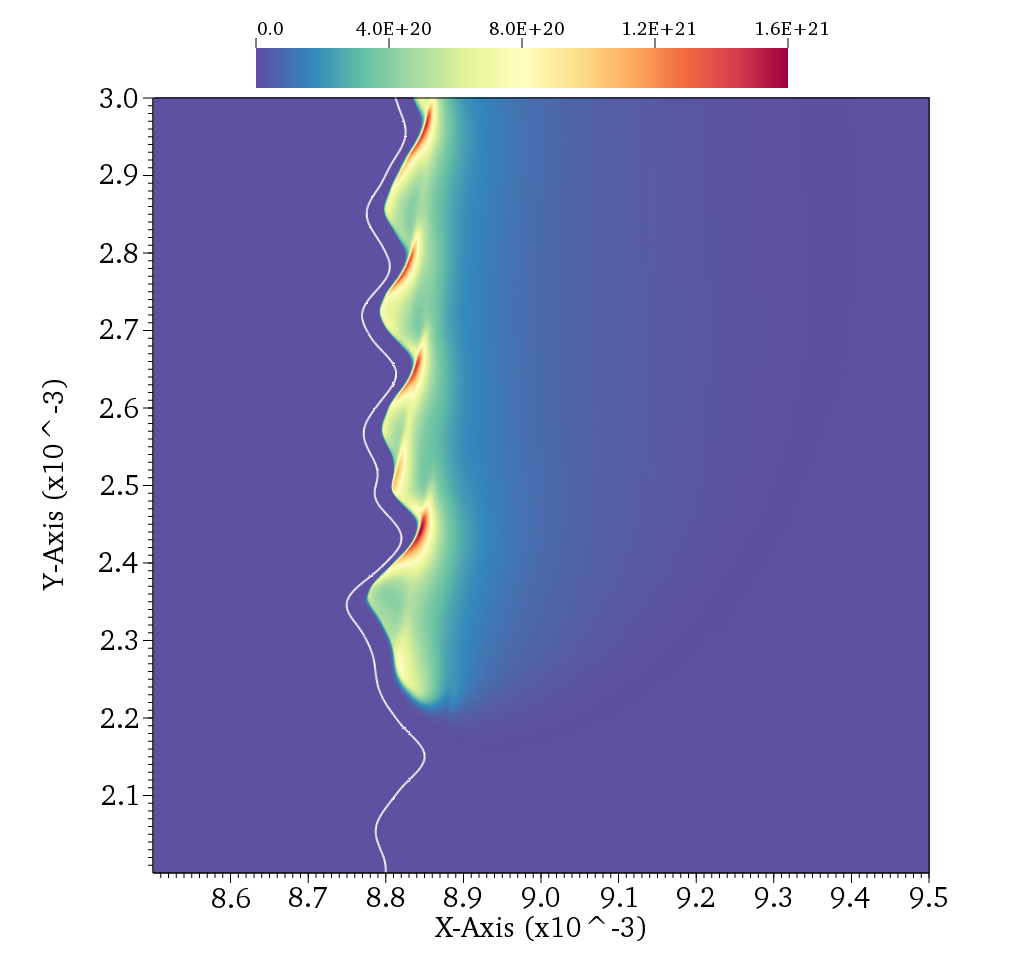}
  \includegraphics[width=.45\textwidth]{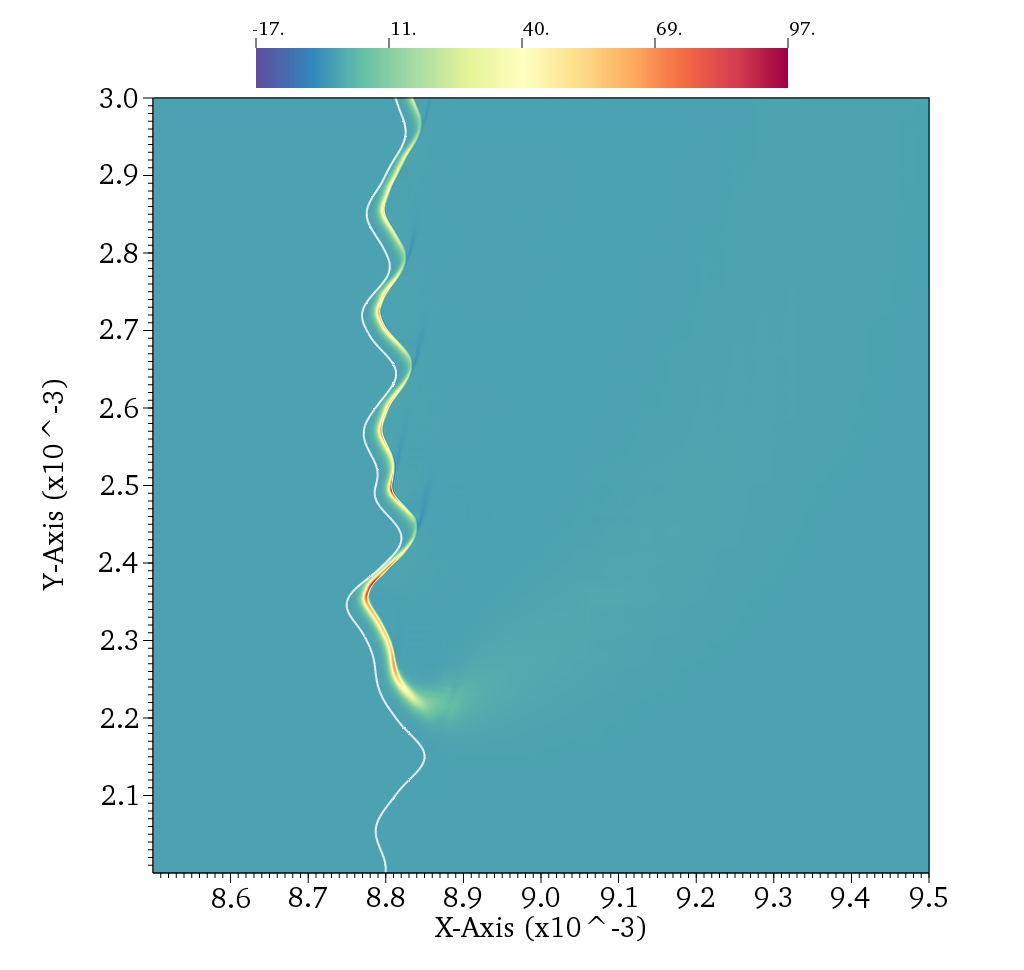}
  \includegraphics[width=.45\textwidth]{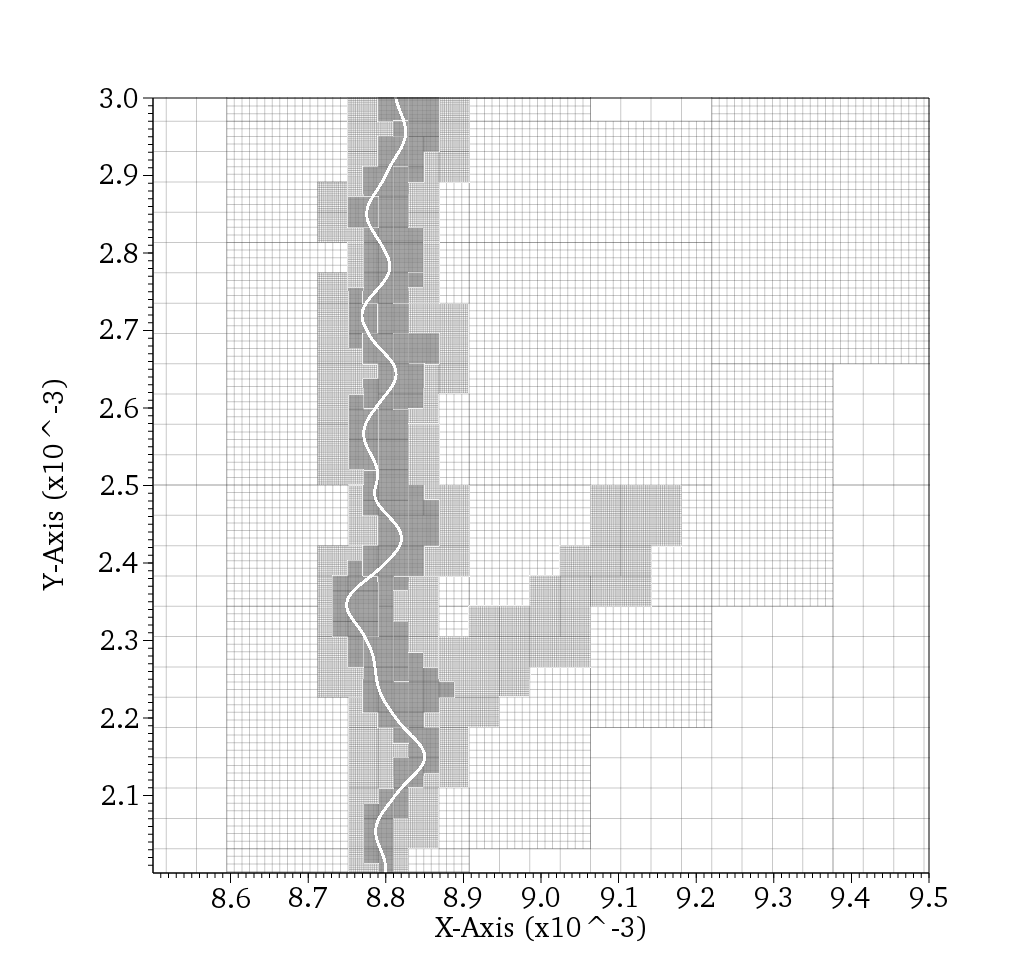}
  \caption{Snapshots of the streamer after $5.25\,\ns$. Top left: Electric field magnitude. Top right: Electron density: Bottom left: Space charge density: Bottom right: Mesh}
  \label{fig:dbd2d_5d25ns}
\end{figure*}

%The inclusion of surface roughness on the insulator is not trivial. Firstly, the changed topology yields has immediate geometric consequences such as increased area of ion bombardment and photoemission. In addition to the increased reactive surface, there are also more subtle features that warrant explicit mention. Firstly, note that the peaks and valleys of the insulating surface act as field enhancing features. Whether or not the peak (valley) amplifies or decreases the field magnitude generally depends on the direction of the applied field. For an applied field that is tangential to the surface there is an increased voltage drop in the valleys. Secondly, geometrically rough surfaces allow for field line re-entry where field lines that first cross into the dielectric re-emerge in one of the valleys on the surface, as seen in Fig.~\ref{fig:dbd_initial}. Thus, two closely-spaced regions on the dielectric can respond differently to the plasma, since one region can act as a cathode and the other as an anode, respectively.

Figure~\ref{fig:dbd_evolution} shows snapshots of the temporal evolution of the electric field, plasma and charge densities during the first few nanoseconds. In general, the streamer follows the topology of the surface and there is no plasma detachment into the gas phase (at least for the surface roughness considered here). We observe a height-of-flight of the positive streamer above the surface and find that the sheath region $l < 30\,\um$ from the surface contains fewer electrons and is incompletely shielded by space and surface charge effects. The topological features that we observe is also observed in planar simulations by others \cite{Celestin2009, Babaeva2009, Soloviev2009, Soloviev2014, Soloviev2017, Hua2018}, also with PIC codes \cite{osti_1367279}, and the plasma sheath thickness that we observe is consistent with analytical estimations \cite{Soloviev2017}.

Through the simulation, we always observe electron impact ionization in the plasma sheath closest to the dielectric, although the ionization rate is modest when compared to the streamer head. Inside the sheath the plasma dynamics occurs in the following way. Emission of secondary electrons from the dielectric appears through ion bombardment or photoemission. In turn, these electrons ionize the gas in the sheath; propagate through the sheath boundary and into the channel where they essentially become immobile due to field screening of a positively charged space charge layer that floats some tens of microns above the surface. For longer integration times, the residual positive ions in the sheath drift toward the dielectric, which increases the surface charge density and decreases space charge effects closest to the dielectric. Likewise, negative ions drift away from the surface. As a result, the sheath remains active until sufficient charge deposition onto the surface has occured. An order of magnitude estimation of the sheath relaxation time can be provided by considering the drift time of ions over the thickness of the sheath. I.e. $t = L/(\mu_{\textrm{ion}} E) \approx 15\,\ns$, where $\mu_{\textrm{ion}} \approx 2\times10^{-4}\m^2/(\textrm{V}\textrm{s})$ is the ion mobility. In practice, this time is a maximum estimate since the choice secondary emission coefficients strongly influence the number of electron-ion pairs that are generated in the sheath, and therefore also the sheath relaxation time.

Furthermore, we find that the plasma essentially follows the topology of the surface throughout the simulation. Figure~\ref{fig:dbd2d_5d25ns} shows the streamer after a time $5.25\,\ns$; the discharge has propagated about $7\,\mm$ in this time, which yields an average velocity of roughly $1.3\,\mm/\ns$. In particular, we observe that the plasma density has local maxima at the protrusions on the dielectric, reaching densities over $10^{21}\,\m^{-3}$. Correspondingly, the space charge density is lower in these regions (see Fig.~\ref{fig:dbd2d_5d25ns}).

\subsection{3D simulations in pin-plane gaps}
In our next example we consider inception and propagation of a positive streamer in a pin-to-plane geometry. The computational domain is a $(2\,\cm)^3$ domain with a cylindrical needle electrode with a $500\,\um$ radius protruding $1\,\cm$ down from the upper $z$-plane. We impose homogeneous Neumann conditions on the potential on the domain faces in the $xz$ and $yz$ planes. The lower $xy$ plane is grounded, and we apply a live potential of $15\,\textrm{kV}$ to the needle and upper $xy$ plane. For initial conditions we take $n_e=n_+=10^{10}\,\m^{3}$. Both the geometry and initial conditions are rotationally symmetric around the needle axis for this case, and this simulation could be run in a 2D cylindrically symmetric configuration, as is done by many other groups. In three dimensions, no currently published results are available for comparison. However, inception and propagation of streamers from a $(1\,\mm)^3$ cubic volume was considered recently \cite{Plewa2018}, and an equivalent setup has also been considered in the past \cite{7295248}. The difference between the simulations in \cite{Plewa2018, 7295248} and ours lies primarily in the volume that is simulated; we consider a volume that is several thousand times bigger. In two-dimensional cylindrically symmetric coordinates, equivalent simulations are too numerous to explicitly mention here. However, the intention of this example is not a comparison with the two-dimensional world, but rather a demonstration of the use of AMR with internal boundaries for moderately large configurations that are of experimental interest, using only a moderate amount of CPU cores. Certainly, deviations in the initial conditions can lead to non-symmetric solutions, as has been for example demonstrated in \cite{Plewa2018}, where plasma spots were used to provoke branching of the streamer. For geometric configurations that lack symmetry, a full 3D description is obviously required.

\begin{figure*}[h!t!b!]
  \centering
  \includegraphics[width=.4\textwidth]{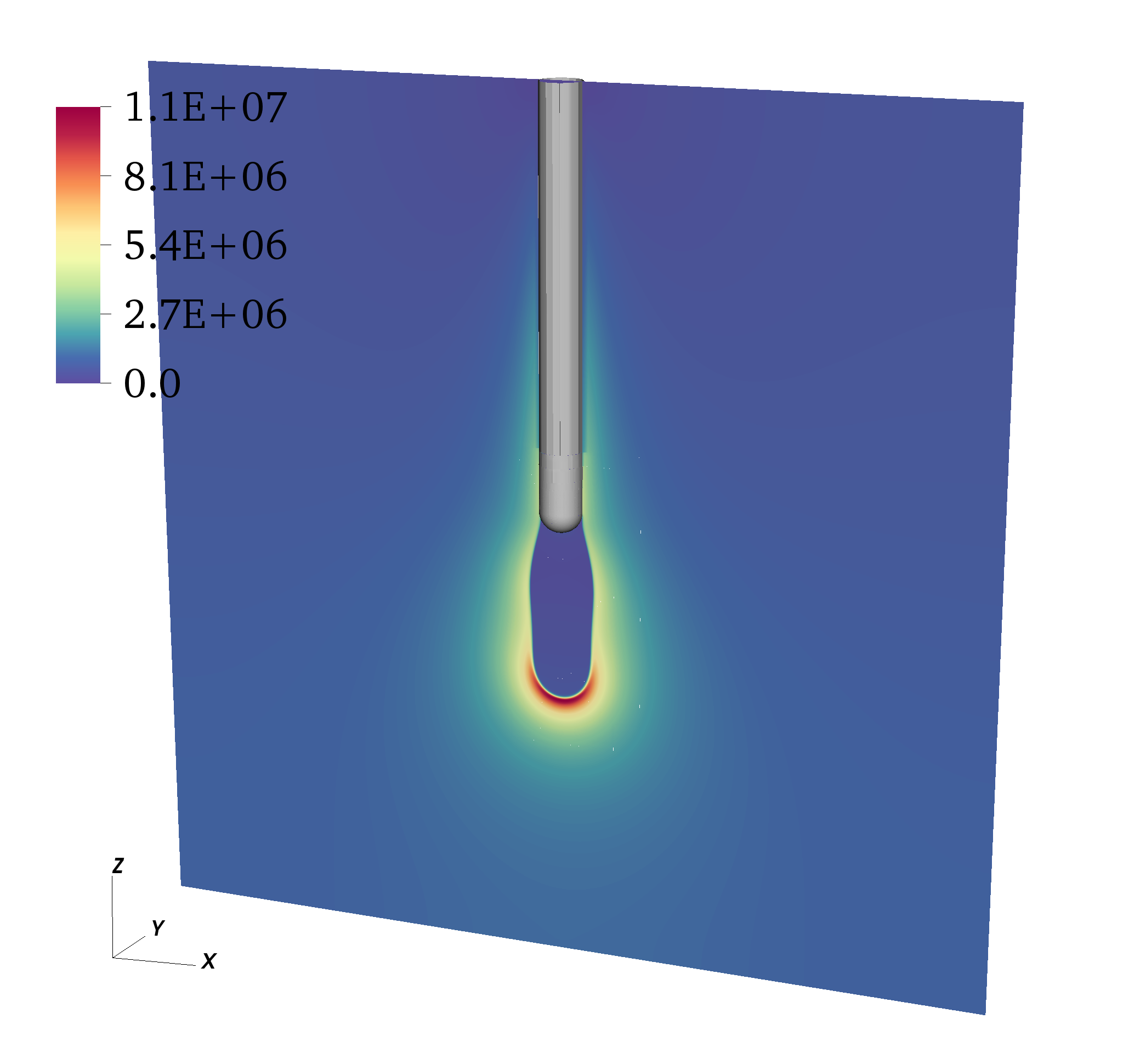}\quad
  \includegraphics[width=.4\textwidth]{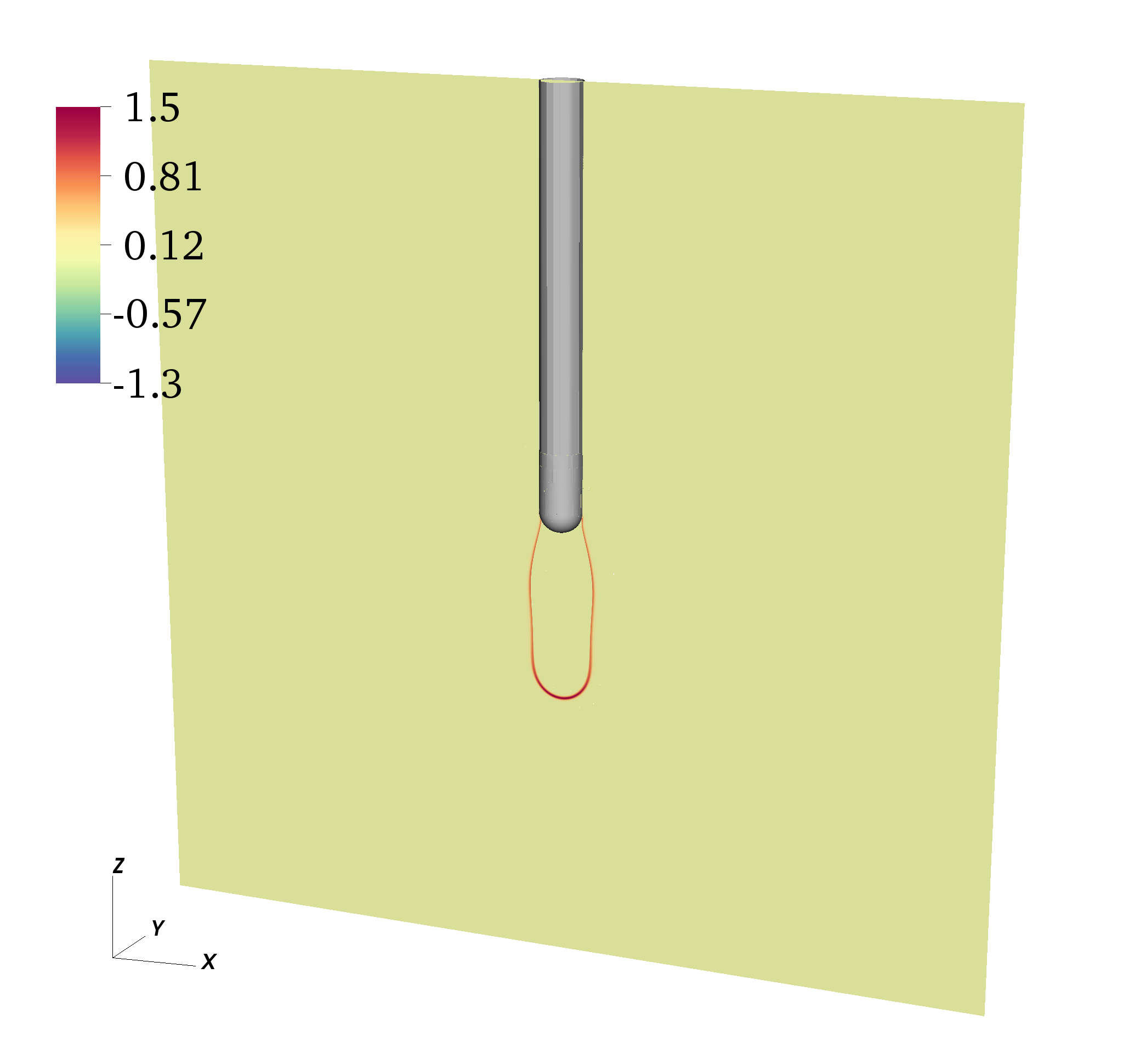} \\ [1em]
  \includegraphics[width=.4\textwidth]{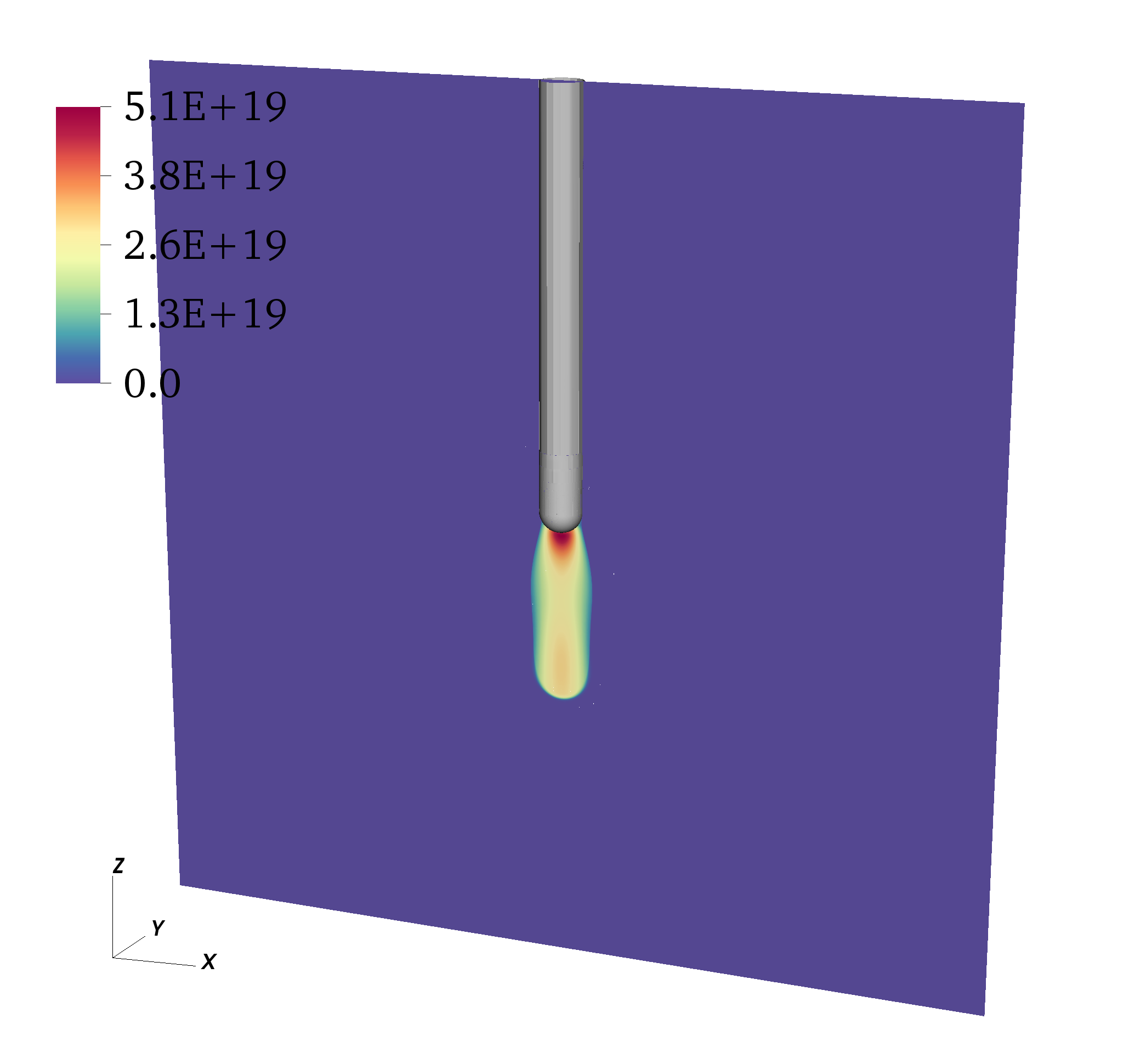}\quad
  \includegraphics[width=.4\textwidth]{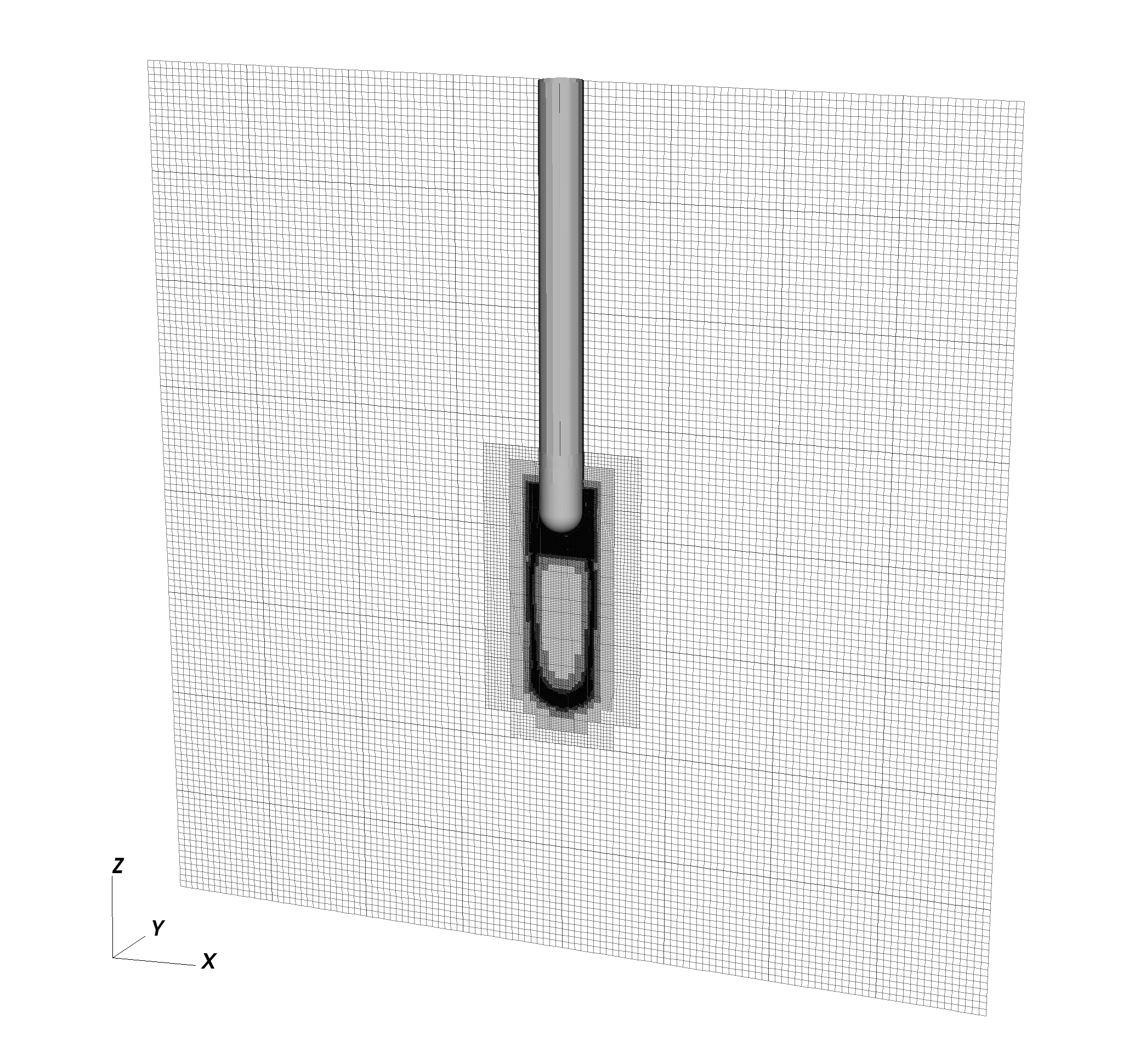} 
  \caption{Simulation state after $10.8\,\ns$. Top left: Field magnitude. Top right: Space charge. Bottom left: Electron density. Bottom right: Mesh.}
  \label{fig:rod3d_10d8ns}
\end{figure*}

For this application we use a base mesh of $128^3$ cells and use 5 levels of refinement with a factor 2 between levels. The effective domain is $(4096)^3$, and the finest level resolution is roughly $5\,\um$. For this application we refine the region around the needle tip, which is where we expect streamer inception to occur. In the bulk gas, we use the refinement and coarsening criteria discussed above. The simulation was run with a CFL number of $0.5$ throughout the simulation, yielding a time step of roughly $5\,\ps$. The strong scaling limit for this simulation yields an executation time per time step of approximately $6$ seconds, depending on the simulation state. The initial avalanche phase is faster to compute since there are no space charge effects, leaving the potential largely unaffected. Once a streamer starts propagating, the execution time per time step increases. We remark that our temporal scheme requires 11 elliptic solves per time step. In addition, unlike \cite{Nijdam2016a, Teunissen2017, Plewa2018, 7295248}, every species is convected so that the slope-limited upwind states that are required for the advection advance must be determined at every Runge-Kutta stage for both electrons, positive ions, and negative ions. The reason for this is that we require ion mobility for modeling of appropriate physical effects near electrodes and dielectrics, examples being surface charging of insulating surfaces or dynamics of cathode sheaths. However, the numerical cost of advancing convection equations is generally less than for elliptic equations. Towards the end of the simulation at $t=20\,\ns$, the mesh is composed of roughly $80$ million cells. 

\begin{figure*}[ht]
  \centering
  \includegraphics[width=.45\textwidth]{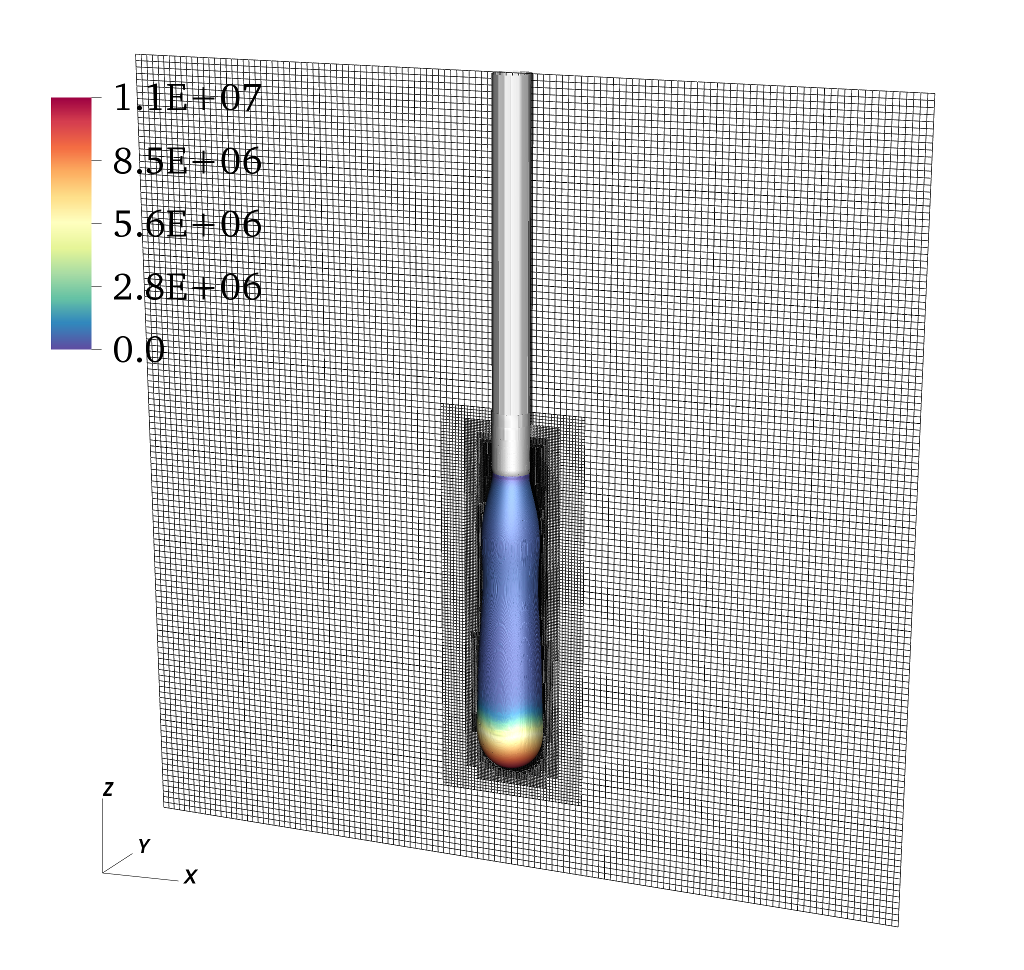} \quad
  \includegraphics[width=.45\textwidth]{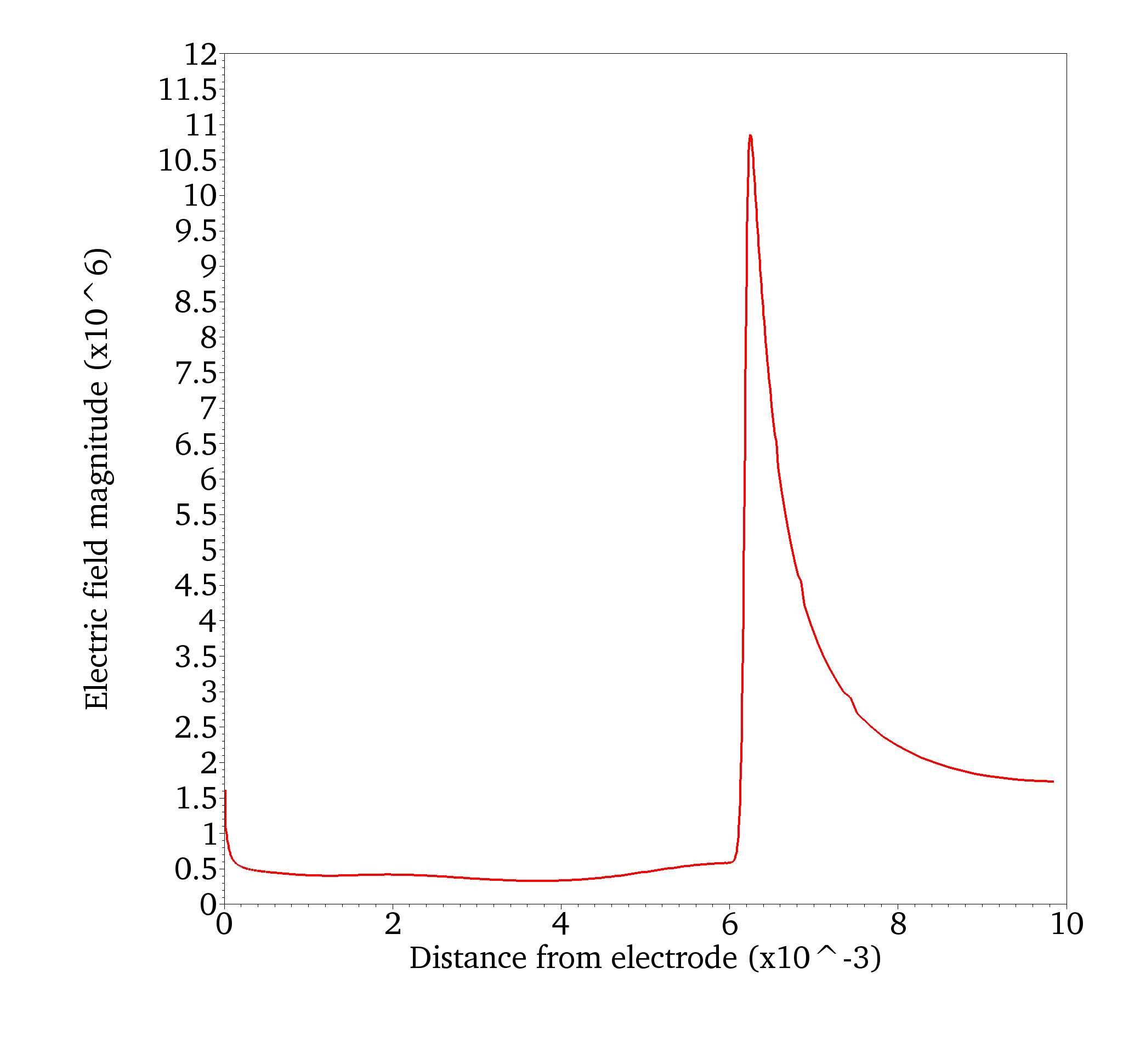} 
  \caption{Left: Three-dimensional evolution of a cylindrically symmetric streamer in a pin-plane gap after a time $17.7\,\ns$. The surface shows the isosurface of the plasma density at $n_e = 10^{18}\,\m^{-3}$ and the color coding shows the electric field magnitude. Right: Field magnitude on the needle axis after a time of $15.8\,\ns$.}
  \label{fig:rod3d_goodconvergence}
\end{figure*}

Figure~\ref{fig:rod3d_10d8ns} shows the simulation state after roughly $10.8\,\ns$. The various panels show the electric field magnitude, space charge layer, electron densitiy, and mesh; all quantities have been sliced through the $xz$ plane that passes through the needle tip. We observe generic features that are associated with propagation of positive streamers, such as a positive space charge layer and screening of the channel interior. More careful evaluation of the data sets shows that the streamer diameter, taken as the width of the space charge layer (i.e. the electrodynamic diameter) at its widest point, measures roughly $1.5\,\mm$. The radiative diameter, however, is smaller and measures less than $1\,\mm$. Furthermore, the streamer has propagated roughly $5\,\mm$ within the first $10.8\,\ns$, yielding an average velocity of $0.5\,\mm/\ns$. Finally, note that our simplified refinement and coarsening criteria resolve the space charge layer without over-resolving the channel volume. In fact, our experience is that appropriate refinement and coarsening criteria are essential for optimium use of computer resources, and can lead to an order of magnitude reductions in computation time or resources. However, we emphasize the need for caution in the grid coarsening stage. Under-resolution of the streamer head, or the region around it, can lead to solution errors that amplify during the propagation process, leading to unphysical solutions (such as numerical branching, for example).

Finally, we remark that the simulation data here is consistent with experiments and two-dimensional rotationally symmetric simulations performed by others. We observe a mean velocity of roughly $0.5\,\mm/\ns$ and an optical diameter less than $1\,\mm$ (the diameter defined by the space charge layer is larger, up to $1.5\,\mm$). \citet{Pancheshnyi2005} has examined the development of positive streamers in slighly larger gaps, and the authors observed a mean velocity of roughly $0.5\,\mm/\ns$ and optical diameters of roughly $0.5\,\mm$, which aligns fairly well with our computational results. However, we remark that this comparison is only indicative. Indeed, streamer velocities and diameters change as a function of the applied voltage, and a broader sampling of simulations and experiments is generally necessary for validating the computational results. Finally, the field magnitude in the channel is not constant, it has an average value of roughly $0.5\,\kV/\mm$ (see Fig.~\ref{fig:rod3d_goodconvergence}), which corresponds to the streamer stability field computed by others \cite{Gallimberti1972,Allen1995,Babaeva1997,Qin2014}, and is also the value that is used by engineering models \cite{Christen2012}. 

\subsection{3D simulations in complex geometries}
\begin{figure*}[h!t!b!]
  \centering
  \includegraphics[width=.9\textwidth]{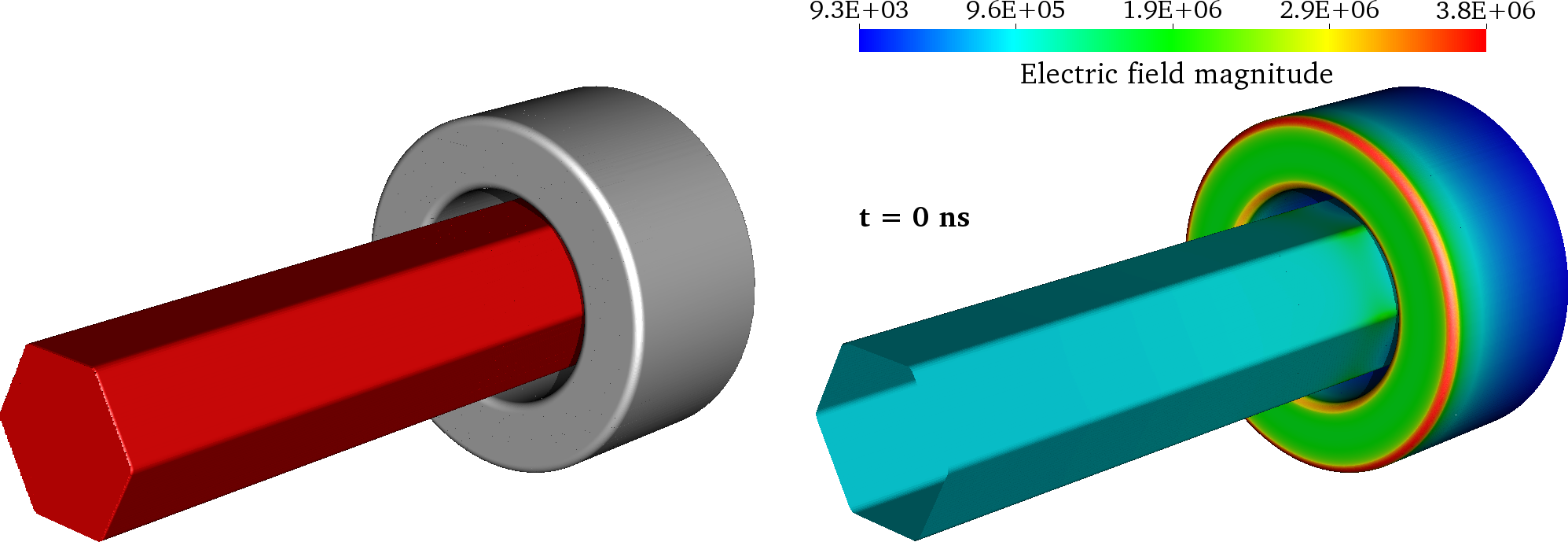}
  \caption{Left: Mechanical shaft simulation geometry. Right: Initial field stress $\left|\bm{E}\right|$ at the boundaries at $t=0$.}
  \label{fig:shaft_geometry}
\end{figure*}

Our final simulation example considers a mock up model of a mechanical switch where a rotating mechanical shaft protrudes through a mechanical bushing. The domain is a $(4\,\cm)^3$ domain with a single electrode and an insulating dielectric shaft, as shown in Fig.~\ref{fig:shaft_geometry}. The electrode is represented by a hollow cylinder that protrudes $1\,\cm$ from the live wall and has inner and outer radii of $6\,\mm$ and $1\,\cm$, respectively. The electrode corners have been rounded with a rounding radius of $500\,\um$. The dielectric is a $4\,\cm$ long rod with a hexagonal cross section and relative permittivity of $4$. The cross-sectional diameter of the insulator is $1\,\cm$, and the corners have been rounded with a $500\,\um$ rounding radius. For fast parametrization, the geometry was generated by means of constructive solid geometry.

\begin{figure}[h!t!b!]
  \centering
  \includegraphics[width=.7\textwidth]{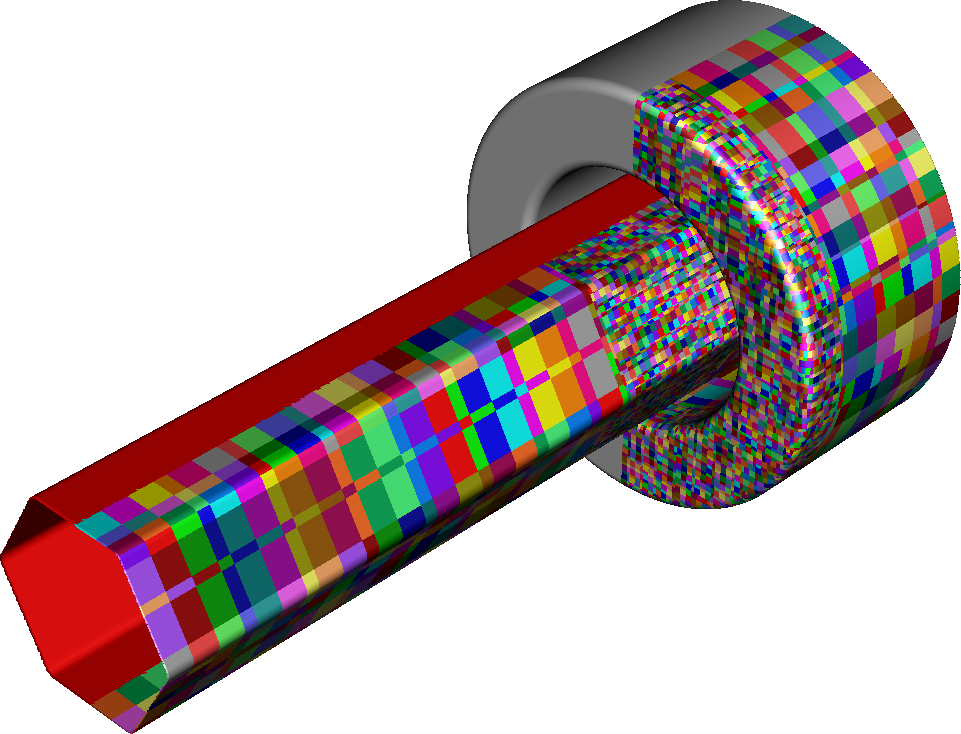}
  \caption{Initial computational patches for the insulating shaft geometry. Each patch represents a unit of computational work and has minimum size $(16)^3$ and maximum size $(32)^3$.}
  \label{fig:shaft_patches}
\end{figure}

For this simulation we have decided to reduce the gas pressure. Current hardware limitations (4096 cores) prevent us from simulating this geometry at atmospheric pressure where the simulation would probably require several billion cells. Because of those limitations, the gas pressure used for this simulation is $p=0.5\,\textrm{atm}$, and we only simulate the inception and initial propagation phases.

We use a base mesh of $(128)^3$ cells and include five levels of refinement. The refinement factor between each level is $2$, which yields a base resolution of $312.5\,\um$ and a finest level resolution of $10\,\um$. We use a reduced EB mesh for this simulation, initially only refining the EB region $z\in[2.5\,\cm, 3.2\,\cm]$ down to the finest AMR level; the remaining part of the EB is refined three times down to a resolution of $39\,\um$. The simulation is run with a maximum patch size of $(32)^3$ and a minimum patch size of $(16)^3$, and the initial mesh contains around $300$ million cells for a total of roughly $2.1$ billion unknowns (potential, three ion densities, and three photon densities). A uniform mesh would require almost $70$ billion cells. For the types of applications that we attempt to simulate here, grids based on unstructured meshes or SAMR are, from both run-time and memory consumption point of views, the only realistic alternatives today. Figure~\ref{fig:shaft_patches} shows the distribution of the initial mesh on the boundary. In this figure, each colored square represents a patch of minimum size $(16)^3$ and maximum size $(32)^3$. 

\begin{figure*}[h!t!b!]
  \centering
  \includegraphics[height=.9\textheight]{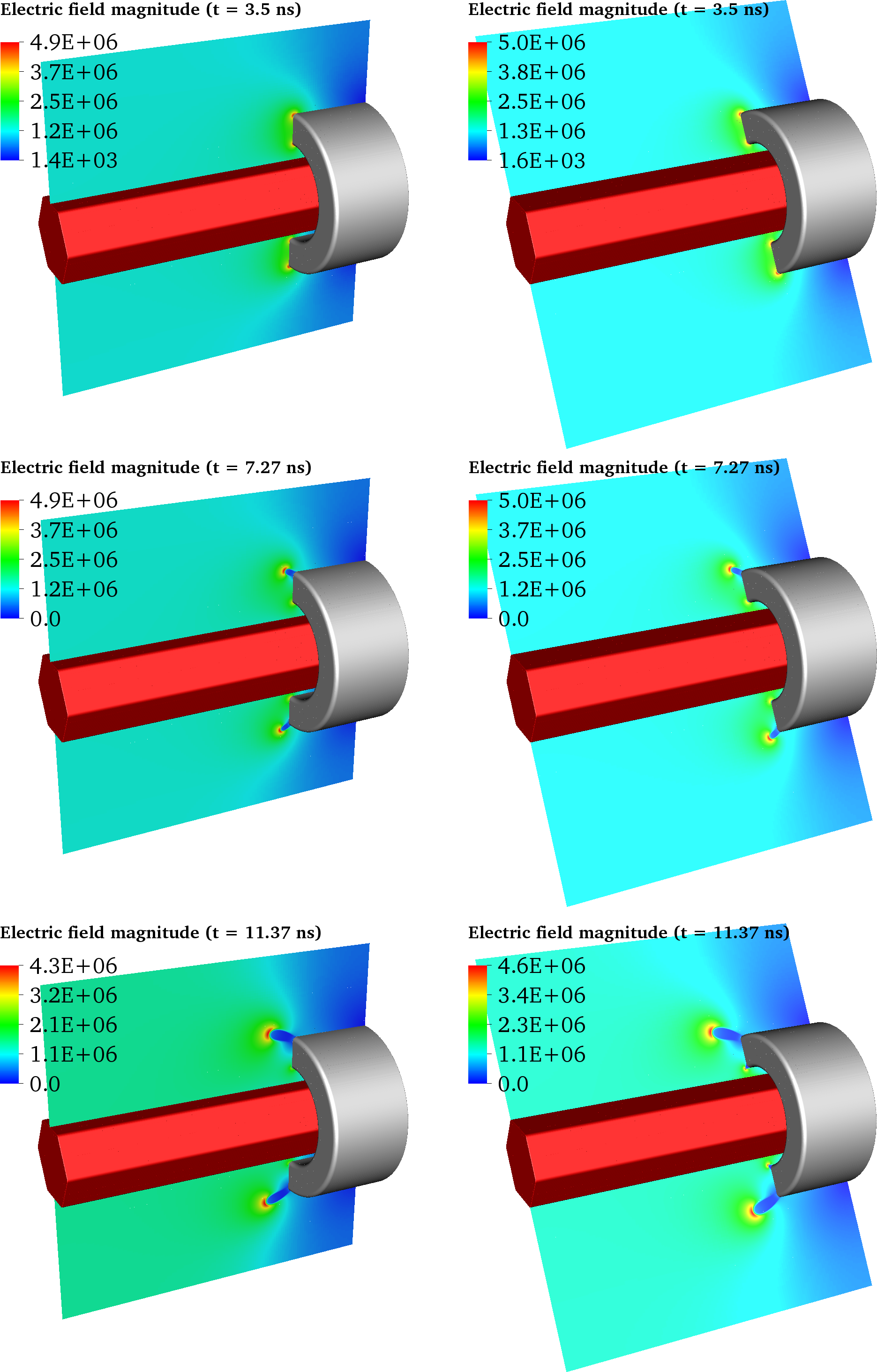}
  \caption{Snapshots of the electric field magnitude in the gas volume. The left column shows the electric field magnitude sliced through the $yz$-plane, and the right column shows the electric field magnitude sliced through one of the corners of the insulator shaft. Times are indicated in each plot.}
  \label{fig:propagation_volume}
\end{figure*}

For boundary conditions on the Poisson equation, we apply a positive step potential of $45\,\kV$ on the electrode and on the wall it protrudes from, the opposite domain face is grounded, and homogeneous Neumann conditions are used on the side walls. The initial charge on the rod is set to zero whereas the ion densities are $n_e = n_+ = 10^{10}\,\m^{-3}$, $n_- = 0$. Figure~\ref{fig:shaft_geometry} shows the initial field stress with these boundary conditions. The maximum field stress is found on the outer edge of the electrode and equals roughly $5.6\,\kV/\mm$, whereas the field stress on the inner edge of the electrode equals roughly $3.5\,\kV/\mm$. For distances greater than $1\,\cm$ away from the electrode and towards the ground plane, the field is otherwise comparatively homogeneous. On the dielectric boundary, we find slightly enhanced field stresses on the corners of the dielectric shaft. Enhancement in these regions are due to a reduced distance between the electrode and insulator. For field lines that penetrate into these regions, polarization of the dielectric enhances the voltage drop in the gas phase, leading to an increased electric field in these positions.

\begin{figure*}[h!t!b!]
  \centering
  \includegraphics[width=.95\textwidth]{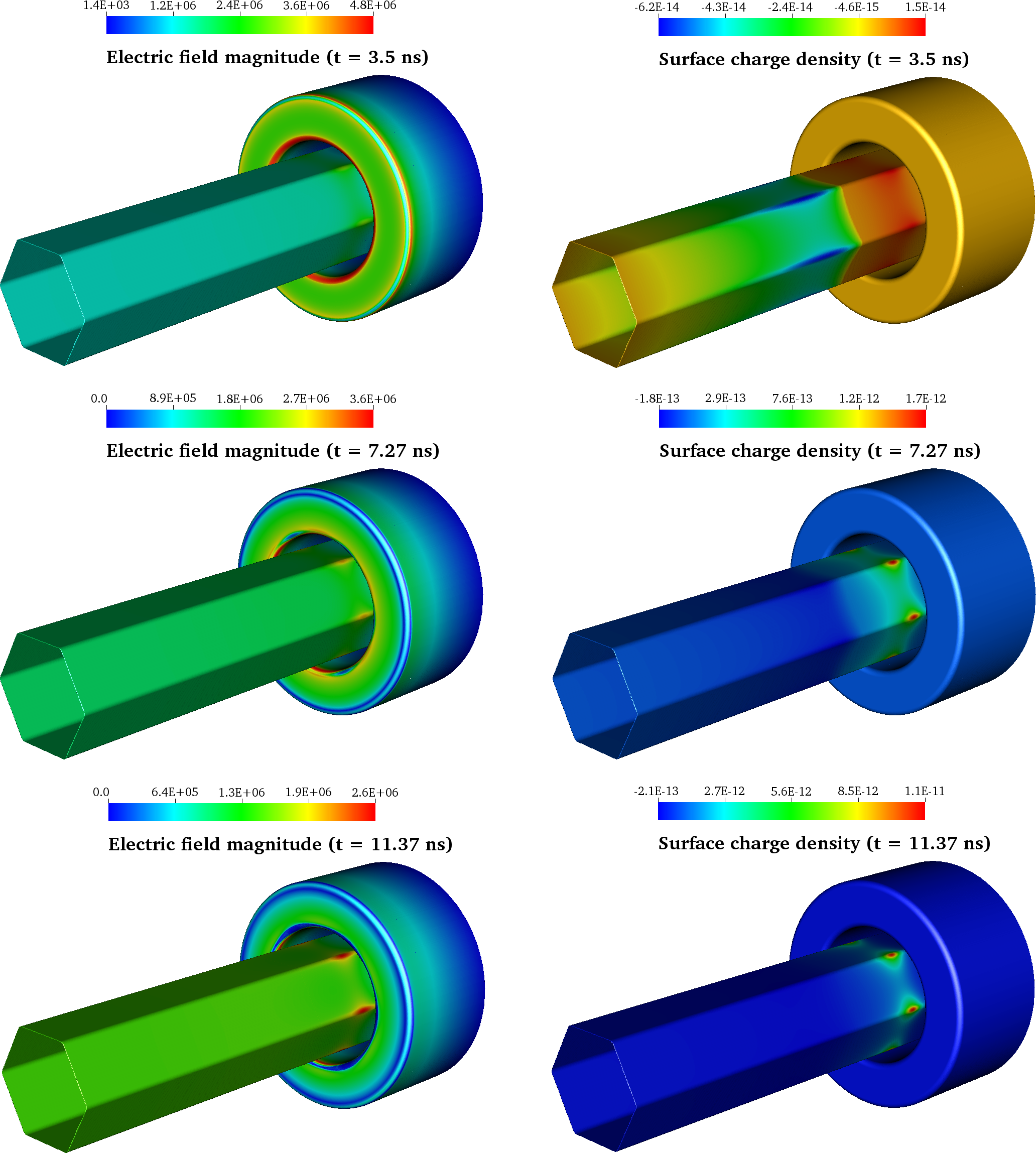}
  \caption{Simulation state after $1.1\,\ns$. The color coded data is in SI units. Left: Field stress at the boundary. Inception has occured on the outer circular edge of the electrode, and is also occuring on the inner edges where the electrode-insulator distance is smallest. Right: Surface charge density on the boundaries.}
  \label{fig:inception_bndry}
\end{figure*}

Figure~\ref{fig:propagation_volume} shows the spatial evolution of the electric field sliced through two symmetry planes in the domain. The left and and right columns in this figure show the electric field magnitude sliced through planes that are rotated $60$ degrees with respect to one another. One of these planes intersect the ''flat'' side of the insulator and the other intersects one of the corners. Simulation times are indicated in each frame. We find that after streamer inception at $t\sim 3.5\,\ns$ the maximum electric field magnitude is roughly $4.9\,\kV/\mm$ for the streamer that initiates from the outer electrode edge. At this point, there are no visible space charge effects on the inner edge, nor do we find significant field screening on that edge (see Fig.~\ref{fig:inception_bndry}). After roughly $t\sim 7.27\,\ns$, a toroidal streamer has disconnected from the outer electrode edge. Streamers have also initiated from the inner electrode edge, but only on the six locations on the electrode that are closest to the insulating surface (Fig.~\ref{fig:propagation_volume}). The maximum field strength are found in the head of the toroidal streamer, with a field magnitude of $\sim5.0\,\kV/\mm$. After $t\sim 11.37$, the maximum field stress is found in the head of the streamers that propagate towards the insulator, with field strengths reaching $4.6\,\kV/\mm$, see bottom row in Fig.~\ref{fig:propagation_volume}. These streamers have not yet reached the surface of the insulator. 

\begin{figure}[h!t!b!]
  \centering
  \includegraphics[width=.8\textwidth]{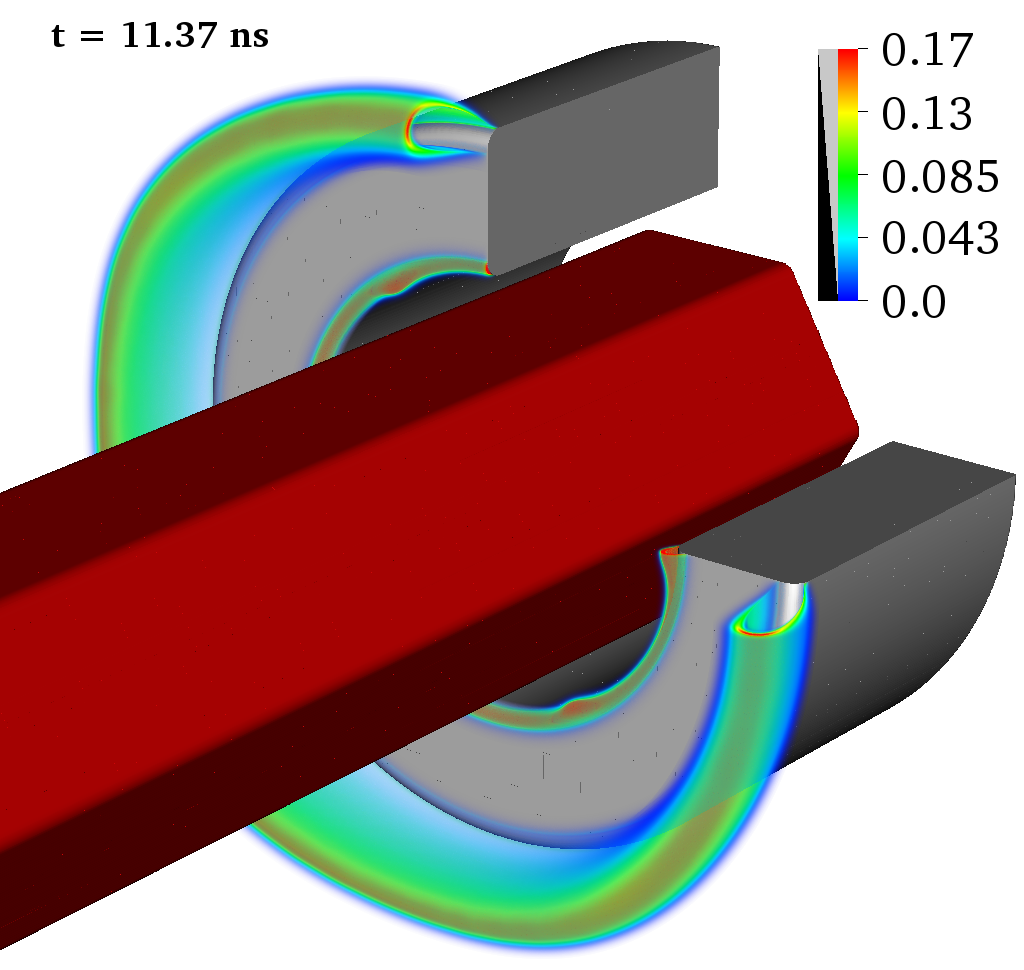}
  \caption{Volume rendered space charge density for the simulation state at $11.37\,\ns$. The data range has been adjusted for improved visibility. }
  \label{fig:spacecharge_volume}
\end{figure}                                                                 

Figure~\ref{fig:inception_bndry} shows the field magnitude and surface charge density on the boundaries for the same times as in Fig.~\ref{fig:propagation_volume}. At the time of inception ($3.5\,\ns$), surface charging of the insulator is insignificant, indicating that non-local photoemission of seed electrons from the insulator leads to negligible charging in the inception stage. For example, Fig.~\ref{fig:inception_bndry} shows that the surface charge density reaches a value of $1.5\times 10^{-14}\,\C/\m^2$ after $3.5\,\ns$. For comparison, the initial surface charge density on the electrode is $\sigma = \epsilon_0E \sim 3.4\times 10^{-5}\,\C/\m^2$. Note that the negative surface charge densities that are observed for $t=3.5\,\ns$ (top right panel in Fig.~\ref{fig:inception_bndry}) are due to electron impact of the initial preionization that was assigned in order to start the simulation. The inflection point where the surface charge goes from positive to negative (approximately $1\,\cm$ from the electrode) is due to electrostatic field effects. The normal field switches sign at this position such that $\bm{E}\cdot\bm{n}$ is negative for distances $\lesssim 1\,\cm$ from the electrode, and positive otherwise. After roughly $7.27\,\ns$, inception of  positive streamer has also occured on the inner electrode edge. In this region, the streamer is not rotationally symmetric but instead initiates on the positions on the electrode that are closest to the six outer corners on the insulator. This can be seen in e.g. Fig.~\ref{fig:inception_bndry} by appearance of a field-screened spot on the inner electrode edge after $7.27\,\ns$. After $11.37\,\ns$, the maximum field stress on the boundary has been transferred to the insulator, predominantly due to space charge effects in the electrode-insulator gap. At this time, both the outer and inner electrode edges are screened by space charges. 

To demonstrate the streamer in greater detail, Fig.~\ref{fig:spacecharge_volume} shows a volume rendering of the space charge density at $t = 11.37\,\ns$. We have cut out a part of the simulation domain and adjusted the data range (only positive space charge densities are plotted) for enhanced visibility. We clearly observe a toroidal space charge density layer that has initiated form the outer edge of the electrode. We remark that this space charge does not have perfect rotational symmetry; the streamer is thinner where domain boundaries are closer (e.g. at the top and right-hand sides in Fig.~\ref{fig:spacecharge_volume}). 

The final mesh for this simulation contains around $800$ million cells, or roughly $1.5$ billion when one accounts for ghost cells. Since there are three species densities and three photon densities that need to be solved in addition to the electric potential, the final mesh contains $5.6$ billion unknowns. The mesh inflation from the initial $300$ million cells is primarily due to the toroidal streamer on the outer electrode edge, which required refinement in an annular ring with a comparatively large radius. Plot files for this simulation ranged in sizes up to $250\,\GB$, which had to be visualized in parallel. 

\section{Final remarks}
\label{sec:remarks}
We have presented an implementation of a low-order plasma fluid model in 2D and 3D, for which several improvements can be made. Our implementation is parallelized with flat MPI and is therefore not expected to scale very well on hetereogeneous HPC architectures. We currently do not support fine-grained parallelism (for example through cache blocking inside each patch), nor do we support the use of GPU accelerators. Ideally, we'd like to reduce memory redundancies across MPI ranks by using inter-node MPI parallelism and on-node fine-grained parallelism by means of loop tiling or GPU accelerators. For the models considered in this paper, the computational bottleneck is due to multiple elliptic solves at every time step, and we expect that we can improve the strong scaling limit quite substantially by use of GPU accelerators. 

Finally, this paper does not provide explicit validation studies and the numerical plasma examples that we compute in this paper only indicate physical correctness. For example, one of our cases (streamer along rough surface) provide velocities and sheath thicknesses that are consistent with other computations, while one of our other examples (3D rod simulation) provide velocities and diameters that are consistent with experimental observations at larger gaps. For our largest simulation, no comparison with experiments is currently available.

\section{Conclusions}
\label{sec:conclusions}
We have presented a parallel code for large-scale simulations of low-temperature filamentary plasmas with inclusion of internal insulators and electrodes in the computational domain. Our code is based on structured adaptive mesh refinement with embedded boundaries for representation of solids, and currently features a multi-fluid Poisson solver; convection-diffusion-reaction solvers, and diffusive radiative transport solvers. We have implemented support for CAD-generated geometries and thus provide a means to perform plasma-fluid simulations in large industrial-grade complex geometries with realistic boundary conditions on dielectric and electrode surfaces. Two- and three-dimensional verification tests show that the overall model is second order accurate for smooth solutions. Weak scalability with parallel efficiencies above 70\% were demonstrated at two different machines (Fram and SuperMUC Phase 2), using up to 8192 cores at one of them. Finally, we presented several two- and three-dimensional simulation examples at up to 4096 cores. These simulation examples include creeping streamers over insulators with surface roughness, pin-plane simulations of plasma filaments, and fully transient 3D simulations in industrial grade complex problems. Our code is flexible with regards to the plasma kinetics, making it applicable to a broad range of research fields, ranging from applied high-voltage engineering, aerodynamics, sterilization of polluted gases, and upper atmosphere lightning (sprites). Future efforts will focus on large scale applications on a subset of these, further scaling, as well as improving the overall efficiency of the code. 

\section*{Acknowledgements}
This work was financially supported by the Research Council of Norway through project 245422 and industrial partner ABB AS, Norway. The computations were performed on resources provided by UNINETT Sigma2 - the National Infrastructure for High Performance Computing and Data Storage in Norway. We acknowledge PRACE for awarding us access to SuperMUC at GCS@LRZ, Germany for the scalability studies. The author expresses his gratitude to D. T. Graves at Lawrence Berkeley National Laboratory (LBNL) for help with the Chombo infrastructure.

\appendix
\section{Discharge model}
\label{sec:discharge_model}
\subsection{Plasma kinetics}
We consider a simplified kinetic scheme for air at standard atmospheric conditions. We solve for three charged species: electrons $n_e$, positive ions $n_+$, and negative ions $n_-$, with only electrons being diffusive: 
\begin{subequations}
  \begin{align} 
    \frac{\partial n_e}{\partial t} &= -\nabla\cdot\left(\bm{v}_en_e\right) + \nabla\cdot(D_e\nabla n_e) + S_e, \\
    \frac{\partial n_+}{\partial t} &= -\nabla\cdot\left(\bm{v}_+n_+\right) + S_+, \\
    \frac{\partial n_-}{\partial t} &= -\nabla\cdot\left(\bm{v}_-n_-\right) + S_-,
  \end{align}
\end{subequations}
where the source terms are
\begin{subequations}
  \begin{align}
    S_e &= \left(\alpha - \eta\right)\left|\bm{v}_e\right|n_e - \beta n_en_+ + S_{\textrm{ph}}, \\
    S_+ &= \alpha\left|\bm{v}_e\right|n_e - \beta n_en_+ - \beta n_-n_+ + S_{\textrm{ph}}, \\
    S_- &= \eta\left|\bm{v}_e\right|n_e - \beta n_+n_-,
  \end{align}
\end{subequations}
Here, $S_{\textrm{ph}}$ is a photonionization source term that is discussed below, $\alpha = \alpha\left(\left|\bm{E}\right|\right)$ and $\eta = \eta\left(\left|\bm{E}\right|\right)$ are impact ionization and attachment coefficients, and $\beta$ is a recombination coefficient. 

The kinetic coefficients are taken from \citet{Morrow1997} and are as follows:
\begin{subequations}
  \label{eq:morrow_lowke_v}
  \begin{align}
    \bm{v}_e &= -\frac{\bm{E}}{\left|\bm{E}\right|}\times\begin{cases}
    \left(6.87\times 10^{24}E + 3.38\times 10^{2}\right), & E \leq 2.6\times 10^{-21}, \\
    \left(7.29\times 10^{23}E + 1.63\times 10^{4}\right), & 2.6\times 10^{-13} \le E \leq 10^{-20}, \\
    \left(1.03\times 10^{24}E + 1.30\times 10^{4}\right), & 10^{-12} \le E \leq 2\times 10^{-19}, \\
    \left(7.40\times 10^{23}E + 7.10\times 10^{4}\right), & E \ge 2\times 10^{-19}
    \end{cases} \\
    \bm{v}_+ &= 2.34\times 10^{-4}\bm{E}, \\
    \bm{v}_- &= -\frac{\bm{E}}{\left|\bm{E}\right|}\times\begin{cases}
    2.7\times10^{-4} & E \leq 5\times 10^{-20}, \\
    1.86\times 10^{-4} & \textrm{otherwise},
    \end{cases}
  \end{align}
\end{subequations}
where $N = 2.45\times 10^{25}\,\m^{-3}$ is the neutral number density and $E = \left|\bm{E}\right|/N$. The electron diffusion coefficient is
\begin{equation}
  D_e = 4.86\frac{\left|\bm{v_e}\right|}{\left|\bm{E}\right|}\times 10^{6}
\end{equation}
The explicit expressions for the ionization, attachment, and recombination coefficients are
\begin{subequations}
  \label{eq:morrow_lowke_alpha}
  \begin{align}
    \alpha &= N \begin{cases}
      6.619\times 10^{-17}\exp\left(-\frac{7.25\times 10^{-11}}{E}\right), & E < 1.5\times 10^{-19}, \\
      2\times10^{-16}\exp\left(-\frac{5.59\times 10^{-11}}{E}\right), & \text{otherwise},
    \end{cases}\\
    \eta &= \eta_1 + \eta_2, \\
    \eta_1 &= N\begin{cases}
    \left(6.09\times 10^{-12}E + 2.983\times 10^{-23}\right), & E \leq 1.05\times 10^{-19}, \\
    \left(8.89\times 10^{-13}E + 2.567\times 10^{-23}\right), &\textrm{otherwise},
    \end{cases}\\
    \eta_2 &= 3.79N^2E^{-1.275}\times 10^{-74}, \\
    \beta &= 2\times 10^{-13}.
  \end{align}
\end{subequations}
In the above, the entries on the right hand side of Eqs.~\eqref{eq:morrow_lowke_v} through \eqref{eq:morrow_lowke_alpha} are dimensionless, whereas the left hand sides are in standard (SI) units. 

\subsection{Photoionization}
The radiative transport module is implemented as a 3-group diffusive model \cite{Bourdon2007}. That is, we consider Eq.~\eqref{eq:sp1} in the stationary approximation with $\kappa_\gamma = \lambda_\gamma p_{\textrm{O}_2}/\sqrt{3}$ for $\gamma=1,2,3$, where $\lambda_\gamma$ is a constant and $p_{\textrm{O}_2} = 0.2\,\text{bar}$ is the oxygen partial pressure. The equations of motion are
\begin{equation}
  \kappa_\gamma \Psi_\gamma - \nabla\cdot\left(\frac{1}{3\kappa_\gamma}\nabla \Psi_\gamma\right)  = \frac{\eta_\gamma}{c}, \quad \gamma=1,2,3.         
\end{equation}
The source terms are the same for all RTE equations and are taken as
\begin{equation}
  \eta_\gamma = \nu_{\textrm{exc}}\frac{p_q}{p + p_q}\alpha n_e\left|\bm{v}_e\right|,
\end{equation}
where the excitation efficient is taken as $\nu_{\textrm{exc}} = 0.6$. The term $p_q/(p + p_q)$ is a collisional quenching term that takes into account collisional quenching of radiative transitions. We take $p_q/p = 0.03947$, whereas $\alpha$ and $\bm{v}_e$ are as given above. The photoionization rate of $\OO$ is then
\begin{equation}
  S_{\textrm{ph}} = \nu_{\textrm{eff}}p_{\textrm{O}_2}\sum_{\gamma = 1, 2, 3}A_\gamma \Psi_\gamma,
\end{equation}
where $\nu_{\textrm{eff}} = 0.1$ is the photoionization efficiency, and $A_\gamma$ are parameters tabulated in Tab.~\ref{tab:photo_model}
\begin{table}[h] 
  \centering
  \begin{tabular}{lll}
    $\gamma$ & $\lambda_\gamma$ & $A_\gamma$ \\
    \hline
    1 & $4.15\times 10^{-2}\,\textrm{m}^{-1}\textrm{Pa}^{-1}$ & $1.12\times 10^{-4}\,\textrm{m}^{-1}\textrm{Pa}^{-1}$ \\
    2 & $1.09\times 10^{-1}\,\textrm{m}^{-1}\textrm{Pa}^{-1}$ & $2.88\times 10^{-3}\,\textrm{m}^{-1}\textrm{Pa}^{-1}$ \\
    3 & $6.69\times 10^{-1}\,\textrm{m}^{-1}\textrm{Pa}^{-1}$ & $2.76\times 10^{-1}\,\textrm{m}^{-1}\textrm{Pa}^{-1}$
  \end{tabular}
  \caption{Parameters use for the photoionization model of oxygen. }
  \label{tab:photo_model}
\end{table}

\subsection{Charged species boundary conditions}
We consider comparatively simple boundary conditions on both dielectrics and electrodes \cite{Hagelaar2000}. The outflow of a species $i$ is
\begin{equation}
  \label{eq:outflow}
  \bm{F}_i\cdot\bm{n} = a_i(\bm{v}_in_i)\cdot\bm{n}
\end{equation}
where
\begin{equation}
  a_i = \begin{cases}
    1 & \bm{v}_i\cdot{\bm{n}} > 0, \\
    0 & \textrm{otherwise}.
  \end{cases} 
\end{equation}

In addition to charged species outflow, we consider injection of electrons from dielectrics due to secondary emission from ion bombardment and photoemission. The electron flux is then
\begin{equation}
  \bm{F}_e\cdot{\bm{n}} \rightarrow \bm{F}_e\cdot{\bm{n}} - \sum_{\sign(Z_i) > 0} \alpha_{i}\bm{F}_i\cdot \bm{n} - \sum_{\gamma } \beta_{\gamma}\bm{F}_\gamma\cdot \bm{n},
\end{equation}
where $\alpha_i = 10^{-6}$ is the second Townsend coefficient for species $i$, and $\beta_\gamma = 10^{-6}$ is the quantum efficiency for photon transition $\gamma$. The condition $\sign(Z_i) > 0$ indicates that only positively charged ions lead to secondary electron emission. 

\bibliography{manuscript}

\end{document}